\documentclass[onecolumn,iop,apj]{aastex63}
\usepackage{amsfonts,amsmath,graphicx,natbib,url,hyperref,nicefrac}
\usepackage{amssymb, verbatim, subfigure, paralist, soul, comment}
\usepackage{graphicx}
\usepackage{wrapfig}
\usepackage{verbatim}
\usepackage{tikz}
\usepackage{color}

\usepackage{overpic}
\usepackage{multirow}
\usepackage{float}

\usepackage[T1]{fontenc}
\usepackage{aecompl}
\usepackage{calligra}
\DeclareMathAlphabet{\mathcalligra}{T1}{calligra}{m}{n}
\DeclareFontShape{T1}{calligra}{m}{n}{<->s*[2.2]callig15}{}

\usepackage{ulem}
\usepackage{cancel}
\usepackage{hyperref}
\hypersetup{
    pdfnewwindow=true,      % links in new window
    colorlinks=true,       % false: boxed links; true: colored links
    linkcolor=blue,          % color of internal links
    citecolor=blue,        % color of links to bibliography
    filecolor=blue,      % color of file links
    urlcolor=blue           % color of external links
}

\def\apjl{ApJL}               % Astrophysical Journal, Letters
\def\apjs{ApJS}               % Astrophysical Journal, Supplement
\def\aap{A\&A}                % Astronomy and Astrophysics
\defcitealias{O99}{O99}
\defcitealias{KK07}{KK07}

\DeclareMathAlphabet{\mathcalligra}{T1}{calligra}{m}{n}
\DeclareFontShape{T1}{calligra}{m}{n}{<->s*[2.2]callig15}{}

\def\Mach{\mathcal{M}}

\hypersetup{colorlinks=true, urlcolor=blue, citecolor=blue}

\usepackage{color}

\defcitealias{Bromberg+2011ApJ}{B11}

\newcommand{\OK}[1]{\textcolor{red}{[OK]}}

\usepackage{comment}

\shorttitle{} 
\shortauthors{O'Neill et al.}

\begin{document}

\title[]{Gaseous Dynamical Friction on Elliptical Keplerian Orbits}

\author[0000-0002-1382-3802]{David O'Neill}
\affiliation{Niels Bohr International Academy, Niels Bohr Institute, Blegdamsvej 17, DK-2100 Copenhagen Ø, Denmark}
\email{david.oneill@nbi.ku.dk}

\author[0000-0002-1271-6247]{Daniel J. D'Orazio}
\affiliation{Niels Bohr International Academy, Niels Bohr Institute, Blegdamsvej 17, DK-2100 Copenhagen Ø, Denmark}
\email{daniel.dorazio@nbi.ku.dk}

\author[0000-0003-0607-8741]{Johan Samsing}
\affiliation{Niels Bohr International Academy, Niels Bohr Institute, Blegdamsvej 17, DK-2100 Copenhagen Ø, Denmark}

\author[0000-0001-8716-3563
]{Martin E. Pessah}
\affiliation{Niels Bohr International Academy, Niels Bohr Institute, Blegdamsvej 17, DK-2100 Copenhagen Ø, Denmark}
\affiliation{School of Natural Sciences, Institute for Advanced Study, 1 Einstein Drive, Princeton, NJ 08540, USA}

\begin{abstract}
    We compute the Gaseous Dynamical Friction (GDF) force experienced by massive perturbers on elliptical Keplerian orbits. In this paper, we investigate the density wake morphology, dynamical friction force, and secular orbital evolution for massive single perturbers as well as equal mass binaries embedded in an homogenous, static background flow. In all cases, the rate-of-change in semi-major axis is found to be negative (as expected), whereas the rate-of-change in eccentricity is negative for strictly-subsonic trajectories and positive for strictly-supersonic trajectories. Transonic orbits can experience both positive and negative torques during the course of an orbit, with some growing in eccentricity and others circularising. We observe all initial orbits becoming highly supersonic and eccentric (over sufficiently long timescales) due to a relentless semi-major axis decay increasing the Mach number and subsequent eccentricity driving. We compare our findings to previous studies for rectilinear and circular motion, while also making our data for orbital decay available.
\end{abstract}

\section{Introduction}
A massive body moving through a medium excites perturbations that carry away energy and momentum. The force exerted on the perturber as a result of the gravitational field produced by density perturbations in the medium is referred to as dynamical friction. This process has the potential to alter the motion of the perturber with important implications for a wide variety of astrophysical settings, including the formation of binary systems and their subsequent orbital decay.\\

Dynamical friction was first formally studied by \cite{Chandrasekhar_DF}, who considered a \emph{stellar collisionless} medium. 
{This classical formula has enjoyed many astrophysical applications and has been applied to model} satellites/dwarf galaxies orbiting inside of a more massive galaxy \citep{ColisionlessGalaxy,DFMagellanicClouds,NumericalSatelliteGalaxy,hashimoto2003circularize}, black holes inside of a young stellar cluster \citep{StarClusters}, globular clusters near a galactic center \citep{kim2003dynamical,mcmillan2003fate,kim2004dynamical} and planetary migration due to the interactions of planets with planetesimals \citep{Del_Popolo_2003}. Chandrasekhar's Dynamical Friction force model was extended to a \emph{gaseous collisional} medium by \cite{O99} (hereafter \citetalias{O99}) whose analytical results agree with numerical studies by \cite{1999ApJ...522L..35S}. The \citetalias{O99} model has therefore been widely utilised across a range of different astrophysical environments. For example, in protoplanetary disks \citep{ProtoplanetaryDisk,Planetesimals1,Planetesimals2}, young gas-enriched globular-clusters \citep{GDF_inGlobularClusters}, common envelopes of massive binary star systems \citep{CEGDF}, supermassive black hole (SMBH) seeds in early galaxies \citep{ETG} and formation of stellar-mass black hole binaries in gaseous disks surrounding active galactic nuclei \citep[AGN;][]{DiskChannel,BinaryFormationAGN,dotti2006laser,GDF_BinaryFormationMechanism,Rowan2023:2309.14433v1} to study so-called ``gas-captures'' and eccentric mergers of black hole binaries.\\

Both force models of \citetalias{O99} and \cite{Chandrasekhar_DF} are valid only for rectilinear, constant-velocity trajectories through a medium. {Many realistic astrophysical scenarios, however, should result from curvilinear, orbital induced motion.} Motivated by this, \cite{KK07} (hereafter \citetalias{KK07}) modelled the gravitational drag force acting on a perturber moving on a circular orbit. The main findings of \citetalias{O99} and \citetalias{KK07} are similar, with the drag force experienced in rectilinear motion being very similar to the azimuthal force component on a circular orbit. However, the circular trajectory also experiences a force in the radial direction, which has no analogue in the \citetalias{O99} model. Typically this radial force is subdominant ($F_r<F_\theta$) for low Mach numbers $\Mach<2.2$ and dominant at higher Mach numbers. We note that, on a circular orbit, the radial force does no work and thereby has no contribution to the 
change in binary semi-major axis. On an eccentric orbit, however, the radial force will contribute to both the change in semi-major axis and eccentricity. A detailed analysis of the evolution of an eccentric orbit experiencing GDF (as well as geometric hydrodynamical drag) was done by \cite{EccentricityEvolution} in an extended gaseous distribution with power-law density profile. Typically they found that supersonic GDF will tend to increase eccentricity whereas geometric drag will decrease it, however this is dependent on the density profile one chooses to use. While the aforementioned study models the evolution of eccentric orbits, they implement the \citetalias{O99} drag force model which is inherently inconsistent with an elliptical trajectory. Indeed such applications of the \citetalias{O99} GDF model to more complicated, non-rectilinear trajectories \citep[]{Planetesimals1,GDF_inGlobularClusters} is a common approach to modelling gas drag due to the simplicity of the O99 model. The aim of this paper is to formulate a consistent semi-analytical treatment of GDF on a bound, elliptical trajectory.\\

A series of single-perturber linear wakes can be superimposed to model the cumulative wake of multiple massive perturbers. This was done by \cite{DoublePerturbers} to model an equal mass, circular binary, in which they found each massive perturber to be dragged back by it's own wake and accelerated by it's companion's wake leading to the possibility of ``dynamical boosts''. In this study we follow a similar approach to create the wake of an equal mass, eccentric binary and thereby compute the gravitational drag force experienced on the binary, allowing us to model it's orbital evolution. This same approach can be extended to unequal mass binaries.\\

This paper is structured as follows: In Section \ref{Methods} we provide a brief overview of the analytical and numerical methods implemented for this study, followed by a presentation of our results in Section \ref{Results}. These include the density wakes created by elliptical trajectories, the drag force experienced over an orbit and the change in orbital semi-major axis, eccentricity and argument of pericenter. We apply these results to a set of initial test trajectories/binaries by integrating their evolution and comparing to previous force models. 
Finally we present our conclusions in Section \ref{Conclusion}.\\

\section{Methods}\label{Methods}
\subsection{Analytical Approach}
We begin by stating the general considerations involved in determining the response of a gaseous medium to a massive perturber moving on a prescribed trajectory. Following \citetalias{O99}, we assume the medium to be homogeneous, infinite, and static. This allows us to express the gas density and its velocity in the form: $\rho_{\mathrm{gas}}(t,\mathbf{x}) = \rho_0[1+\alpha(t,\mathbf{x})]$ and $\mathbf{v}_{\mathrm{gas}}(t,\mathbf{x}) = \mathbf{\beta}(t,\mathbf{x}) c_{\rm s}$. Here, $c_{\rm s} = \sqrt{\partial P_0/\partial \rho_0}$ is the adiabatic sound speed and 
$\rho_0$ and $P_0$ are the density and pressure of the homogeneous background gas density. Expanding the continuity and momentum equations to linear order in the perturbed quantities, $|\alpha|, |\beta| \ll 1$, we can decouple the $\alpha$ and $\beta$ dependencies to find
\begin{equation}
    \partial_t^2\alpha - c_\mathrm{s}^2\nabla^2\alpha = \nabla^2\Phi_{\mathrm{pert}}.
    \label{PerturbedFluid}
\end{equation}
Here, the gravitational potential seen by the medium satisfies the Poisson equation as sourced by the mass density of the perturber, i.e., 
 \begin{equation}
    \nabla^2\Phi_{\mathrm{pert}} = 4\pi G \rho_{\mathrm{pert}}(t,\mathbf{x}).
    \label{PoissonEq}
\end{equation}
The physical content of these equations describes the propagation of sound waves into the medium as excited by the perturber. Eq.~(\ref{PerturbedFluid}) admits a Green's function solution \citep[§6.6]{jackson1999classical} of the form 
\begin{equation}
    \alpha(t,\mathbf{x})=\frac{4\pi{G}}{c_\mathrm{s}^2}\int{dt'd^3x'}\left[\frac{\delta(t'-t+|x'-x|/c_{\rm s})}{4\pi|x'-x|}\right]\rho_{\mathrm{pert}}(t',\mathbf{x}')H(t').
    \label{GeneralAlphaSolution}
\end{equation}
Here, the Heaviside function $H(t')$ ensures that the perturber only influences the gaseous medium after the time $t=0$ allowing us to understand the formation of the wake structure.\\

This general solution determines the (dimensionless) wake structure $\alpha(t,\mathbf{x})$ if the density of the perturber $\rho_{\mathrm{pert}}$ is provided. The rectilinear trajectory of a point mass perturber with mass $m$ moving with constant speed $v$ as given by $\rho_{\mathrm{pert}} = m\delta(z-vt)\delta(x)\delta(y)$ was studied by \citetalias{O99} while \citetalias{KK07} focused on circular orbits with constant radius $r_{\rm p}$ and constant angular frequency $\Omega_{\rm p}$, for which $\rho_{\mathrm{pert}} = m\delta[r_{\rm p}(\theta-\Omega_{\rm p} t)]\delta(r-r_{\rm p})\delta(z)$.
\subsection{Elliptical Trajectories}
In what follows, we consider a point mass perturber moving on a prescribed bound-Keplerian trajectory with semi-major axis $a$ and eccentricity $0\leq e<1$. This trajectory is most conveniently described by employing an elliptic cylindrical coordinate system centered at the origin $\mathbf{x}=0$ such that Cartesian and elliptic coordinates are related via
\begin{equation}
    (x,y,z) = (a\cos{\theta},~ a\sqrt{1-e^2}\sin{\theta},~ z),
\end{equation}
where the angle $\theta$ is such that $\tan{\theta} = y/(x \sqrt{1-e^2})$.
This allows us to prescribe the world-line density of the perturber as
\begin{equation}
    \rho_{\mathrm{pert}}(t',\mathbf{x}') = m\delta(a'-a)\delta[a'(\theta-E(t))]\delta(z'),
    \label{Trajectory}
\end{equation}
with $E$ denoting the eccentric anomaly, which satisfies Kepler's Equation, 
\begin{equation}
    E(t)-e\sin{E(t)} = \Omega t,
    \label{Kepler}
\end{equation}
where $\Omega = \sqrt{G\mu/a^3}$ is the mean angular motion for a perturber under the gravitational influence of a point mass $\mu$. By inserting the perturber's world-line density given by Eq.~(\ref{Trajectory}) into the general Green's function solution, Eq.~(\ref{GeneralAlphaSolution}), we find an expression for the density perturbation $\alpha(t,\mathbf{x})$, 
\begin{equation}
    \alpha(t,\mathbf{x})=\frac{Gm}{c_\mathrm{s}^2}\int{dE'} \delta\left[E'-E(t_R)\right]\frac{H(t_R)}{|\mathbf{x}'-\mathbf{x}|},
\label{roots}
\end{equation}
where $t_{\rm R} := t-|\mathbf{x}'-\mathbf{x}|/c_{\rm s}$ is the retarded time. The integral can be computed 
by defining $F(E') = E'-E(t_{\rm R})$ and 
recalling that 
\begin{equation}
\delta\left[F(E')\right]
=
\delta\left[E'-E(t_{\rm R})\right]
= \sum_i \frac{\delta(E'-E_i)}{\left| \left.\frac{\partial F(E')}{\partial E'}\right|_{E'=E_i}\right|}
= \sum_i \frac{\delta(E'-E_i)}{\left|1-\left(\frac{\partial{E}(t_{\rm R})}{\partial{t}}\frac{\partial t}{\partial t_{\mathrm R}}\frac{\partial t_\mathrm R}{\partial{E'}}\right)_{E'=E_i}\right|}
=\sum_{i}\frac{\delta(E'-E_i)}{\left|1-\frac{\partial{E(t_\mathrm R)}}{\partial t}\frac{\partial\left|\mathbf{x}_i-\mathbf{x}\right|}{c_\mathrm{s}\partial{E_i}}\right|}.
\label{ExpandingTheDelta}
\end{equation}
In the denominator of Eq.~(\ref{ExpandingTheDelta}) we denote the retarded position $\mathbf{x}_i = (a,E_i)$ and the retarded time $t_r = t-|\mathbf{x}_i-\mathbf{x}|/c_\mathrm{s}$. The eccentric anomaly evaluated at the retarded time $E(t_r)$ must equal the root value $E_i$. Hence, we combine Eqs.~(\ref{roots}) and (\ref{ExpandingTheDelta}) to obtain
\begin{equation}
    \alpha(t,\mathbf{x}) = \frac{Gm}{ac_\mathrm{s}^2}\int dE' ~{\sum_{i}\frac{\delta(E'-E_i)}{\left|1-\frac{\partial{E_i}}{\partial t}\frac{\partial\left|\mathbf{x}_i-\mathbf{x}\right|}{c_\mathrm{s}\partial{E_i}}\right|}} \frac{H(t_\mathrm{R})}{\left|\mathbf{x}'/a-\mathbf{x}/a\right|} = \frac{Gm}{ac_{\mathrm{s}}^2}\sum_i \frac{1}{\left|1-\frac{\partial E_i}{\partial t}\frac{\partial|\mathbf{x}_i-\mathbf{x}|}{c_\mathrm{s}\partial E_i}\right|}\frac{H(t_\mathrm{R})}{\left|\mathbf{x}_i/a-\mathbf{x}/a\right|}.
\label{AlphaSolution}
\end{equation}
\begin{figure}
    \centering
    \includegraphics[width = 0.7\linewidth]{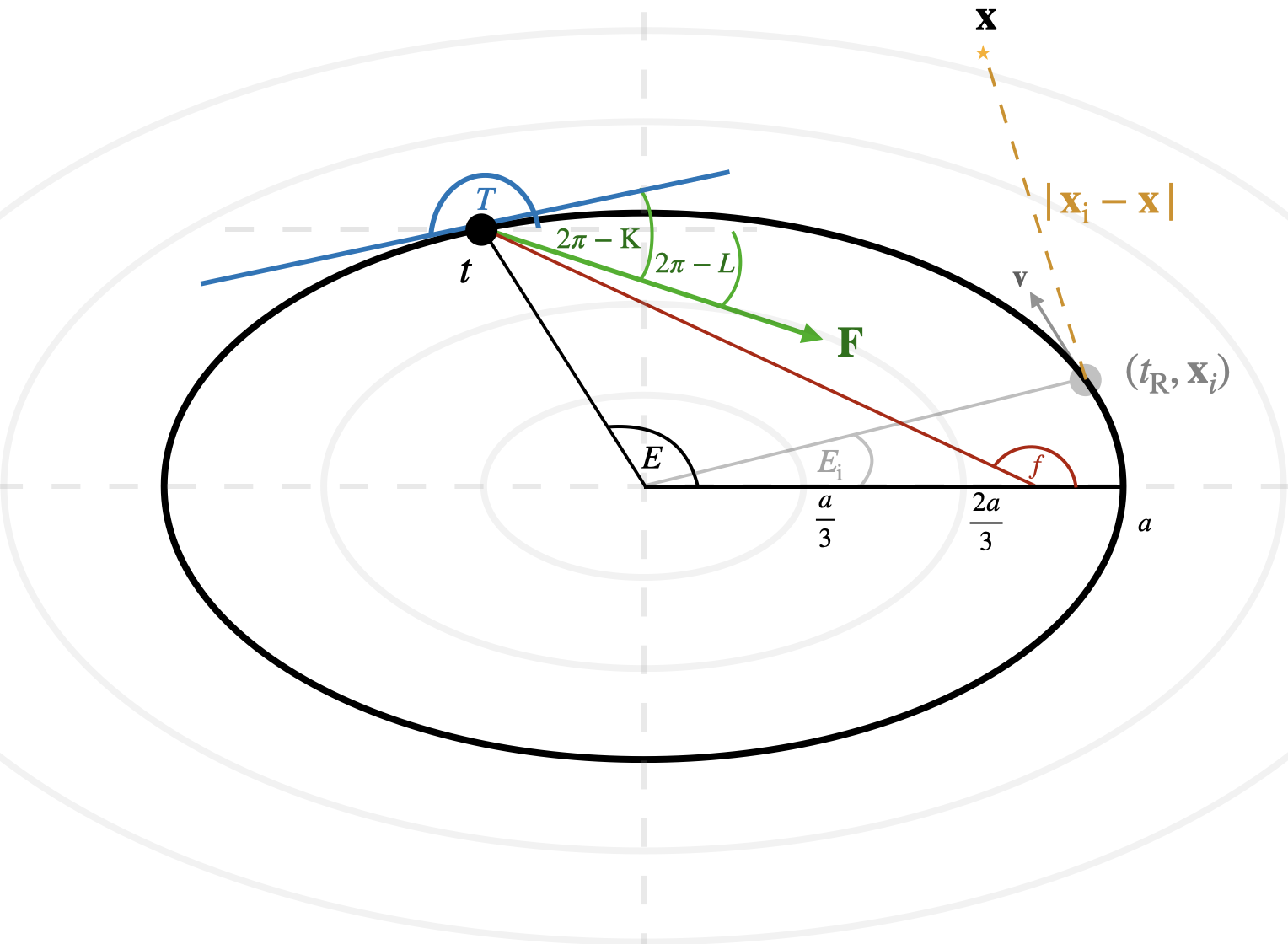}
    \caption{Schematic diagram illustrating the relevant coordinates and angles in our study. The grey lines represent the elliptical coordinate system, while the solid black ellipse is the orbital trajectory along which the perturber (black circle) is in motion. At any given point in the medium $\mathbf{x}$ we can consider the density perturbation as being sourced from a sound wave which was emitted at some earlier time $t_\mathrm{R}$ and position $\mathbf{x}_i$. The gravitational force exerted by these density perturbations on the perturber is illustrated by $\mathbf{F}$ (green). The blue tangent line, and the angles $T, K, G$ are necessary to coordinate transform the force into the relevant basis (Eqs.~\ref{CoordTransCartesian},~\ref{CoordTransParallelPerp}), while $E, f$ are the eccentric and true anomalies respectively, describing the current position of the perturber.}
    \label{GeneralPicture}
\end{figure}
\subsection{Secular Orbital Evolution Due to Gravitational Drag Force}
Given the structure of the density wake, it is desirable to understand the effects of the gravitational force exerted on the perturber. For an object moving on a Keplerian elliptical trajectory, with semi-major axis $a$ and mean angular motion $\Omega$, the time rate of change of the semi-major axis $\dot{a}$, orbital eccentricity $\dot{e}$ and precession of the longitude of pericenter $\dot\omega$ can be found via the perturbed orbital element equations from Eq.~(28) \cite{Orbital_Elements} and \cite{Erratum}
\begin{equation}
     \frac{1}{a}\frac{da}{dt} = \frac{2}{a\Omega \sqrt{(1-e^2)}}
     \times\left[\frac{{F}_r}{m}e\sin{f}+\frac{{F}_\mathrm{\theta}}{m}(1+e\cos{f})\right],
    \label{FirstBurnsA}
\end{equation}
\begin{equation}
     \frac{de}{dt} = \frac{\sqrt{1-e^2}}{{a\Omega}}\times\left[\frac{F_r}{m}\sin{f}+\frac{F_\theta}{m}(\cos{f}+\cos{E})\right].
    \label{FirstBurnsE}
\end{equation}
\begin{equation}
    \frac{d\omega}{dt} =  \frac{\sqrt{1-e^2}}{{a\Omega}e}\left[-\frac{{F}_r}{m}\cos{f}+\frac{{F}_\mathrm{\theta}}{m}\sin{f}\left(\frac{2+e\cos{f}}{1+e\cos{f}}\right)\right].
    \label{FirstBurnsOmega}
\end{equation}
Here, $F_r$ and $F_\theta$ are the radial and azimuthal components of the total drag force on the perturber when its true anomaly $f$ corresponds to the location of the perturber $\mathbf{x}_{\mathrm p}$, i.e.,
\begin{equation}
    \mathbf{F}_{\mathrm{GDF}} = Gm \rho_0\int d^3x \frac{\alpha(t,\mathbf{x})(\mathbf{x}-\mathbf{x}_\mathrm{p})}{|\mathbf{x}-\mathbf{x}_\mathrm{p}|^3}.
    \label{Integration}
\end{equation}
It proves to be convenient to compute directly the Cartesian components of the force $F_\mathrm{x}, F_\mathrm{y}$ and then coordinate transform into the polar/azimuthal basis $F_r, F_\mathrm{\theta}$ via
\begin{equation}
    \mathbf{F}_{\mathrm{GDF}} = F\cos(L-f)\hat{r}+F\sin(L-f)\hat{\theta},
    \label{CoordTransCartesian}
\end{equation}
where $F$ is the magnitude of the force and $L = \arctan(F_\mathrm{y}/F_\mathrm{x})$ is the angle made by the force vector as measured clockwise from the positive $x$-axis (Figure \ref{GeneralPicture}). In anticipation of creating rectilinear and circular proxy models, we also introduce the coordinate transformation from the basis parallel/perpendicular to the velocity vector, to the polar (radial/azimuthal) basis
\begin{equation}
    \mathbf{F}_{\mathrm{GDF}} = F\cos(T-f-K)\hat{r}-F\sin(T-f-K)\hat{\theta},
    \label{CoordTransParallelPerp}
\end{equation}
where $K = \arctan\left({F_\perp}/{F_{\scriptscriptstyle \parallel}}\right)$
is the angle the force vector makes with the tangent line and $T$ is the angle of the tangent line measured anti-clockwise from the positive $x$-axis such that it is always parallel to the velocity vector (see Fig.~\ref{GeneralPicture}).
\pagebreak

\subsection{Rectilinear and Circular Proxies}
\label{sec:proxies}
It is of interest to connect our findings to previous works that have employed the results obtained for rectilinear \citepalias{O99} and circular trajectories \citepalias{KK07} to describe perturbers moving on elliptical orbits. To do this, it is convenient to define what we refer to as ``proxy'' forces. These are the forces that can be computed from the results obtained by \citetalias{O99} and \citetalias{KK07} but considering that the Mach number involved is not fixed but depends on time (orbital location) as though the perturber were moving on an elliptical trajectory according to
\begin{equation}
    \Mach(t) 
    = \frac{a\Omega}{c_{\rm s}}  \sqrt{\frac{1+e\cos{E(t)}}{1-e\cos{E(t)}}}.
    \label{ProxyDefinitions}
\end{equation}
At each point in time, one can determine the steady state, Mach-number dependent drag force as predicted by \citetalias{O99} and \citetalias{KK07}. Note that the \citetalias{O99} and \citetalias{KK07} drag forces are known \emph{along the direction of motion}-- meaning when adapting to an elliptical orbit, these forces are understood to be in the tangential basis. Therefore we require Eq.~(\ref{CoordTransParallelPerp}) to express our forces in the polar basis. This procedure provides us with the radial and azimuthal force components of the \citetalias{O99} and \citetalias{KK07} force models in the case that $\Mach(t)$ varies along an elliptical orbit. We refer to these as the ``rectilinear'' and ``circular'' proxies for the drag force, respectively.

\subsection{Unit System and Dimensionless Variables}
\label{sec:Scalings}
In an elliptical trajectory, it is natural to use the semi-major axis $a$ to define the unit length and the reciprocal mean angular frequency $\Omega^{-1}$ to define the unit time. Because of our choice of units, distinct trajectories are specified only by the Mach number at pericenter $\Mach_p$ and the eccentricity $e$. We find the pericenter Mach number by evaluating Eq.~(\ref{ProxyDefinitions}) at $E(t) = 0$ giving \begin{equation}
    \Mach_p = \frac{a\Omega}{c_\mathrm{s}}\sqrt{\frac{1+e}{1-e}}.
\label{PericenterMachNumber}
\end{equation}  
In order to obtain the gravitational drag force, we first 
define the dimensionless density perturbation $\mathcal{D}(\Omega t, \mathbf{x}/a, e, \Mach_p)$ as
\begin{equation}
    \mathcal{D}(\Omega t, \mathbf{x}/a, e, \Mach_p) \equiv \sum_i \frac{1}{\left|1-\frac{\partial E_i}{\partial \left(\Omega t\right)}\frac{\partial|\mathbf{x}_i/a-\mathbf{x}/a|}{\partial E_i}\Mach_p\sqrt{\frac{1-e}{1+e}}\right|}\frac{H(\Omega t_\mathrm{i})}{\left|\mathbf{x}_i/a-\mathbf{x}/a\right|},
\label{CodeUnitDensityPerturbations}
\end{equation}
where we have used Eq.~(\ref{PericenterMachNumber}) to relate the Mach number on a circular orbit $a\Omega/c_\mathrm{s}$ to the pericenter Mach number on an eccentric orbit. For a given Mach number and eccentricity, we can use Eq.~(\ref{AlphaSolution}) to write the density perturbation $\alpha(t,\mathbf{x})$ as
\begin{equation}
\alpha(t,\mathbf{x})= \frac{Gm}{a c_\mathrm{s}^2}\times \mathcal{D}(\Omega t,\mathbf{x}/a, e, \Mach_p).
\label{AlphaDimensions}
\end{equation}
In this form, the drag force in Eq.~(\ref{Integration}) can be expressed as
\begin{equation}
   \mathbf{F}_{\mathrm{GDF}} = 4\pi \left(Gm \rho_0 a\right) \left(\frac{Gm}{a c_\mathrm{s}^2}\right) \mathbf{\mathcal{F}}(\Omega t, e, \Mach_p),
\label{ForceDimensions}
\end{equation}
with the dimensionless gravitational drag force $\mathbf{\mathcal{F}}$ defined as
\begin{equation}
    \mathcal{F}\equiv \frac{1}{4\pi a} \int d^3\mathbf{x}\frac{\mathcal{D}(\Omega t, \mathbf{x}/a,e,\Mach_p)}{\left|\mathbf{x}-\mathbf{x}_p\right|^3}\left(\mathbf{x}-\mathbf{x}_p\right).
    \label{DimensionlessForce}
\end{equation}
The physical scales relevant to GDF can be encoded into two dimensionless ratios, which we define as
\begin{equation}
    \tau \equiv \frac{G\rho_0}{\Omega^2},
    \label{TauDef}
\end{equation} 
\begin{equation}   
    \mathcal{A} \equiv \frac{Gm}{ac_\mathrm{s}^2}.
    \label{ADef}
\end{equation}
Firstly, we recognise $\sqrt{\tau}$ as the ratio between the perturber's orbital period and the gas' free fall timescale. This must be small to justify the assumption that the gaseous medium is not self-gravitating-- enabling the perturber to undergo long-term orbital evolution inside a homogeneous medium. Secondly, $\mathcal{A}$ can be considered as the non-linear parameter representing the ratio of the Bondi radius to the semi-major axis of the orbit, $a$. If this quantity is small, then the gas can be considered to be weakly perturbed on the scale of the semi-major axis. In what follows, we assume both $\tau, \mathcal{A} \ll 1$.\pagebreak

Naturally, this allows us to write the dimensionless changes in semi-major axis $a$, eccentricity $e$ and precession frequency $\omega$ from Eqs.~(\ref{FirstBurnsA}), (\ref{FirstBurnsE}) and (\ref{FirstBurnsOmega}) as
\begin{equation}
     \frac{1}{a}\frac{da}{d\left(\Omega t\right)} =4\pi\tau\mathcal{A}  
     ~\dot{\mathcal{I}}_a,
\label{EvolutionScalingsA}
\end{equation}
\begin{equation}
     \frac{de}{d(\Omega t)} = 4\pi\tau\mathcal{A} ~\dot{\mathcal{I}}_e,
\label{EvolutionScalingsE}
\end{equation}
\begin{equation}
     \frac{d\omega}{d(\Omega t)} = 4\pi\tau\mathcal{A} ~\dot{\mathcal{I}}_\omega,
\label{EvolutionScalingsOmega}
\end{equation}
where we have introduced the dimensionless quantities $\dot{\mathcal{I}}_a, \dot{\mathcal{I}}_e$, and $\dot{\mathcal{I}}_\omega$. Explicitly, these can be expressed as
\begin{align}
    \dot{\mathcal{I}}_a &= \frac{2}{\sqrt{1-e^2}}\left[\mathbf{\mathcal{F}}_re\sin{f}+\mathbf{\mathcal{F}}_{\mathrm{\theta}}(1+e\cos{f})\right],\\
    \dot{\mathcal{I}}_e &= \sqrt{1-e^2}\left[\mathbf{\mathcal{F}}_r\sin{f}+\mathbf{\mathcal{F}}_{\mathrm{\theta}}(\cos{f}+\cos{E})\right],\\
    \dot{\mathcal{I}}_\omega &= \sqrt{\frac{1-e^2}{e^2}}\left[-{\mathcal{F}_r}\cos{f}+{\mathcal{F}_\mathrm{\theta}}\sin{f}\left(\frac{2+e\cos{f}}{1+e\cos{f}}\right)\right].
    \label{IDefinitions}
\end{align}

\subsection{Numerical Considerations}
The force computation consists primarily of evaluating the density wake $\mathcal{D}(t,\mathbf{x})$ from Eq.~(\ref{CodeUnitDensityPerturbations}). The integral of $\mathcal{D}(t,\mathbf{x})$ via the trapezoidal rule provides us with the dimensionless gravitational drag force, as defined in Eq.~(\ref{DimensionlessForce}), which is coordinate transformed from the Cartesian to polar basis via Eq.~(\ref{CoordTransCartesian}). We choose to implement a logarithmic, spherical 3D grid centered at the perturber over which we integrate the wake\footnote{We puncture a small sphere of radius $r_{\mathrm{min}} = 0.1a$ around the perturber in line with previous studies \citepalias{KK07} to ensure the force is convergent and finite.}. As time is in units of the orbital period, we effectively change the Mach number by varying the sound speed. We note that the sound crossing time $t = 2a/c_\mathrm{s}$ is typically the timescale which determines when the force on the perturber reaches a  quasi-steady state (as it is expected to experience natural periodicity during any elliptical orbit). Therefore the time it takes for the force to reach steady state is different depending on the Mach number, requiring us to adapt the temporal and spatial grids over which we evaluate the force. For subsonic perturbers, the sound speed is larger than the orbital velocity and so the wake is smoother and larger in extent requiring a large spatial domain ($\sim 70 a$) to capture the non-negligible contributions from far away. In contrast, the supersonic wakes are much more local and discontinuous, requiring higher spatial resolution, yet only up to $\sim 8 a$ (Appendix~\ref{sec:Convergence}). We compute the dimensionless drag force $\mathcal{F}$ from Eq.~(\ref{DimensionlessForce}) between $20$ to $30$ different times per orbit (depending on the eccentricity and Mach number) to accurately capture the time evolution of the force over the course of an orbit. Given this data, we interpolate over time before evaluating Eqs.~(\ref{EvolutionScalingsA}),(\ref{EvolutionScalingsE}) and (\ref{EvolutionScalingsOmega}) to give the change in orbital elements for an orbit characterised by $(e,\Mach_p)$. We demonstrate numerical convergence of the orbital elements in time and spatial resolution in Appendix~\ref{sec:Convergence}.\\

It is necessary to find the eccentric anomaly as a function of time, for which we solve Eq.~(\ref{Kepler}) using the Newton-Raphson method. This allows us to find the roots to Eq.~(\ref{roots}) by employing a numerical bisection method.\footnote{We note that there also exist analytical methods for evaluating the density perturbations for the circular case \citep{Analytical}.} In Eq.~(\ref{CodeUnitDensityPerturbations}) these roots are then summed over (denoted by $i$) to find $\alpha(t,\mathbf{x})$. Our numerical method is simply parallelisable allowing us to optimise our code by splitting the integral domain into a number of independent sub-domains.\vfill
\pagebreak

\section{Results}\label{Results}
\subsection{Gas Wake Morphology}
\label{SingleDensities}
\begin{figure}[H]
\centering
\begin{overpic}[width=.295\textwidth]{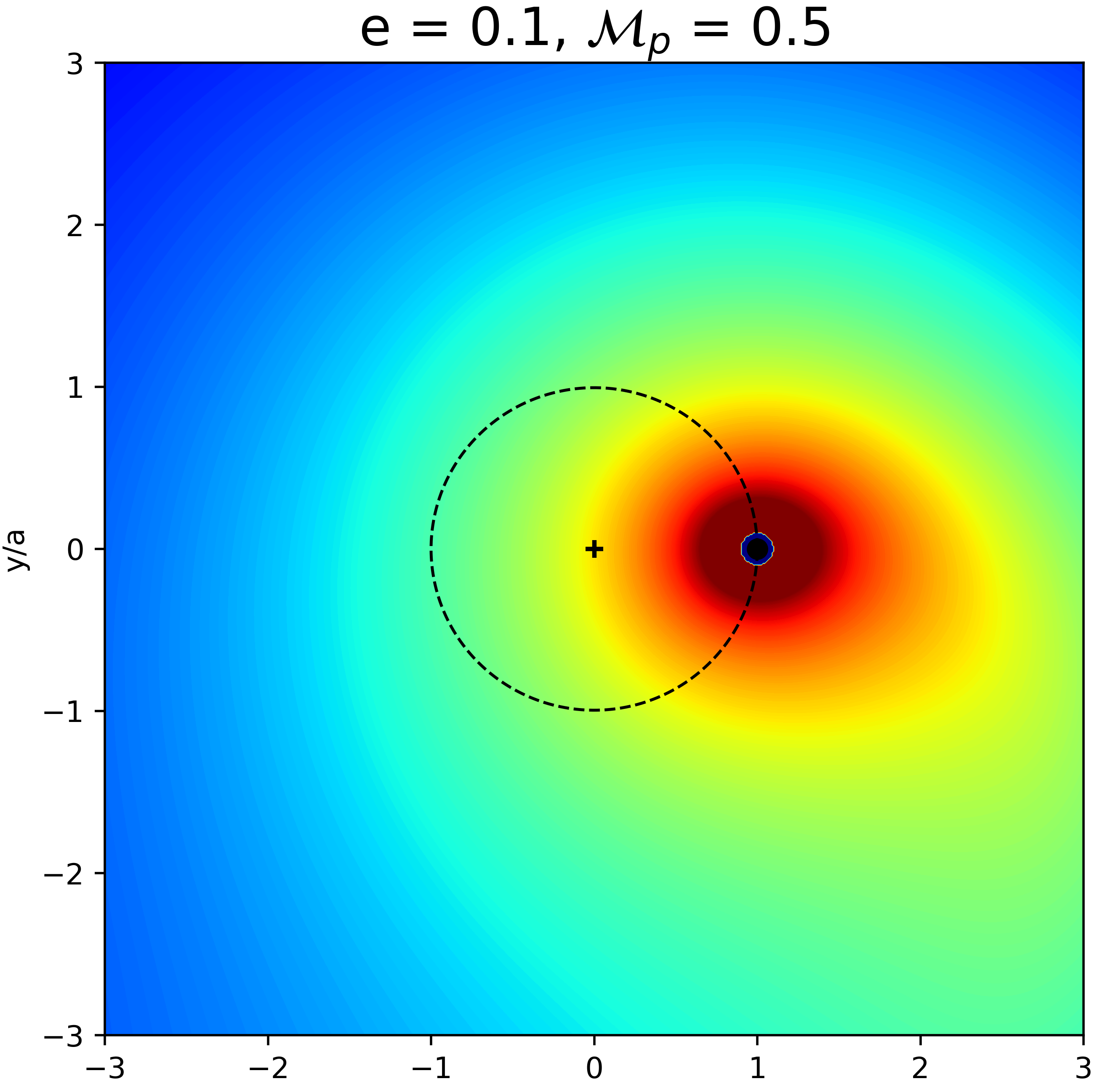}
  \put(36,86){\makebox(0,0){\color{white}\textbf{(a) -- Subsonic}}}
\end{overpic}\quad
\begin{overpic}[width=.288\textwidth]{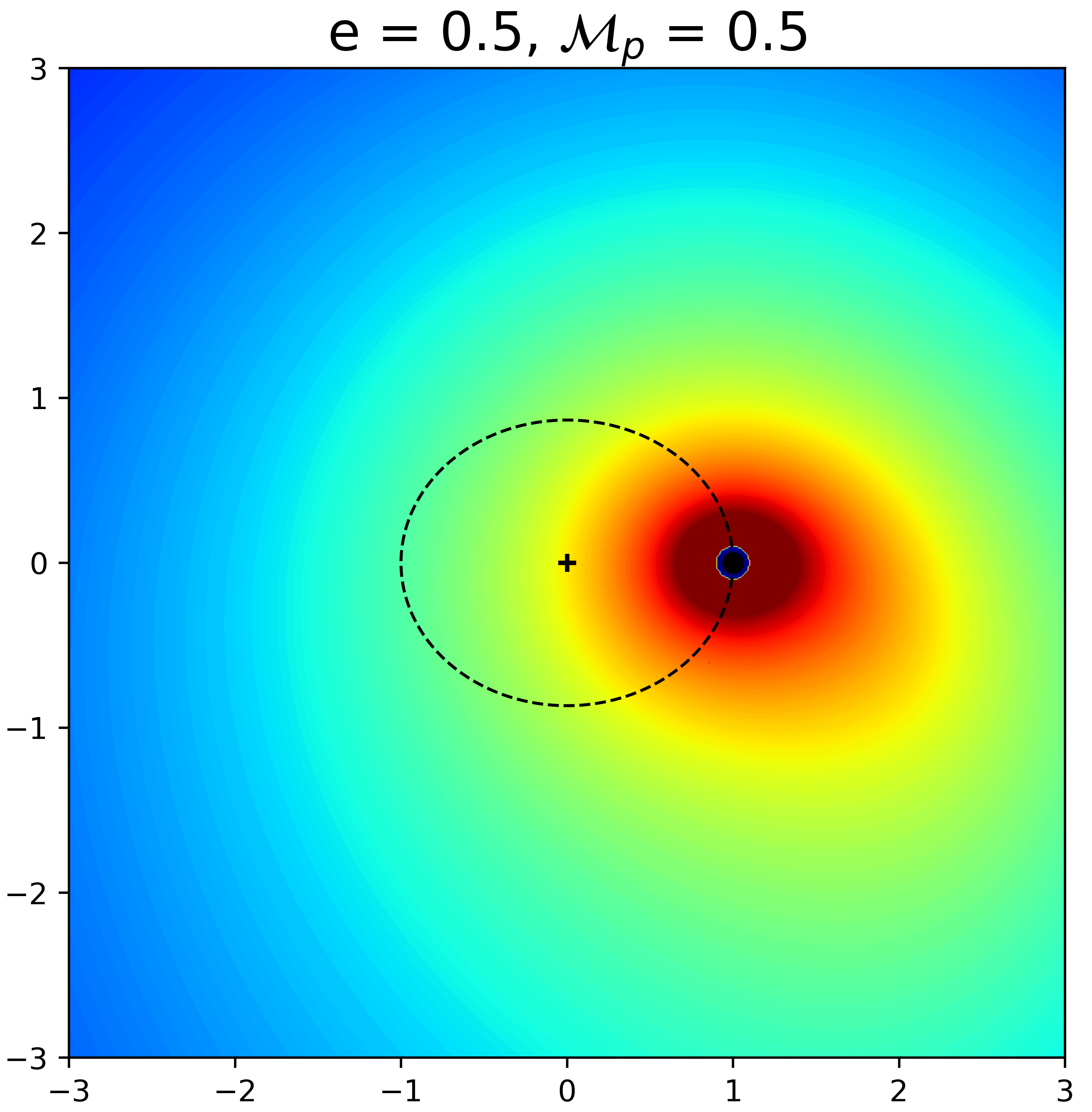}
  \put(33,87){\makebox(0,0){\color{white}\textbf{(b) -- Subsonic}}}
\end{overpic}\quad
\begin{overpic}[width=.356\textwidth]{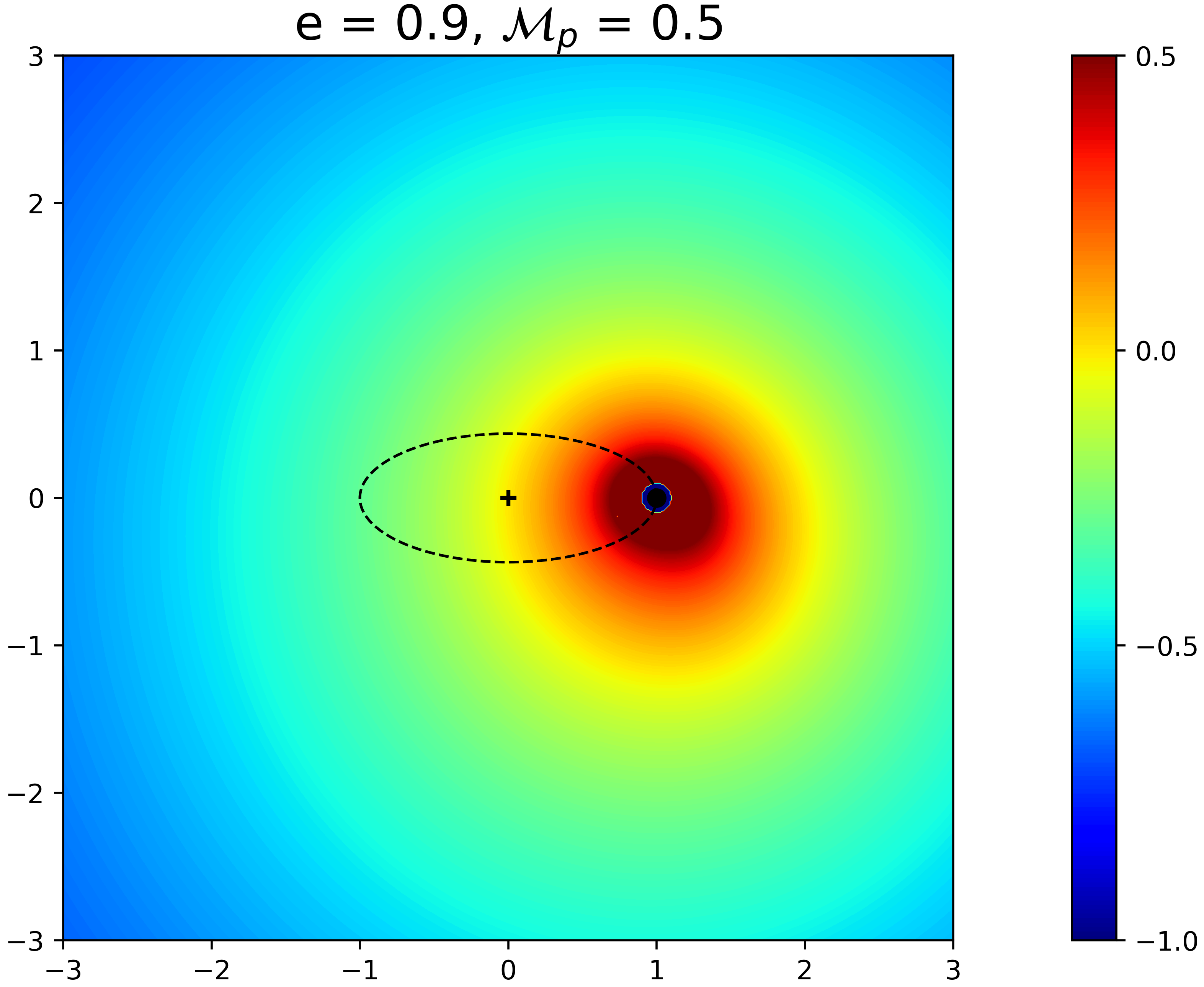}
  \put(29,72){\makebox(0,0){\color{white}\textbf{(c) -- Subsonic}}}
\end{overpic}
\medskip
\begin{overpic}[width=.3\textwidth]{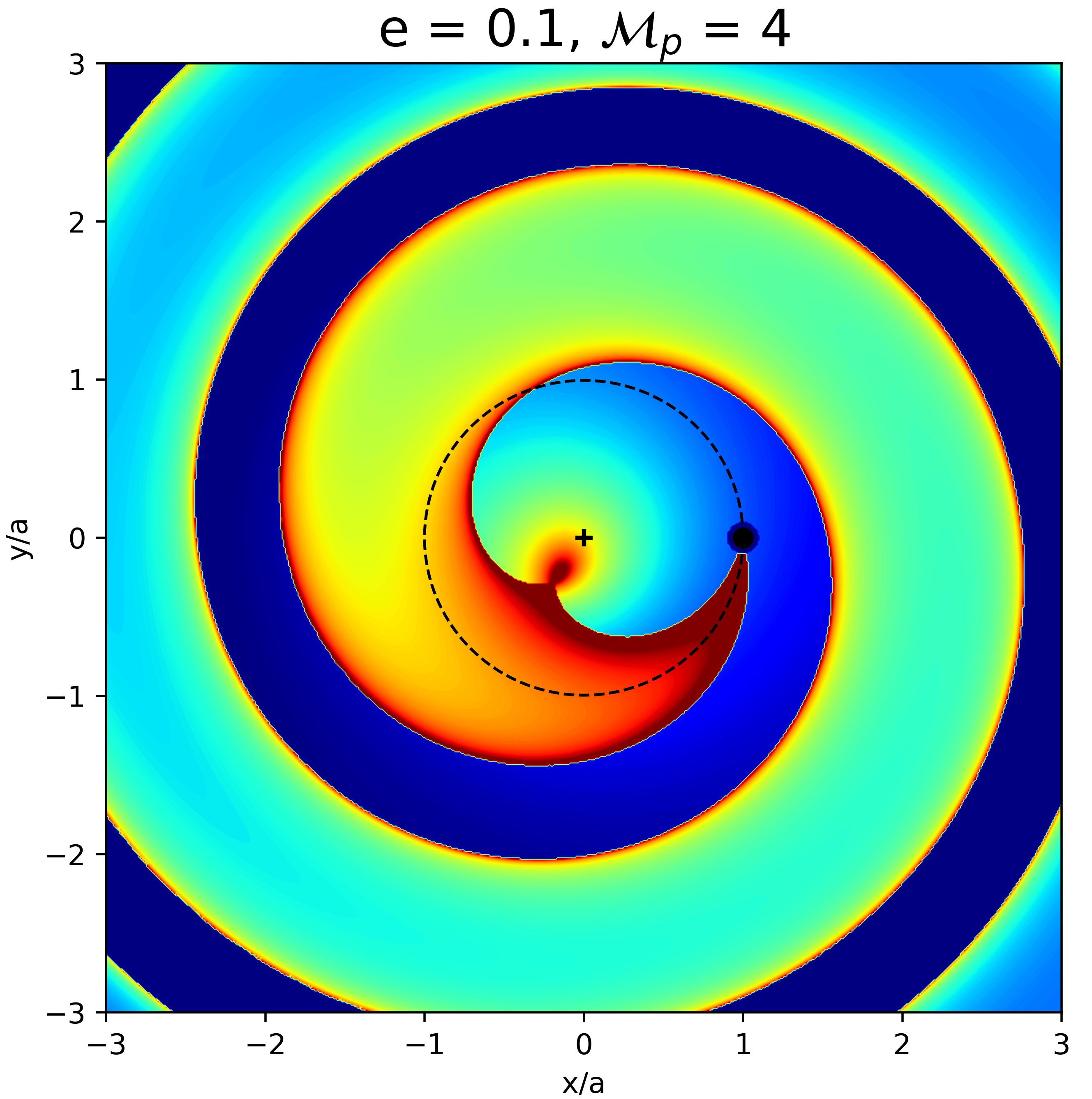}
  \put(39,87){\makebox(0,0){\color{white}\textbf{(d) -- Supersonic}}}
\end{overpic}\quad
\begin{overpic}[width=.284\textwidth]{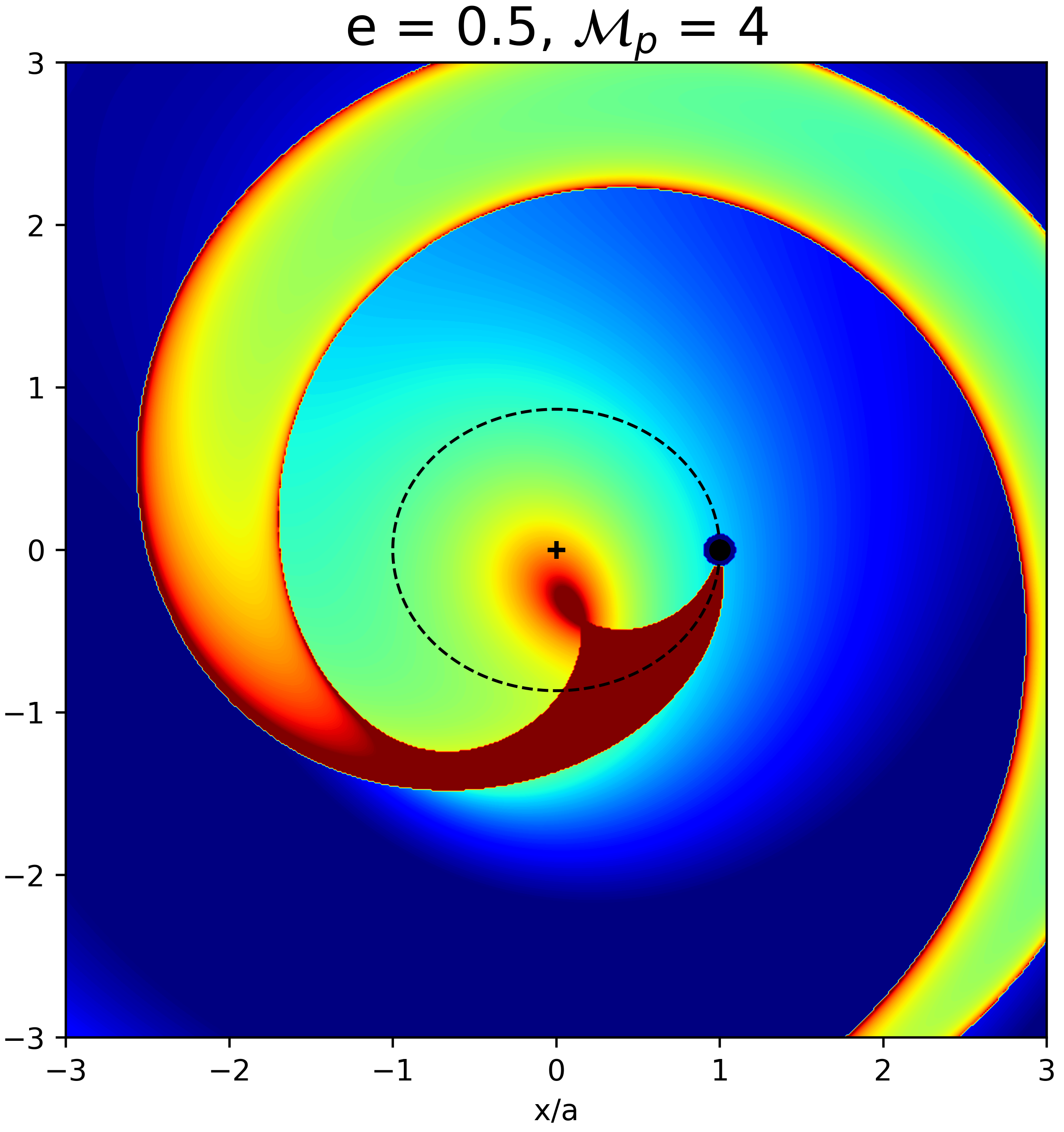}
  \put(33,87){\makebox(0,0){\color{white}\textbf{(e) -- Supersonic}}}
\end{overpic}\quad
\begin{overpic}[width=.354\textwidth]{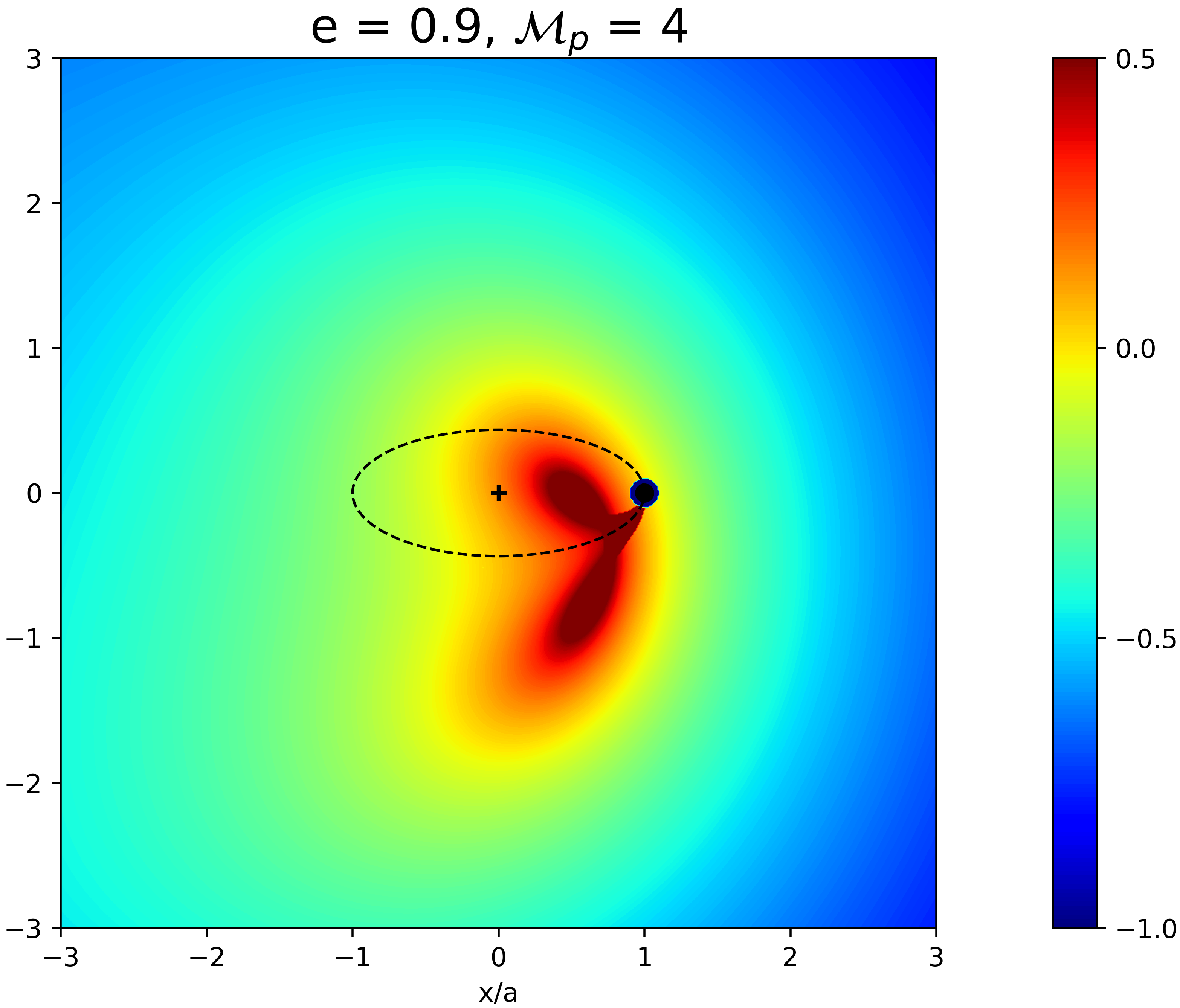}
  \put(29,76){\makebox(0,0){\color{white}\textbf{(f) -- Transonic}}}
\end{overpic}
\caption{A logarithmic plot of density perturbations $\mathcal{D}(t,\mathbf{x}) = \alpha(t,\mathbf{x})/\mathcal{A}$ at pericenter for a perturber moving with pericenter Mach numbers $\mathcal{M}_p = 0.5$ (top) $\mathcal{M}_p = 4$ (bottom). The dashed black line corresponds to the fixed orbital path of the perturber and is centered at $x = 0, y = 0$ (black `$+$'). We see sharp density discontinuities for supersonic motion, while subsonic motion leads to a more continuous density perturbation profile. The figures above are plotted over a Cartesian domain with regular grid spacing of $0.01a$}
\label{Densitypericenter}
\end{figure}

\begin{figure}[h]
\centering
\begin{overpic}[width=.295\textwidth]{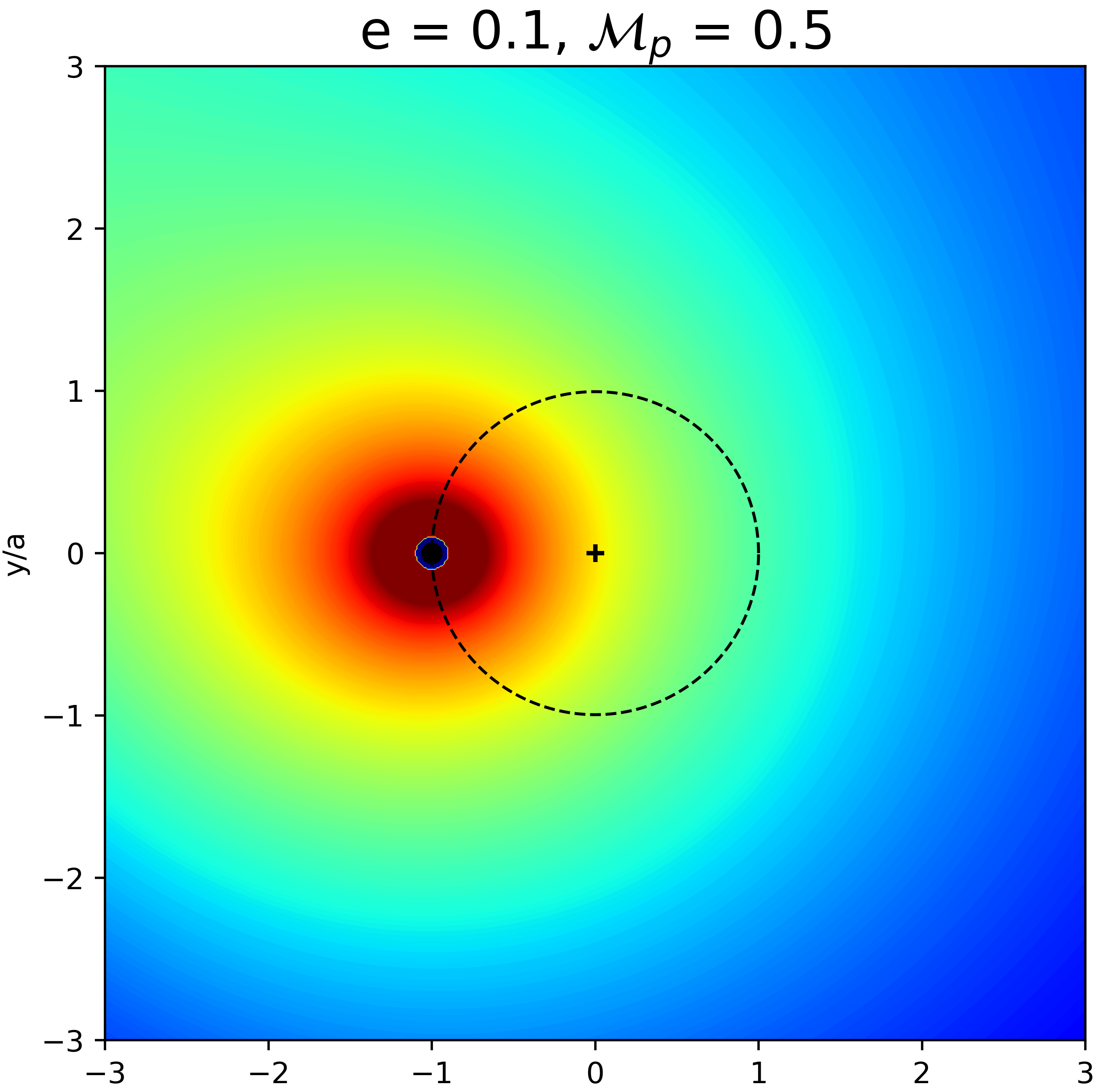}
  \put(37,86){\makebox(0,0){\textbf{\color{white}(a) -- Subsonic}}}
\end{overpic}\quad
\begin{overpic}[width=.285\textwidth]{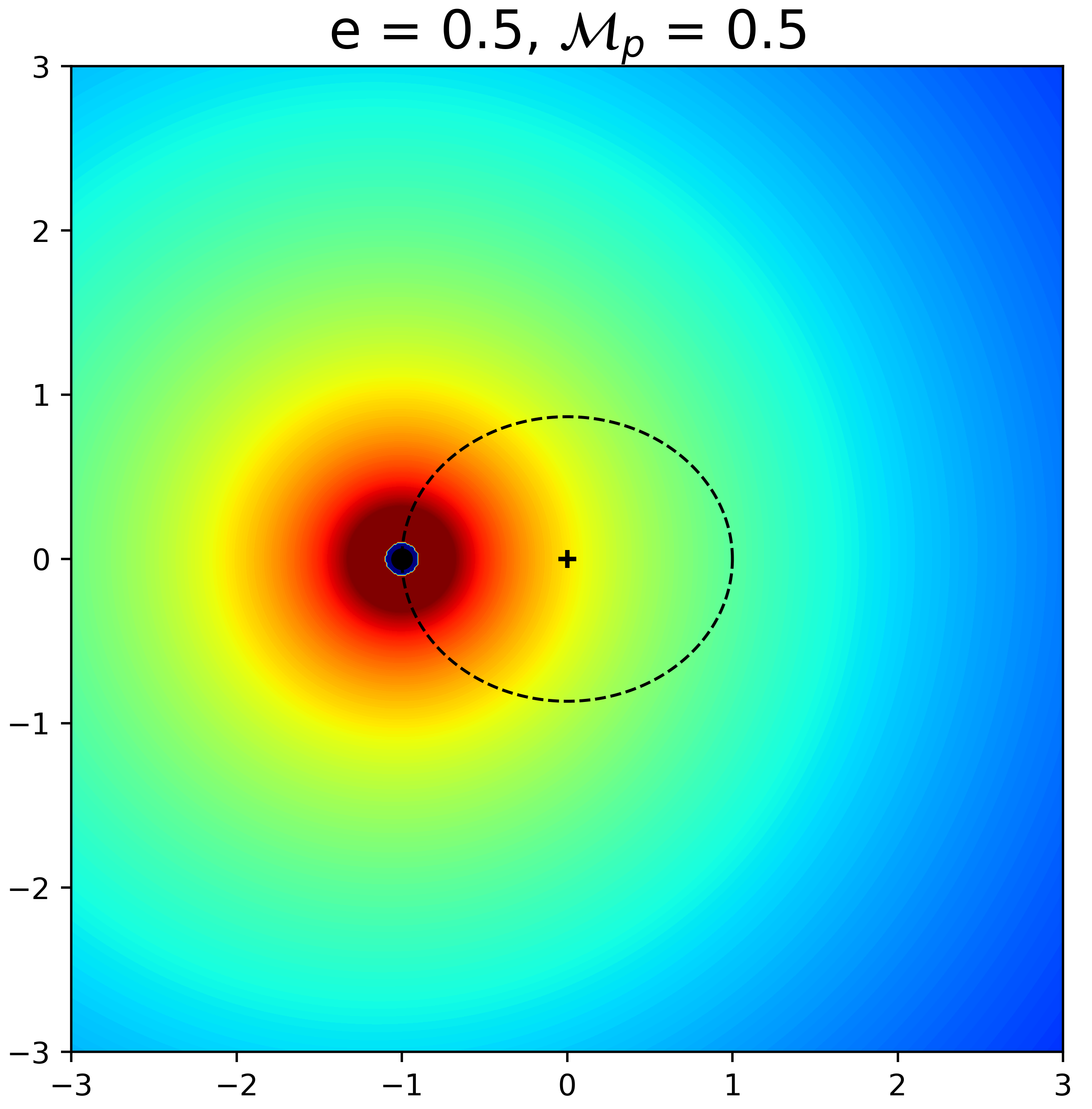}
  \put(36,87){\makebox(0,0){\textbf{\color{white}(b) -- Subsonic}}}
\end{overpic}\quad
\begin{overpic}[width=.352\textwidth]{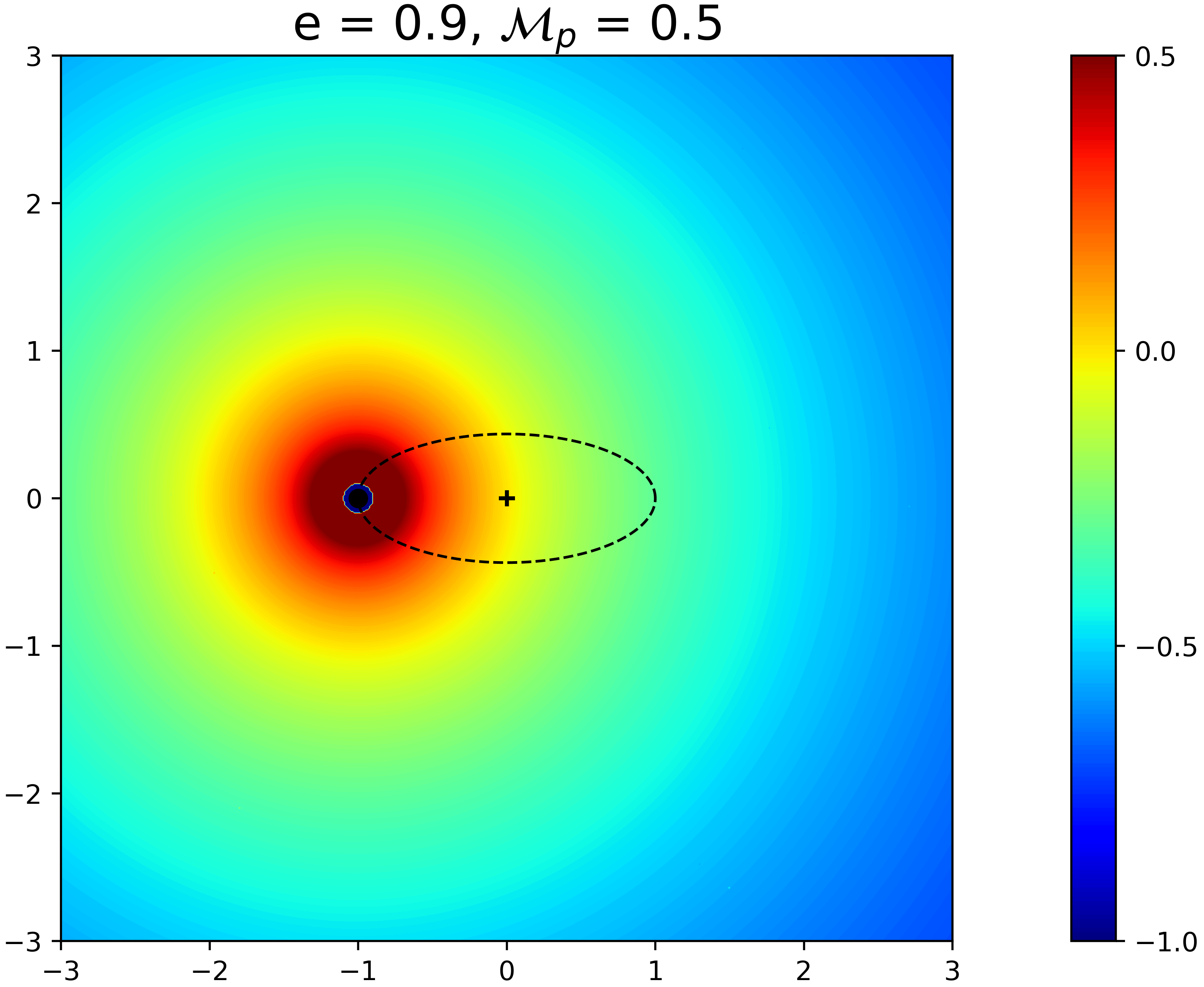}
  \put(32,72){\makebox(0,0){\textbf{\color{white}(c) -- Subsonic}}}
\end{overpic}
\medskip
\hspace{1mm}\begin{overpic}[width=.298\textwidth]{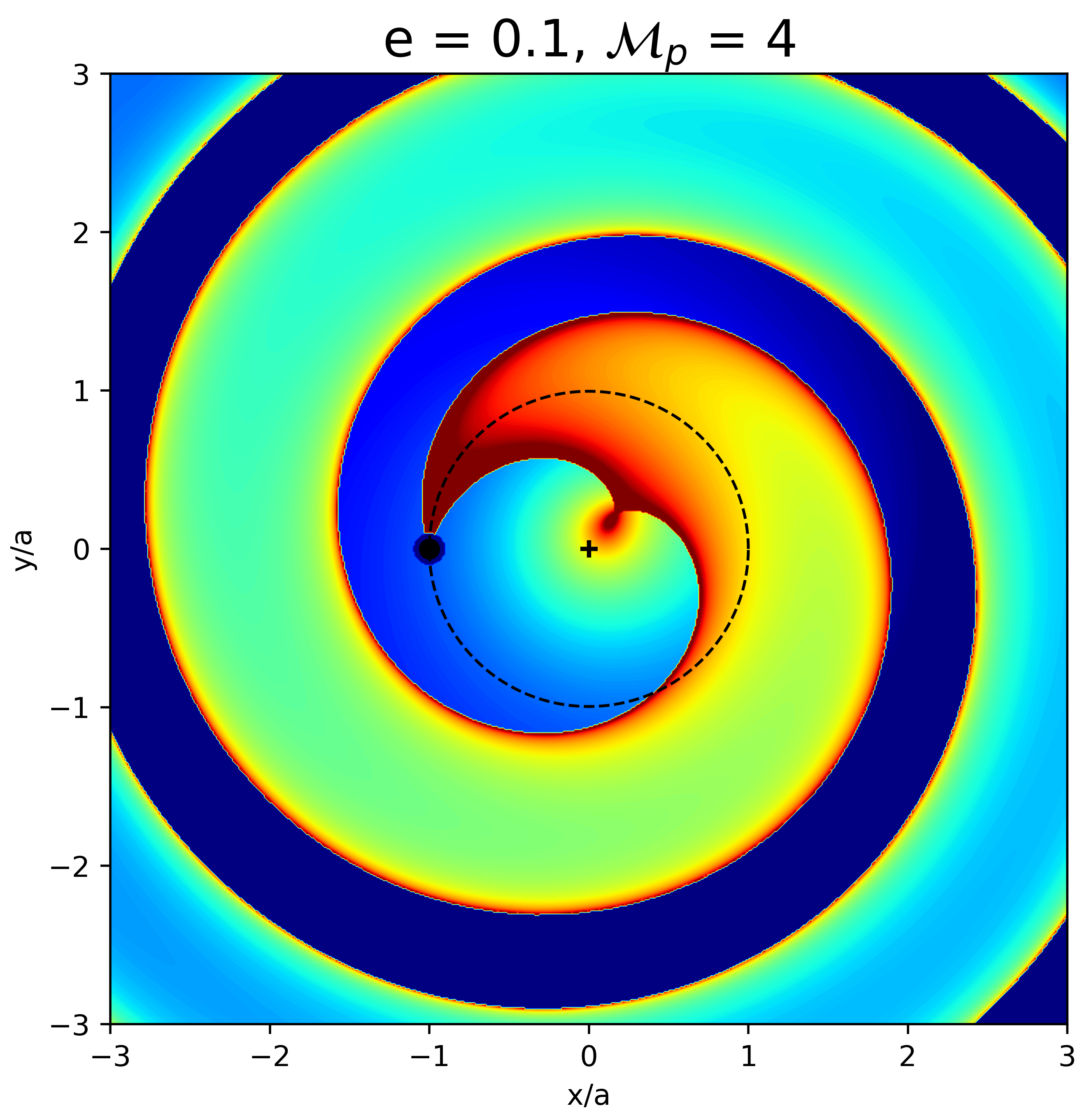}
  \put(39,87){\makebox(0,0){\textbf{\color{white}(d) -- Supersonic}}}
\end{overpic}\quad
\begin{overpic}[width=.285\textwidth]{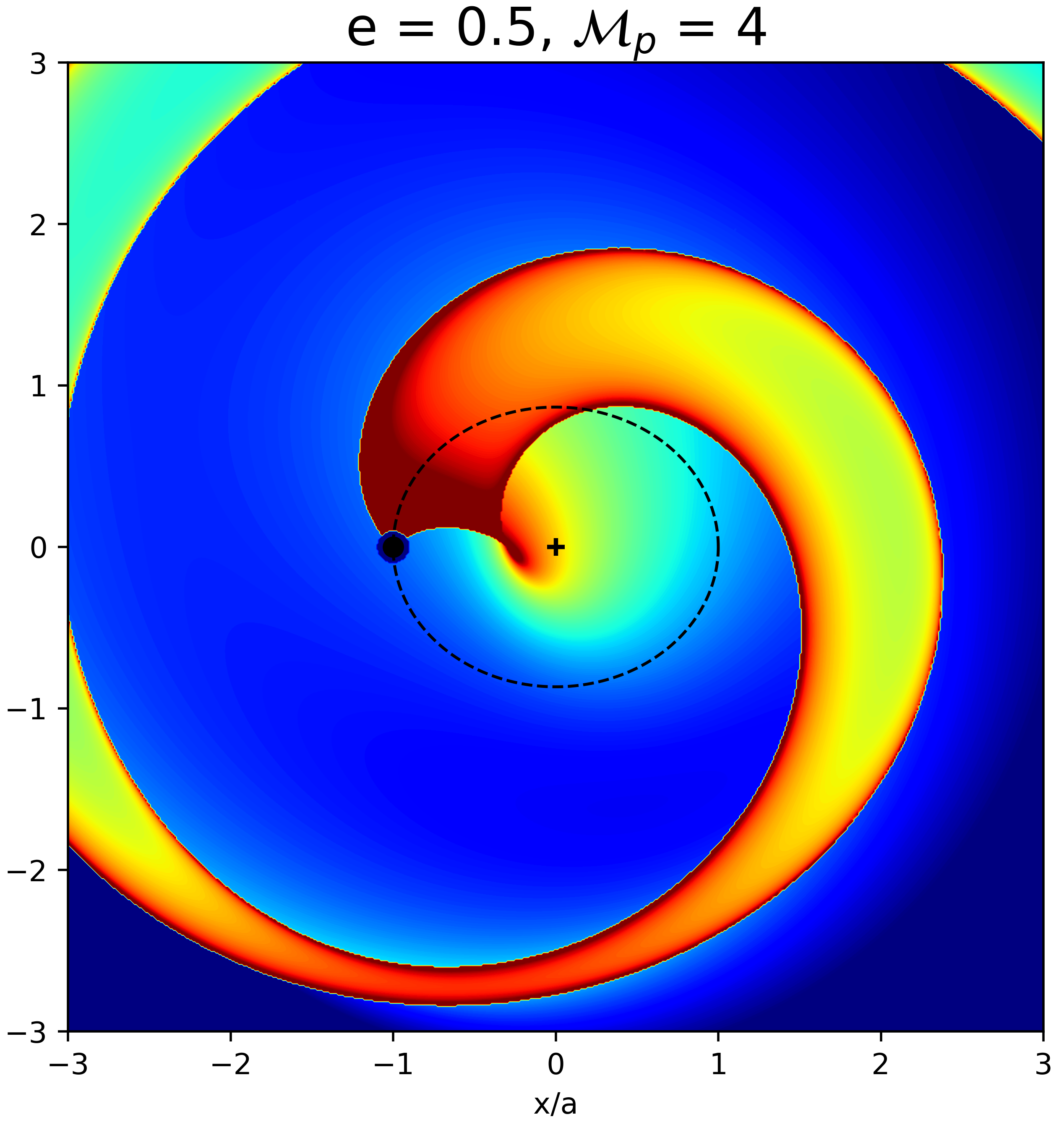}
  \put(33,88){\makebox(0,0){\textbf{\color{white}(e) -- Supersonic}}}
\end{overpic}\quad
\begin{overpic}[width=.354\textwidth]{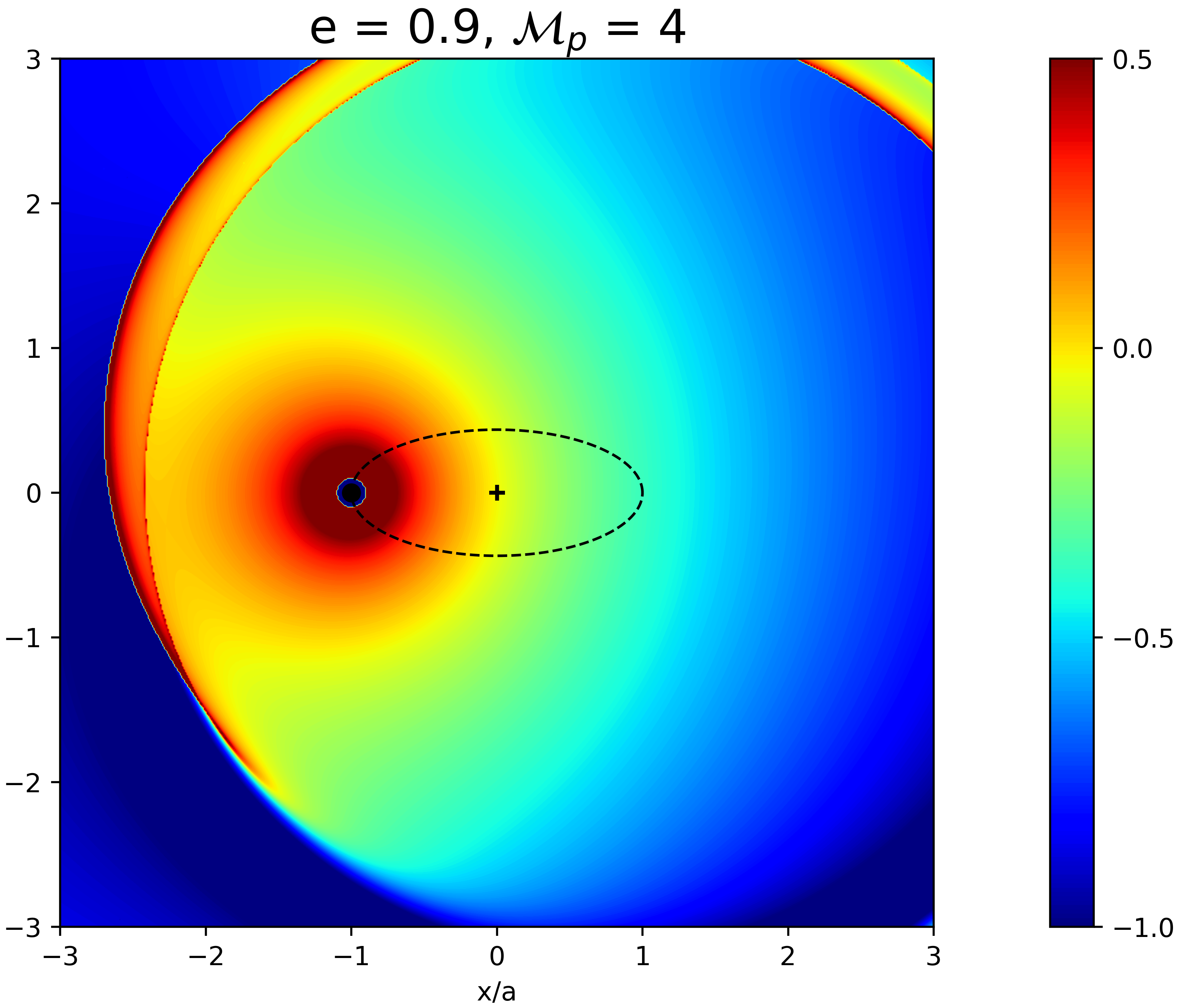}
  \put(29,76){\makebox(0,0){\textbf{\color{white}(f) -- Transonic}}}
\end{overpic}
\caption{The same as Fig.~\ref{Densitypericenter} except here the wake snapshots are taken at apocenter. At apocenter the lower mach number has the effect of: (a) widening the Mach cone and producing a broader tail for supersonic perturbers and (b) producing more spherically symmetric density perturbations around the subsonic perturber. }
\label{Densityapocenter}
\end{figure}

Circular trajectories are either subsonic or supersonic, however, elliptical orbits can exhibit transonic behaviour. In order to build intuition before computing forces and orbital evolutions, we first analyse the morphology of the wakes. We present snapshots of our solutions for the gas wake structure for a range of different Mach numbers and eccentricities. We show a slice of the orbital plane ($z=0$) with pericenter at $x/a = 1, y/a = 0$ in of each of the panels in Fig.~\ref{Densitypericenter}, while in Fig~\ref{Densityapocenter} we show the same systems at their apocenter approach. Each orbit belongs in one of three regimes; strictly subsonic (a, b, c), strictly supersonic (d, e) or transonic (f). It is in the transonic case where the most striking feature occurs, the appearance of an advancing shock front. The orbital speed of an elliptical trajectory changes over the course of an orbit, with a minimum at apocenter and maximum at pericenter. This creates the possibility of the sound speed lying between the two in which case there will be two sound barrier crossings per orbit. The bottom right panels of Fig.~\ref{Densitypericenter} and \ref{Densityapocenter} illustrate this situation. In Fig.~\ref{Densitypericenter}, the perturber is supersonic at pericenter and exhibits a Mach cone density wake, while in Fig.~\ref{Densityapocenter} the perturber is moving subsonically on the same orbit at apocenter and tails behind an expanding, high-density sonic wave. The perturber's deceleration across the sound barrier creates a thin, dense shell propagating at the speed of sound and is emitted once per orbit. This effect is due to the tangential acceleration (acceleration in the direction of the motion) of a perturber in a gaseous medium and, therefore, has not been present in the studies of rectilinear \citepalias{O99} or circular motion \citepalias{KK07}.\footnote{Accelerated rectilinear motion has been studied by \cite{2010MNRAS.401..319N}. The gravitational drag force was found to be weaker than uniform rectilinear motion \citepalias{O99} around $\Mach \sim 1$, but larger for $\Mach\gtrsim 1.5$.}\\
In the subsonic cases of Figures \ref{Densitypericenter} and \ref{Densityapocenter} the wake structure becomes more symmetric for higher eccentricities. The wake created by a perturber on rectilinear motion is comprised of concentric ellipsoids with eccentricities $e_{\mathrm{wake}}= \mathcal{M}$ as seen in \citetalias{O99}. This suggests that the wakes are comprised of concentric thin ellipsoidal shells with varying eccentricities and densities, depending on the Mach number of the perturber at the location where the perturbations were emitted. Therefore at any given point, the wake's local structure (the dark red region) will be determined by it's Mach number at that point in the orbit. For example in the supersonic panels of Fig.~\ref{Densitypericenter} the Mach number is higher at pericenter, creating a more eccentric local wake structure which we can observe in the dark red region bulging/elongating. We note that, as orbital eccentricity increases, the perturber spends less time at pericenter and more time at apocenter causing the wake structure to more closely resemble that of a slowly moving perturber. For highly eccentric subsonic orbits, the majority of the wake has been emitted from an orbital location where the Mach number was very low, thereby the density contours will be more circular and exhibit a greater symmetry. The only asymmetry comes from a very short time during the pericenter passage when the Mach number is larger, and elliptical density contours are produced.\\

\begin{figure}[t]
    \centering
    \includegraphics[width = 0.34\linewidth]{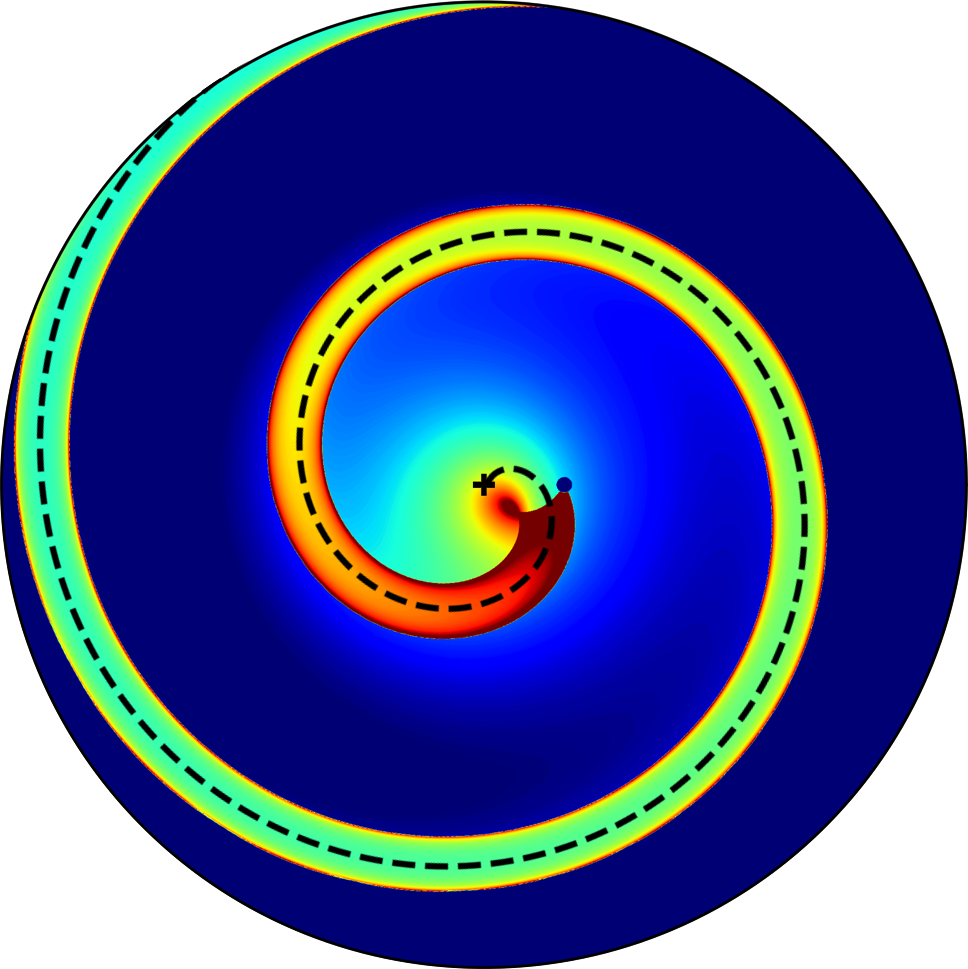}
    \hspace{15mm}
    \includegraphics[width = 0.45\linewidth]{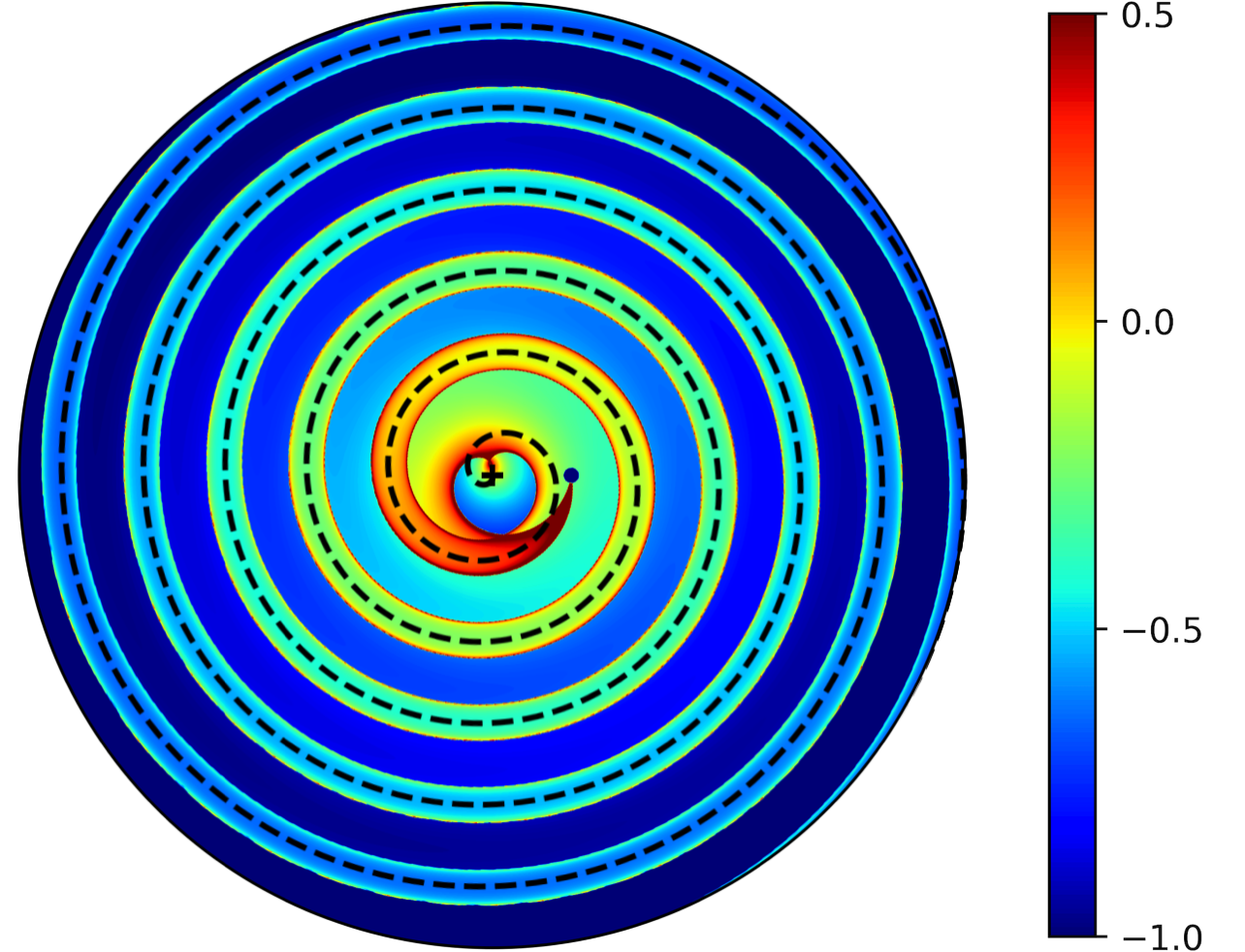}
    \put(-400,195){\rotatebox{0}{$\mathrm{e} = 0, \mathcal{M}_p = 2$}}
    \put(-170,195){\rotatebox{0}{$\mathrm{e} = 0, \mathcal{M}_p = 6$}}
    \put(-75,134){\textcolor{white}{$6$}}
    \put(-305,134){\textcolor{white}{$6$}}
    \put(-55,195){\rotatebox{0}{$\log_{10}\mathcal{D}(t,\mathbf{x})$}}
\caption{Archimedean spirals observed in the wake structure for circular, supersonic perturbers. Both figures are plotted over the same spatial domain of radius $6a$, demonstrating that larger Mach numbers create more tightly wound density wake tails.
}
\label{ArchimedeanSpiral}
\end{figure}
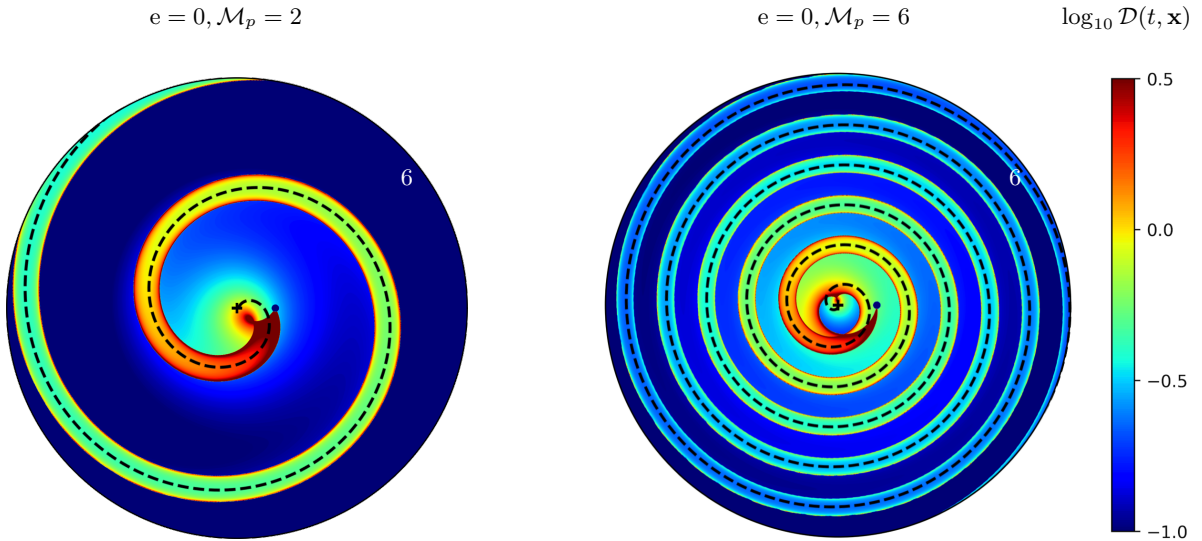

In the supersonic orbits of Figures~\ref{Densitypericenter},~\ref{Densityapocenter} the wake structure is more symmetric for lower eccentricities. In fact, the wake created by a circular trajectory can be fitted exactly by an Archimedean spiral. We illustrate this for two circular, supersonic trajectories in Fig.~\ref{ArchimedeanSpiral}. There are indeed parallels between a perturber moving with constant angular speed, emitting sound waves that propagate outward, and an Archimedean spiral delineating the trajectory of a point moving radially outward, along a line rotating with a constant angular speed. For a given domain of radius $r_\mathrm{d}$, the winding number the spiral created by a perturber on a circular orbit can be expressed in terms of the Mach number
\begin{equation}
    N_\mathrm{wind} = \frac{\Mach_p r_\mathrm{d}}{4\pi^2}
    \sim \frac{1}{c_\mathrm{s}}.
\label{WindingNumber}
\end{equation}

\pagebreak

\begin{figure}[t]
    \centering
    \includegraphics[width = 0.39\linewidth]{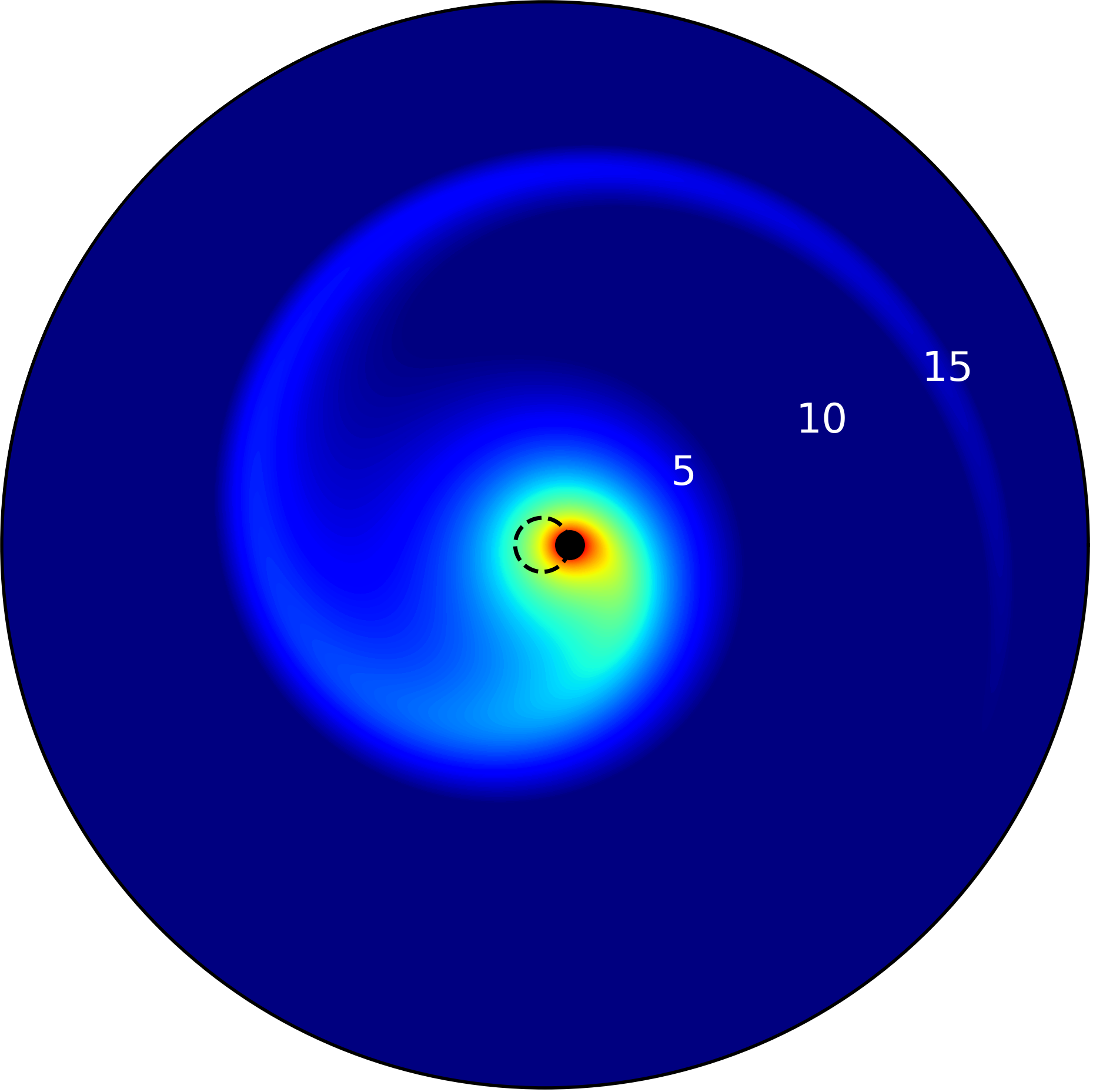}\hspace{15mm}
    \includegraphics[width = 0.5\linewidth]{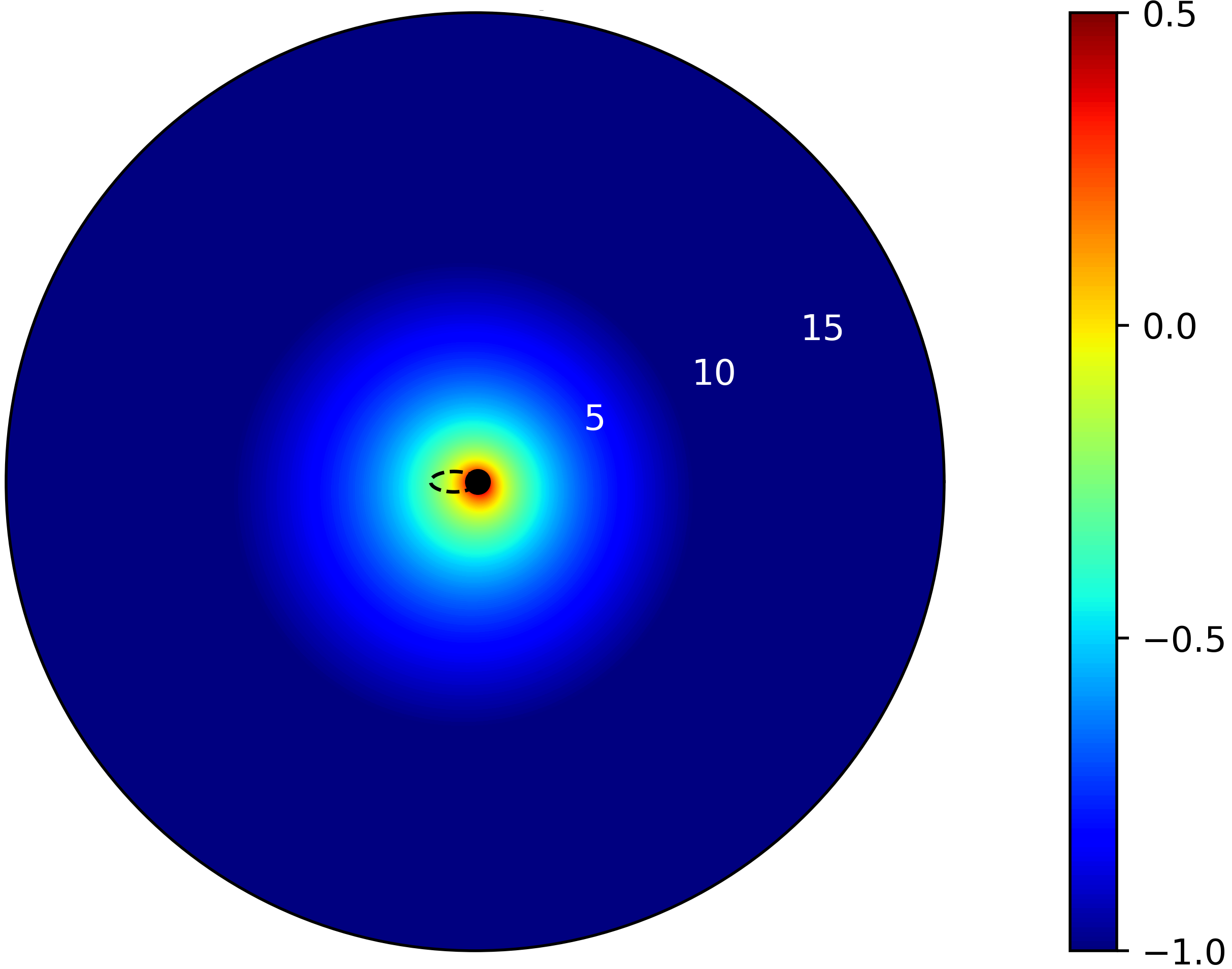}
    \put(-68,215){\rotatebox{0}{$\log_{10}\mathcal{D}(t,\mathbf{x})$}}
    \put(-435,215){\rotatebox{0}{$\mathrm{e} = 0.1, \mathcal{M}_p = 0.5$}}
    \put(-195,215){\rotatebox{0}{$\mathrm{e} = 0.9, \mathcal{M}_p = 0.5$}}

\caption{ 
Logarithmic plots of the dimensionless density perturbation $\mathcal{D}(t,\mathbf{x}) = \alpha(t,\mathbf{x})/\mathcal{A}$, over large spatial domain. There is a clear density tail in the low eccentricity trajectory, which is lacking in the highly eccentric one. These figures correspond to the upper left and upper right panels of Fig.~\ref{Densitypericenter} extended out to a radius of $20a$ with resolution of $0.02a$ in radius and $\pi/120$ in azimuth.}
\label{WiderSubsonic}
\end{figure}

For the near circular case, both \cite{2001MNRAS.322...67S} and \cite{NearCircular3dHydro} found a trailing spiral tail to be present in 3D numerical simulations. However, for the highly eccentric cases the motion of the perturber can be considered approximately one-dimensional, from which no asymmetry would be expected perpendicular to the direction of motion-- and hence no trailing spiral tails. Consequently, the highly eccentric, subsonic trajectory must display a more symmetric outer wake. In Fig.~\ref{WiderSubsonic} we illustrate the wakes created by two perturbers with eccentricities $e = 0.1$ and $0.9$, both with Mach numbers $\Mach_p = 0.5$. The wake structure far away from the perturber confirms the trailing spiral density tail to be present in the near circular trajectory and absent in the highly eccentric one.\\

Thus far we have introduced only a single perturber, on a fixed orbit, without any background gas motion -- and thus a single time/frequency scale. Therefore it may be surprising to observe spiral density waves, a typically resonant phenomenon \citep{SpiralWaveResonances}. The background gas is at rest, so there are neither angular nor epicyclic frequencies to induce Lindblad resonances-- instead the kinematic density waves can be attributed to the aligning and overlapping of density contours. This is because at each point in the orbit the perturber creates density perturbations which are eccentric ellipsoids oblate to the direction of motion. As the direction of motion changes, these shells precess and over time they grow in size. Similarly to Fig.~6.12 in \cite{GalacticDynamics}, this series of precessing ellipsoids creates a non-axisymmetric spiral pattern (with the spiral being where the density contours overlap). In contrast, when the motion is highly elliptical, concentric contours have very little relative orientation to each other, and expand more concentrically leaving no spiral tail (right panel, Fig.~\ref{WiderSubsonic}).\\

Near the origin we can see high density lobes for supersonic trajectories (the bottom left and bottom middle panels of Figs.~\ref{Densitypericenter},\ref{Densityapocenter}). This effect is caused by the curvature of the orbit where sound waves travelling in opposite directions overlap with each other. The position of this lobe is observed to vary over the course of an orbit dependent on the Mach number and will overlap with the perturber exactly when $\Mach = 1$. For a strictly supersonic trajectory it is always interior to the orbit, but transonically it becomes launched as part of the shock front when the trajectory becomes subsonic. Between this lobe and the current position of the perturber, we see a curved Mach cone (eg. (d) and (e) in Figs.~\ref{Densitypericenter},\ref{Densityapocenter}). We might naively expect this to be true for the entire wake due to the curvature of the orbit, however, beyond this localized region, the cone structure disappears and the wake has an almost constant width -- starkly different to a rectilinear motion perturber. In the steady-state, rectilinear motion Mach cone, there are always two causal roots $E_+$ and $E_-$ corresponding to sound waves propagating backwards ($E_+$) and forwards ($E_-$). On a bounded orbit, the leg of the Mach cone interior to the orbit is shorter than the leg on the exterior meaning earlier-launched sound waves $E_-$ will overshoot this region leaving no rear wide Mach cone interior to the orbit.\\
\\

\subsection{Force Evolution Over a Period}
\label{sec:ForceOverPeriod3.2}
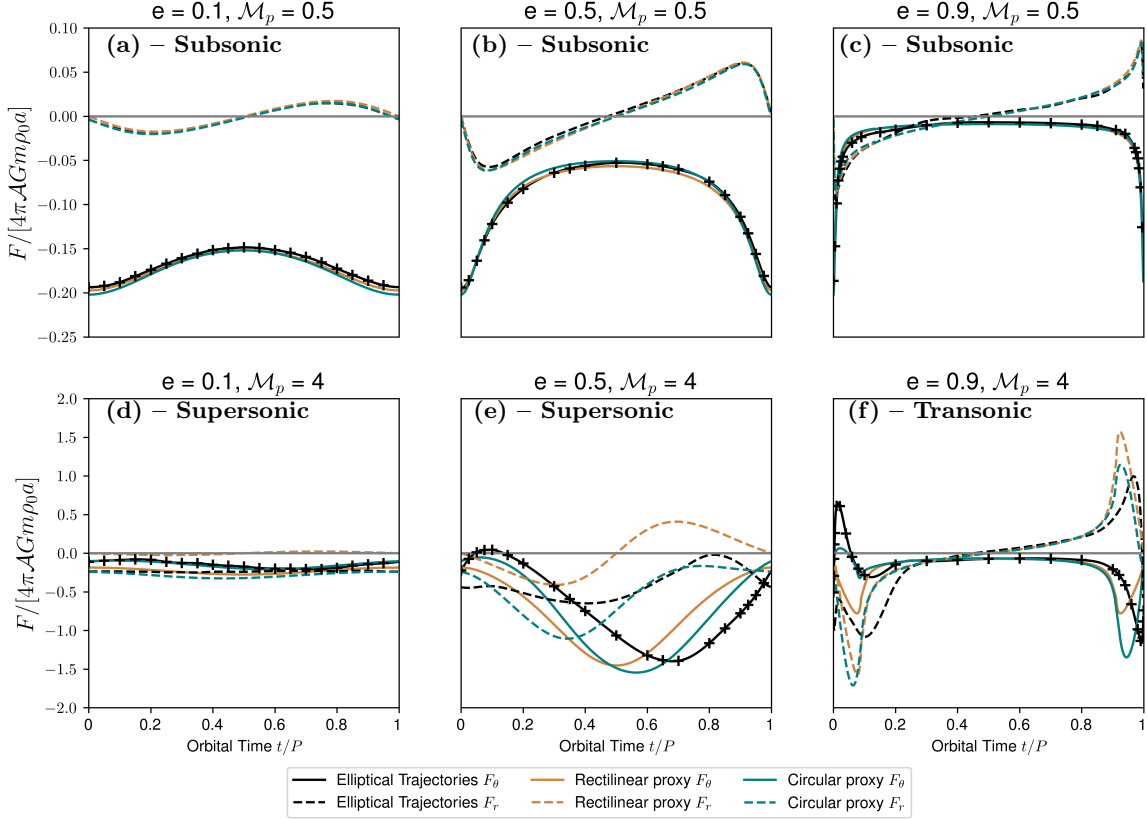
\begin{figure}[H]
\centering
\begin{overpic}[width=\textwidth]{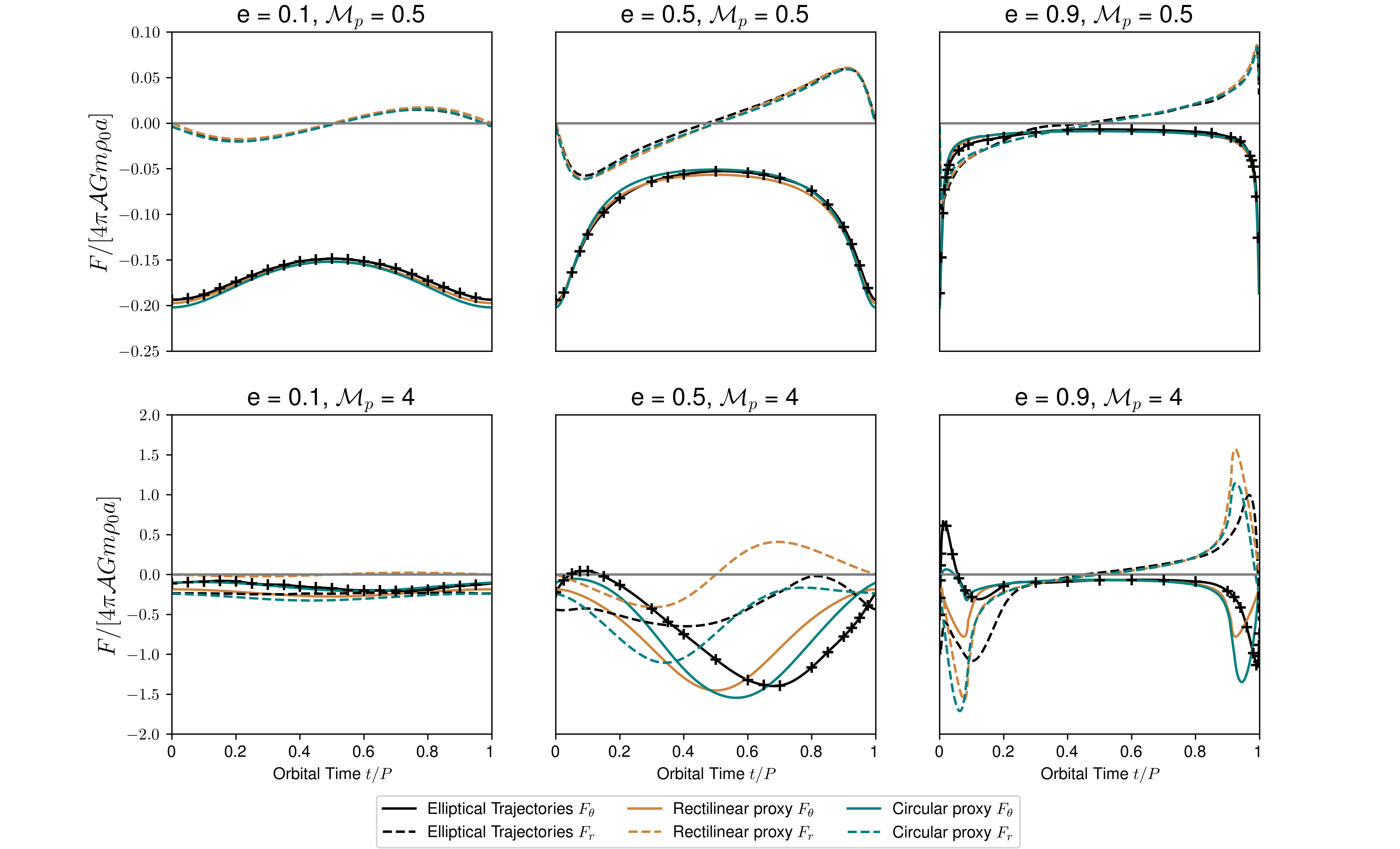}
  \put(20,57){\makebox(0,0){\textbf{(a) -- Subsonic}}}
  \put(47,57){\makebox(0,0){\textbf{(b) -- Subsonic}}}
  \put(74,57){\makebox(0,0){\textbf{(c) -- Subsonic}}}
  \put(21,30){\makebox(0,0){\textbf{(d) -- Supersonic}}}
  \put(48,30){\makebox(0,0){\textbf{(e) -- Supersonic}}}
  \put(75,30){\makebox(0,0){\textbf{(f) -- Transonic}}}
\end{overpic}
\vspace{-4mm}
\caption{The variation of the drag force on the perturber over the course of an orbit for the same cases as in Figs.~\ref{Densitypericenter} and \ref{Densityapocenter} corresponding to subsonic (a, b, c), supersonic (d, e) and transonic (f) trajectories. Solid lines indicate the interpolated azimuthal component of the force and dashed lines correspond to interpolated radial components. We compare our forces with the rectilinear (orange) and circular (teal) proxy forces as described in Section \ref{sec:proxies}. In this figure, the positive y-axis describes a driving force in the direction of motion, while a negative y-axis describes a drag force. We have included a `$+$' over the azimuthal curve to illustrate our time-sampling with $t/P=0$ corresponding to pericenter.} 
\label{ForceTime}
\end{figure}

An instantaneous force is computed at each time using Eq.~(\ref{Integration}), given a pericenter Mach number and eccentricity. A smooth and robust characterisation of the force over the course of an orbit requires convergence in the number of time samples, and good spatial resolution throughout the orbit (see Appendix \ref{sec:Convergence} for details). We perform an interpolation of the force over time and compare our results to the rectilinear and circular proxy values (Section \ref{sec:proxies}) in Fig.~\ref{ForceTime}.\\

There is good agreement between our findings and the proxies for subsonic trajectories (a, b and c of Fig.~\ref{ForceTime}), while for transonic and supersonic trajectories with intermediate eccentricities (panels e and f of Fig.~\ref{ForceTime}), there are noticeable differences from the proxy models. For example, after pericenter passage, the azimuthal component of the force is significantly positive in panel (f), which is not captured in the proxy forces. In the limit of high orbital eccentricity, some discrepancies might be expected, as the trajectory experiences significant tangential acceleration which stands in contrast to the steady-state proxy models. In Fig.~\ref{ForceTime}, the forces exhibit a slight delay in the peaks/dips from what is seen in the rectilinear and circular proxy forces. This is because the proxies effectively jump from one Mach number steady-state force to the next Mach number steady state, implying an instantaneous change in the wake structure. In reality, accelerations take time to propagate outwards creating a time delay before the perturber actually feels this affect from the wake, which is evident in Fig.~\ref{ForceTime}. In summary, while the proxy forces are accurate for subsonic and circular trajectories, they are less accurate for eccentric transonic and eccentric supersonic trajectories.\\

\subsection{Period-Averaged Change in Orbital Elements}
\label{sec:OrbitalEvolutionSection}
\begin{figure}[htp]
    \centering
    \vspace{3mm}
    \hspace{1mm}\textbf{\fontsize{13}{21.6}\selectfont $\frac{\dot{a}}{a}~ \Big[4\pi \tau \mathcal{A} \Omega\Big]$}
    \hspace{35mm}\textbf{\fontsize{13}{21.6}\selectfont $\dot{e}~ \Big[4\pi \tau \mathcal{A} \Omega\Big]$}
    \hspace{35mm}\textbf{\fontsize{13}{21.6}\selectfont $\dot{\omega}~ \Big[4\pi \tau \mathcal{A} \Omega\Big]$}\par\medskip
        
    \includegraphics[width=0.323\textwidth]{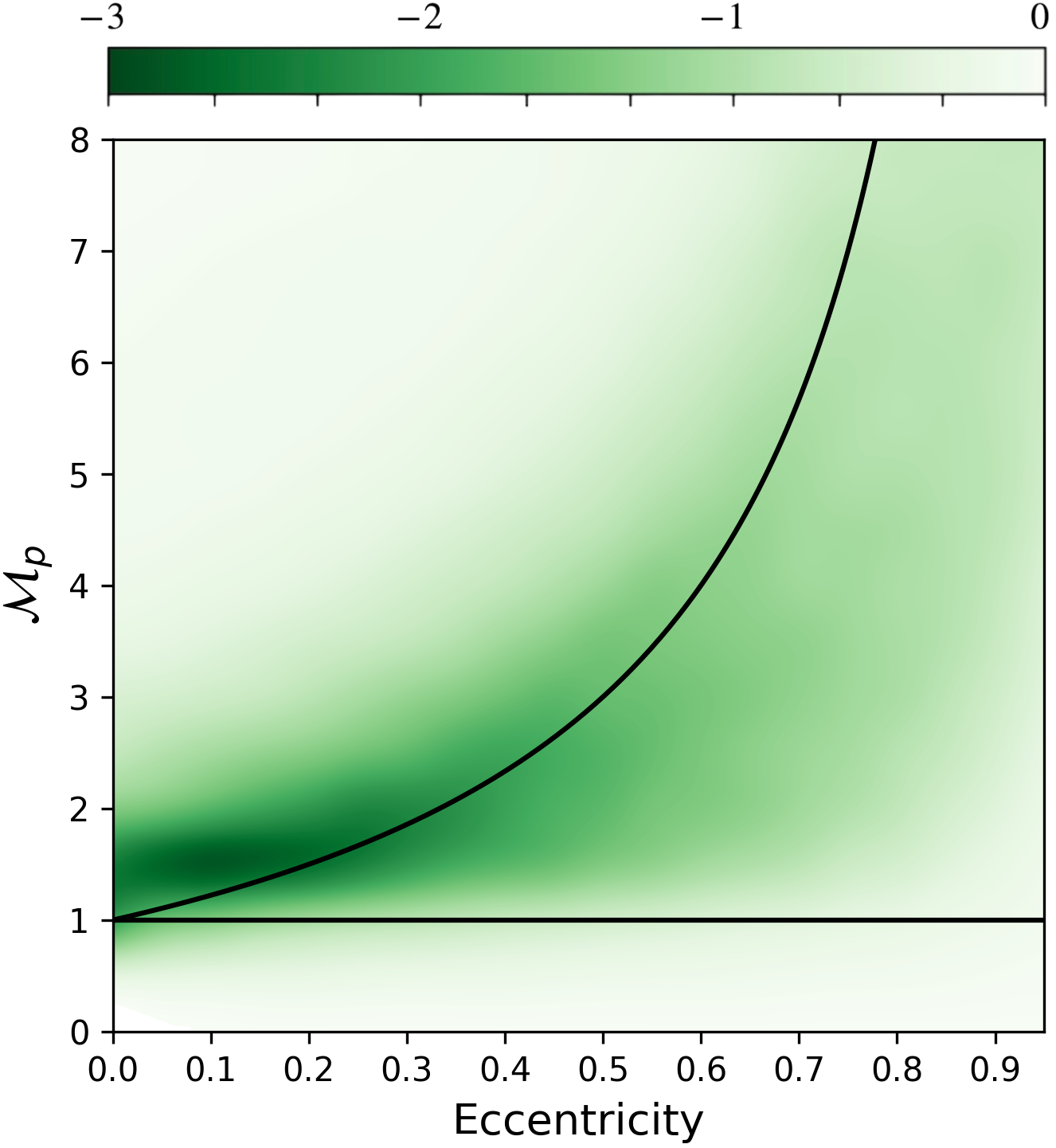}\quad
    \includegraphics[width=0.308\textwidth]{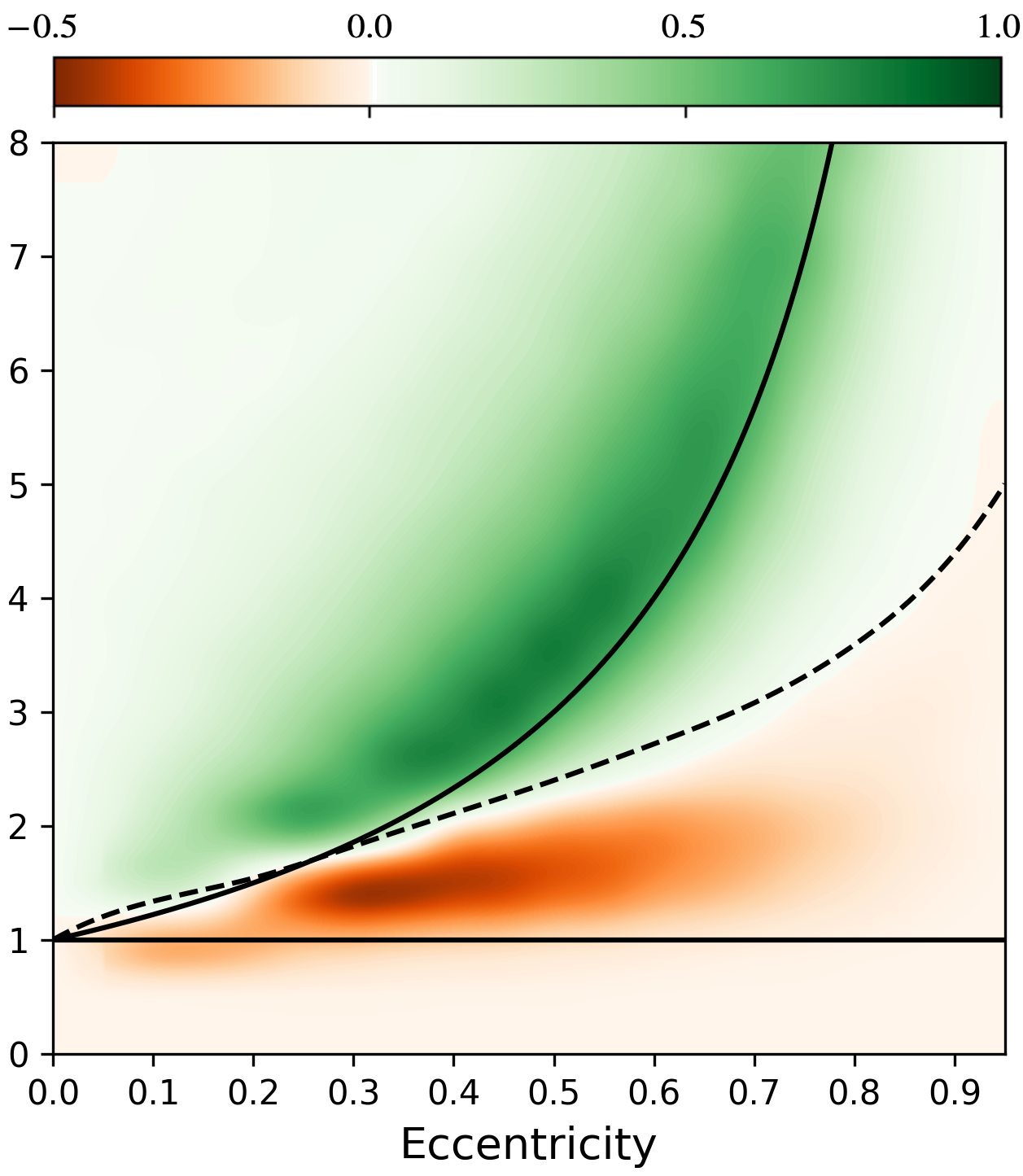}\quad
    \includegraphics[width=0.322\textwidth]{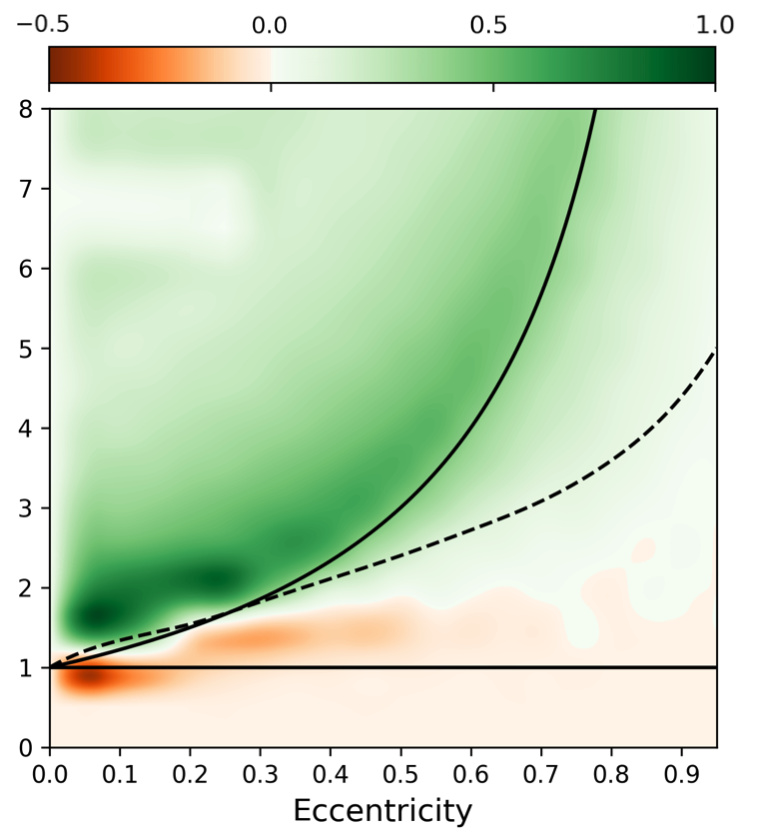}\quad

    \caption{The normalised rate of change of the semi-major axis $\dot{a}/a$ (left), eccentricity $\dot{e}$ (center) and pericenter-precession frequency $\dot\omega$ (right) as a function of pericenter Mach number $\mathcal{M}_p$ and eccentricity. Our data is interpolated over a regular grid of spacing $0.5$ in $\mathcal{M}_p$ and $0.05$ in eccentricity. The solid black curves separate the domain into strictly supersonic (above the upper solid line), transonic (between both solid black lines) and strictly subsonic (below the lower black line). In addition to this we include a dashed black line in the center panel illustrating the orbits in which the force at apocenter is equal to the force at pericenter, acting as an indicator for when the change in orbital eccentricity is zero. In all cases, the timescale associated with these rates is set by the product $\tau\mathcal{A}\Omega$. Data available at: \url{https://github.com/davidoneill00/Dynamical-Friction-Elliptical-Keplerian-Orbits-Data}
    }
\label{OrbitalChanges}
\end{figure}

On secular timescales, GDF will result in evolution of the massive perturber's orbital elements. The interpolated forces enable us to calculate this evolution by integrating Eqs.~(\ref{EvolutionScalingsA}), (\ref{EvolutionScalingsE}) and (\ref{EvolutionScalingsOmega}) over one orbital period assuming that $a$ and $e$ are fixed during that timescale. In this process, we make use of the (interpolated) forces as a function of $t$. We perform these evaluations for a wide range of Mach numbers and eccentricities in Fig.~\ref{OrbitalChanges}. On the left panel, we plot the rate-of-change in semi-major axis $\dot{a}/a$ due to the drag force. In the center panel, we plot the rate-of-change of eccentricity $\dot{e}$, while on the right we have the rate of pericenter precession $\dot\omega$. Depending on the region of parameter space, $\dot{e}$ and $\dot\omega$ can be either positive (green) or negative (orange), whereas $\dot{a}/a$ is always negative.\\

A diagonal band is seen in all panels of Fig.~\ref{OrbitalChanges} where GDF is most greatly affecting the perturber's orbital motion. This behaviour is expected, corresponding to the transonic region of parameter space with mean velocity near the sound speed. The most rapid semi-major axis decay occurs around $e=0,~\Mach_p= 1.5$ (left panel of Fig.~\ref{OrbitalChanges}) which is the low eccentricity tail of this diagonal band. As has been shown in \citetalias{KK07}, the strongest force experienced on a circular orbit occurs at $\Mach_p = 1.3$ and is, therefore, in the same region of the parameter space (Fig. 8 in \citetalias{KK07}). At the low eccentricity tail, there is little variance in the Mach number over the course of an orbit and, therefore, a perturber will spend long periods of time, experiencing the strongest drag resulting in the most rapid semi-major axis decay. As the eccentricity increases, the variance in Mach number is higher and the perturber no longer experiences the same maximal drag throughout an orbit. This causes the strong diagonal band to  diffuse at higher eccentricities.\\

In addition to the green diagonal band, we see strong eccentricity damping (orange) occurring in the center panel of Fig.~\ref{OrbitalChanges}. We identify this as the subsonic and (lower) transonic regions of parameter space for which pericenter gas forces dominate. When the motion is strictly subsonic, the force increases monotonically with the Mach number (cf. \citetalias{O99}), achieving a maximum at pericenter and minimum at apocenter. The orbital speed at pericenter is greater than that of a circular orbit with the same semi-major axis $v_p\propto\sqrt{(1+e)/(1-e)}$ meaning the application of a strong drag force will decrease the pericenter speed of an eccentric orbit until it matches the circular value. Ultimately this will cause a decrease in eccentricity for subsonic/lower transonic orbits. In contrast, for a strictly supersonic trajectory the force decreases monotonically with the Mach number,
creating a peak at apocenter and a minimum at pericenter. After passing apocenter, the particle must return to the same point forbidding a change in energy -- but not angular momentum. Thus the particle falls back on a more radial approach and a higher eccentricity orbit which explains this supersonic region of eccentricity driving in Fig.~\ref{OrbitalChanges}. The same conclusion can be reached by considering an instantaneous drag force occurring at either pericenter $f = E = 0$ or apocenter $f = E = \pi$ in Eq.~(\ref{FirstBurnsE}). Finally, the white region separating eccentricity pumping (green) and damping is well fit by orbits for which the drag force at apocenter is equal to the drag force at pericenter, representing a balanced orbital eccentricity.\\

The gas-induced precession frequency (right panel of Fig.~\ref{OrbitalChanges}) demonstrates similar behaviour to the rate-of-change in orbital eccentricity (center panel of Fig.~\ref{OrbitalChanges}). By including the separation line delineating balanced apocenter and pericenter forces, we observe prograde precession typically above this line, whereas retrograde precession always occurs below. This suggests that forces dominant at pericenter cause prograde precession, whereas apocenter dominating forces lead to retrograde precession, as explained in \cite[Section 3,][]{Tiede+2023}. The characterisation of apsidal precession, along with the changes in semi major axis and orbital eccentricity, culminate the effect of GDF on a Keplerian orbit.\pagebreak

\subsection{Secular Orbital Evolution}
\begin{figure}[H]
\vspace{1mm}
\hspace{68mm}\textbf{\fontsize{13}{17}\selectfont Single Perturber}
\vspace{3mm}

\minipage{0.297\textwidth}
  \includegraphics[width=\linewidth]{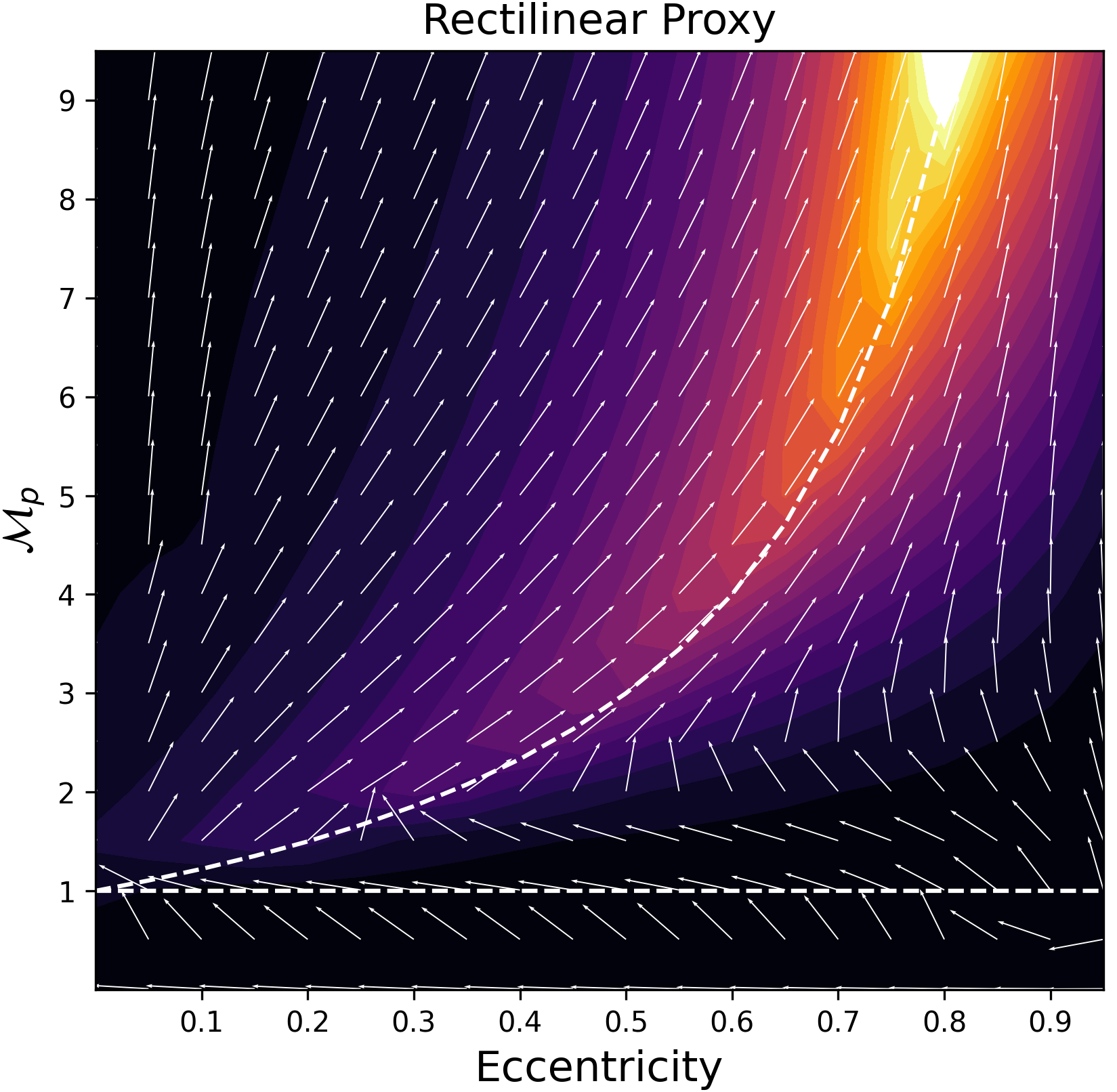}
\endminipage
\minipage{0.285\textwidth}
  \includegraphics[width=\linewidth]{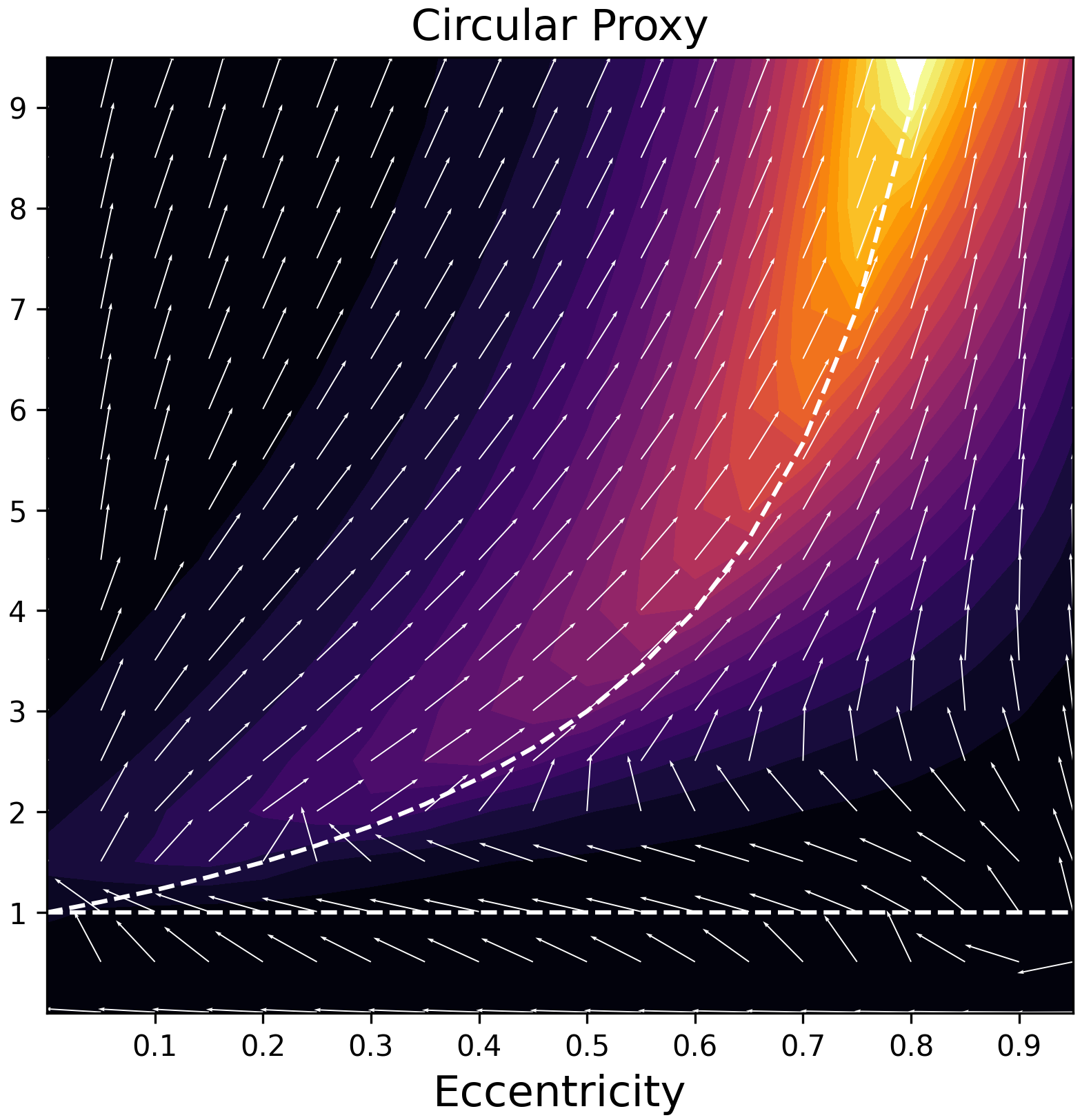}
\endminipage
\minipage{0.34\textwidth}
  \includegraphics[width=\linewidth]{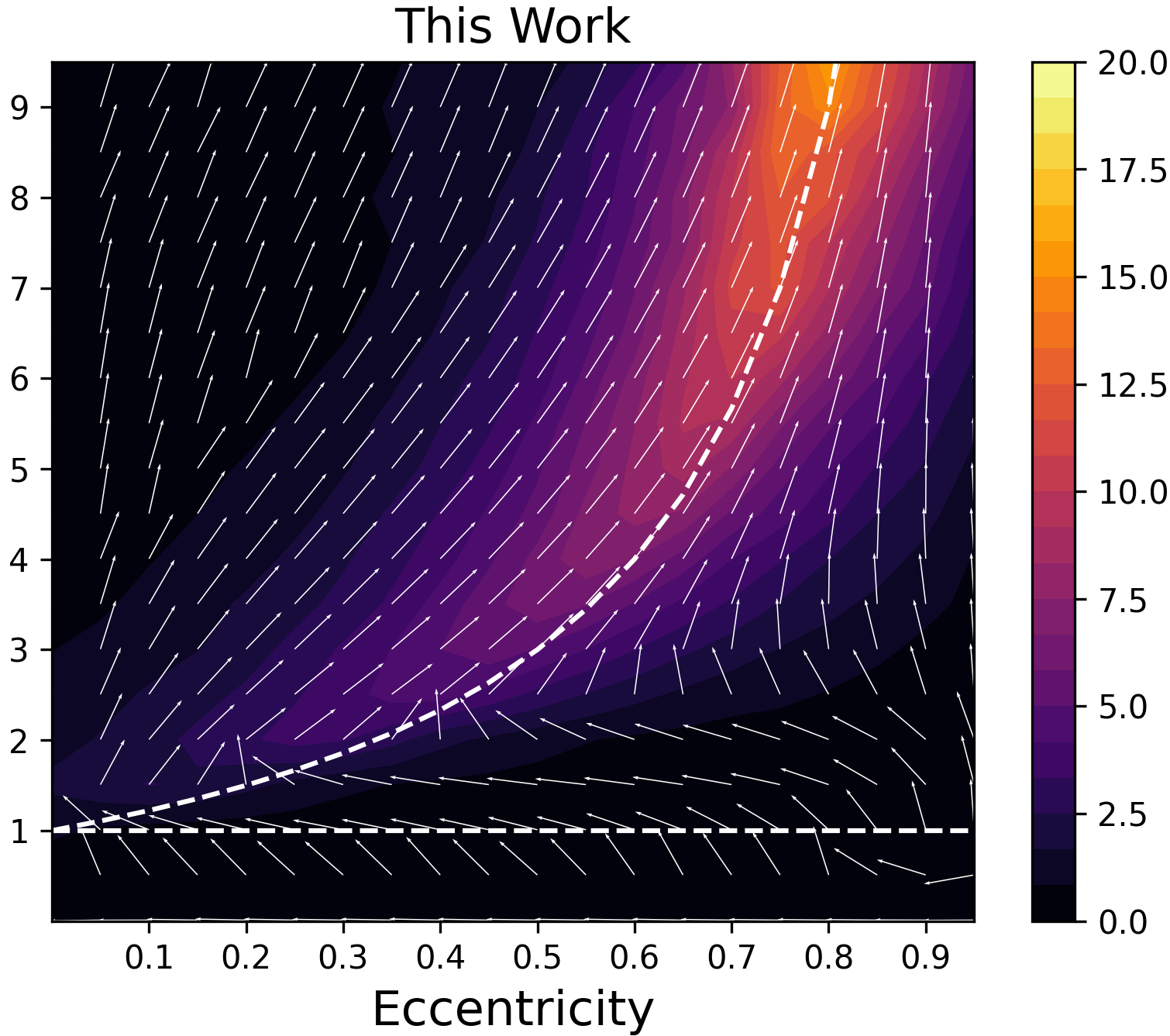}
\endminipage
\minipage{0.345\textwidth}
\hspace{1mm}
\rotatebox{90}{\textbf{\fontsize{8}{8}\selectfont $\sqrt{\dot{e}^2+\dot{\Mach}_p^2}~ \Big[4\pi \tau \mathcal{A} \Omega\Big]$}}
\endminipage
\caption{The evolutionary flow of trajectories experiencing GDF as a function of eccentricity and the Mach number at pericenter. The white arrows are unit vectors illustrating the direction of evolution at a given point while the color scheme represents the magnitude of the vector. This magnitude is inversely proportional to the evolution timescale, meaning the dark regions indicate a slowly evolving orbit, whereas the bright orange strip describes a more rapid orbital evolution. We compare the results of the rectilinear (left) and circular (middle) proxies to our results (right). The dashed white lines separate the parameter space into supersonic (upper left), transonic (right) and subsonic (bottom). Data available at: \url{https://github.com/davidoneill00/Dynamical-Friction-Elliptical-Keplerian-Orbits-Data}}
\label{OrbitTracks}
\end{figure}

We next characterise the orbital evolution according to changes in the two relevant parameters, $\Mach_p$ and $e$. To translate our results in Fig.~\ref{OrbitalChanges} into an orbital evolution, we calculate $~\dot{e}$ and 
\begin{equation}
    \dot{\Mach}_p = \frac{\Mach_p}{(1-e)^2}\dot{e} - \frac{\Mach_p}{2}\frac{\dot{a}}{a},
    \label{MachDot}
\end{equation}

which illustrates the increase in pericenter orbital velocity for both tighter orbits and higher eccentricities. In Fig.~\ref{OrbitTracks} we use our solutions for $\dot{a}/a$ and $\dot{e}$ to describe the local flow of orbital evolution, by plotting the vector $\mathbf{f}(e,\Mach_p)$, 
\begin{equation}
    \mathbf{f}(e,\Mach_p) = \dot{e}~\hat{e}+\left[\frac{\Mach_p}{(1-e^2)}\dot{e}-\frac{\Mach_p}{2}\frac{\dot{a}}{a}\right]~\hat{\Mach}_p.
    \label{VectorDirection}
\end{equation}
Fig.~\ref{OrbitTracks} illustrates the local evolution of the Mach number and the eccentricity of an orbit experiencing GDF. The strong band from Fig.~\ref{OrbitalChanges} is present once again in Fig.~\ref{OrbitTracks}, wherein a more rapid orbital evolution occurs. Likewise, the darker regions represent longer timescales necessary for GDF to change the orbital elements. For highly eccentric, low-Mach-number orbits there is a competition between decreasing semi-major axis and the decreasing eccentricity Eq.~(\ref{MachDot}). As the semi-major axis decreases the orbital speed increases leading to $\dot{\Mach}_p>0$, while as the eccentricity decreases the orbit circularises and $\dot{\Mach}_p<0$. Once trajectories eventually become supersonic, the eccentricity is always expected to increase, along with the Mach number at pericenter.\\

\subsection{Double Perturbers: An Equal Mass Binary System}
\label{sec:BinarySection}
\begin{figure}[!htb]
\hspace{15mm}\textbf{$\mathrm{e} = 0.5, \Mach_p=0.9$}
\hspace{30mm}\textbf{$\mathrm{e} = 0.5, \Mach_p=4$}
\hspace{30mm}\textbf{$\mathrm{e} = 0.5, \Mach_p=2$}

\vspace{3mm}

\minipage{0.28\textwidth}
  \includegraphics[width=\linewidth]{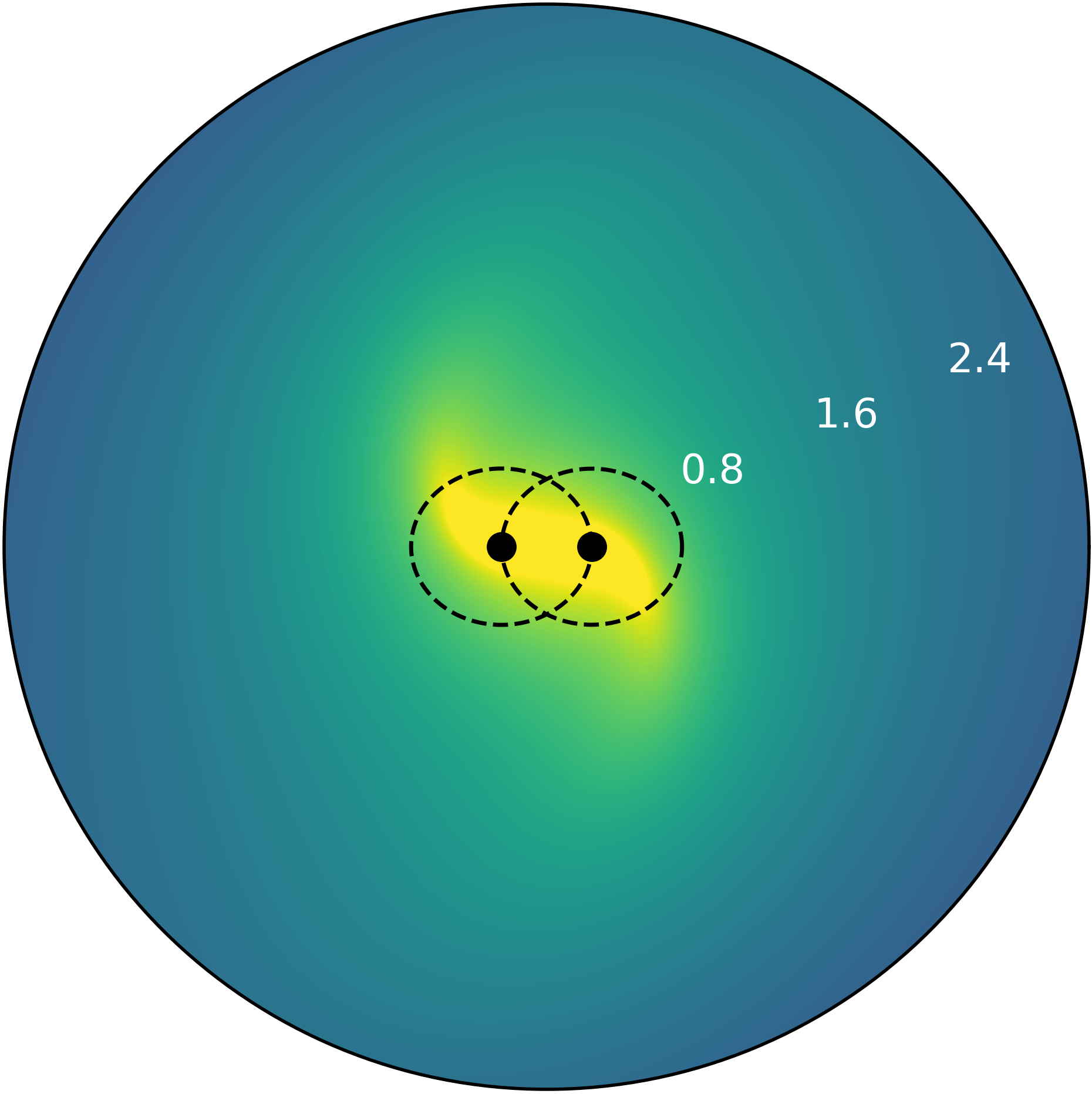}
\endminipage\hfill
\minipage{0.28\textwidth}
  \includegraphics[width=\linewidth]{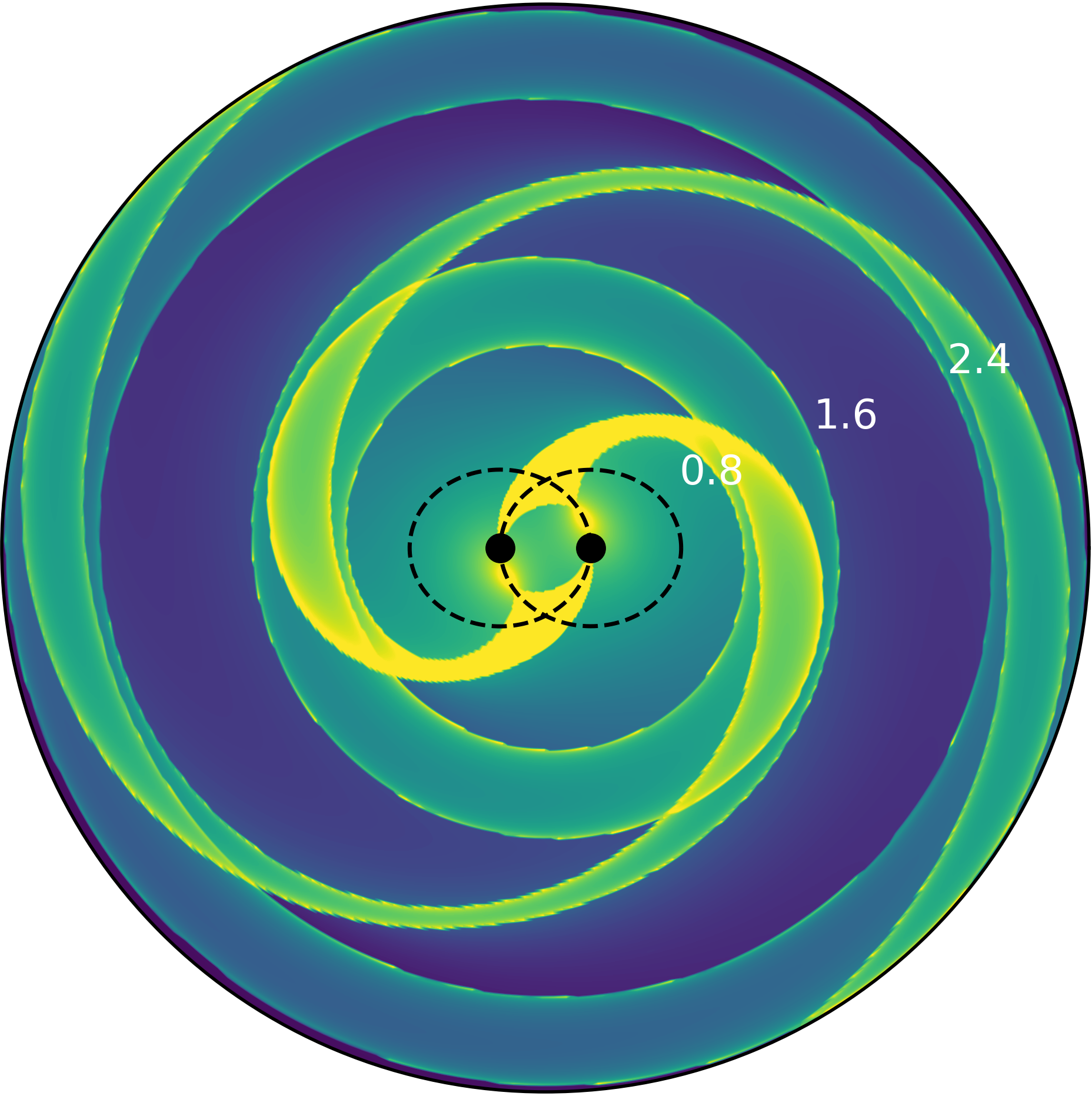}
\endminipage\hfill
\minipage{0.37\textwidth}%
  \includegraphics[width=\linewidth]{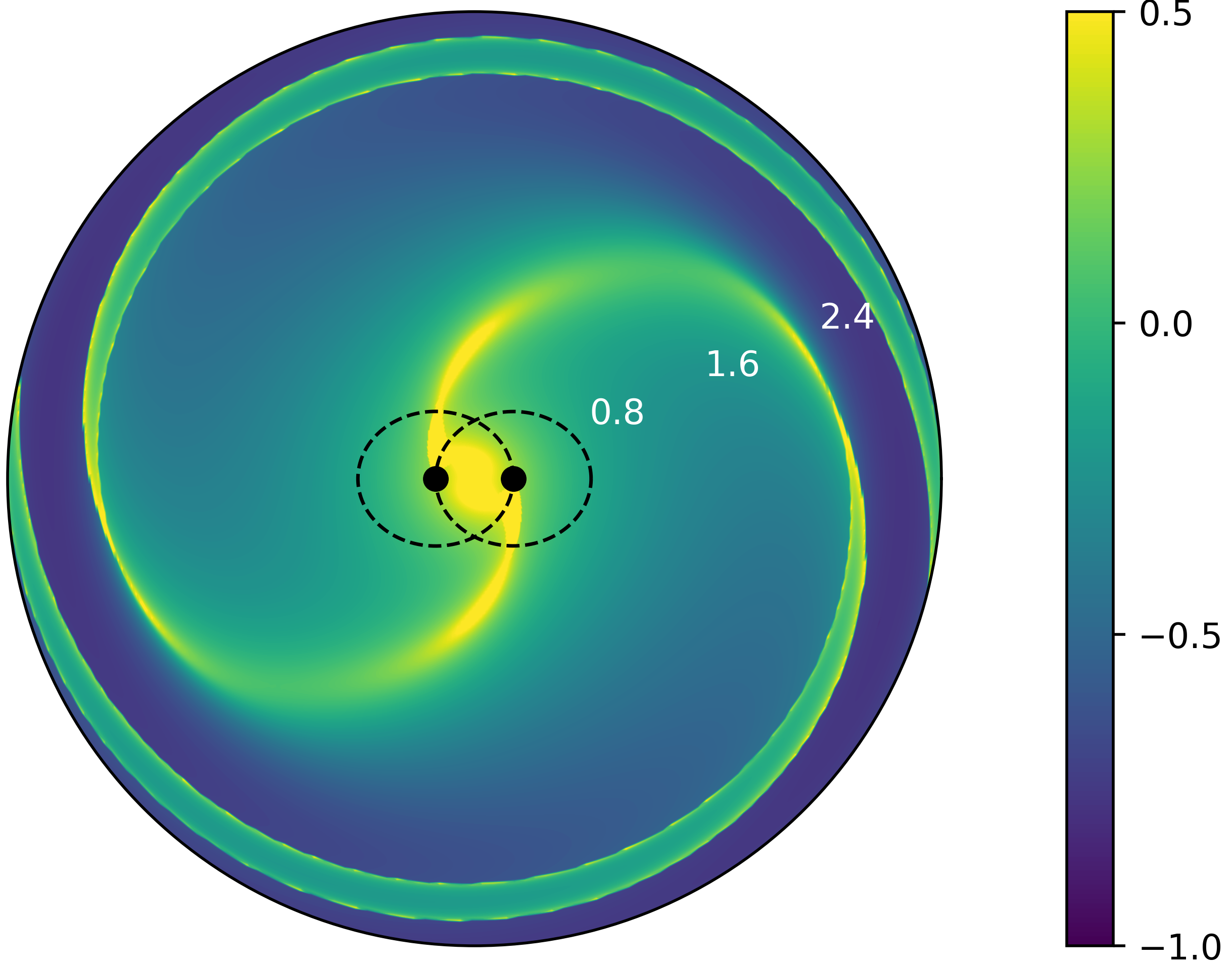}
\endminipage

\caption{Logarithmic contour plots of the dimensionless density perturbations $\mathcal{D}_\mathrm{b}$ (as defined in Eq.~(\ref{DimensionlessForBinary})) for a eccentric equal mass binaries in the barycenter frame. We consider three cases: strictly subsonic (left), strictly supersonic (middle) and transonic (right). Each binary has the same orbital eccentricity of $e = 0.5$, and each of the panels above is plotted over a region of radius $r = 3a_\mathrm{b}$, with resolution of $0.02a$ in radius and $\pi/120$ in azimuth.}
\label{BinaryDensities}
\end{figure}

As yet we have only considered the density wakes of a single massive perturber. Due to the linear nature of our approach, it is possible to superimpose a second density perturbation of an equal mass companion, creating the wake of a binary  $\alpha_{\mathrm{b}}$. We assume that the single perturber wake $\alpha_\mathrm{s}(t,\mathbf{x})$ (centered at the barycenter) is known from Eq.~(\ref{CodeUnitDensityPerturbations}), from which, the companion wake is a simple point reflection through the origin,
\begin{align}
    \alpha_{\mathrm{b}}(t,\mathbf{x}) &= \alpha_\mathrm{s}(t,\mathbf{x})+\alpha_\mathrm{s}(t,-\mathbf{x}), \nonumber \\
    &= \frac{Gm}{ac_\mathrm{s}^2}\left[\mathcal{D}_\mathrm{s}(\Omega t,\mathbf{x}/a, e, \Mach_p)+\mathcal{D}_\mathrm{s}(\Omega t,-\mathbf{x}/a, e, \Mach_p)\right],
    \label{DoubleWakeAnalytical}
\end{align}
where $\mathcal{D}_\mathrm{s}$ is the dimensionless density perturbation of a single perturber as defined in Eq.~(\ref{AlphaDimensions}). In this scenario we have decomposed the two-body problem into two independent one-body problems, with the origin of our coordinates being the binary's barycenter. Consequently, our coordinate system is at rest with respect to the medium, and the Mach number is given relative to the gas. In the barycentric frame of an equal mass binary, both mass components will orbit with semi-major axis $a$, eccentricity $e$ and orbital frequency $\Omega$. We can, however, choose to define our unit length in terms of the binary semi-major axis $a_\mathrm{b} \equiv 2a$ (for an equal mass binary with $M_{\mathrm{b}} = 2m$). With these considerations, Eq. (\ref{DoubleWakeAnalytical}) can be written as
\begin{equation}
    \alpha_\mathrm{b}(t, \mathbf{x}) \equiv \frac{GM_{\mathrm{b}}}{a_\mathrm{b}c_\mathrm{s}^2} \mathcal{D}_\mathrm{b}(\Omega t,\mathbf{x}/a_\mathrm{b}, e, \Mach_p),
\end{equation}
where,
\begin{equation}
     \mathcal{D}_\mathrm{b}(\Omega t,\mathbf{x}/a_\mathrm{b}, e, \Mach_p) = \mathcal{D}_\mathrm{s}(\Omega t,2\mathbf{x}/a_\mathrm{b}, e, \Mach_p)+\mathcal{D}_\mathrm{s}(\Omega t,-2\mathbf{x}/a_\mathrm{b}, e, \Mach_p).
\label{DimensionlessForBinary}
\end{equation}
In Fig.~\ref{BinaryDensities} we illustrate the density perturbations of three eccentric equal mass binaries, each with $e = 0.5$ but varying Mach numbers of $\Mach_p = 0.9,4,2$ (from left to right).\\ 

\begin{figure}[!htb]
\hspace{55mm}\textbf{Binary system with $\mathrm{e} = 0.9, \Mach_p=2$}\par\medskip
\vspace{2.3mm}
\minipage{0.28\textwidth}
  \includegraphics[width=\linewidth]{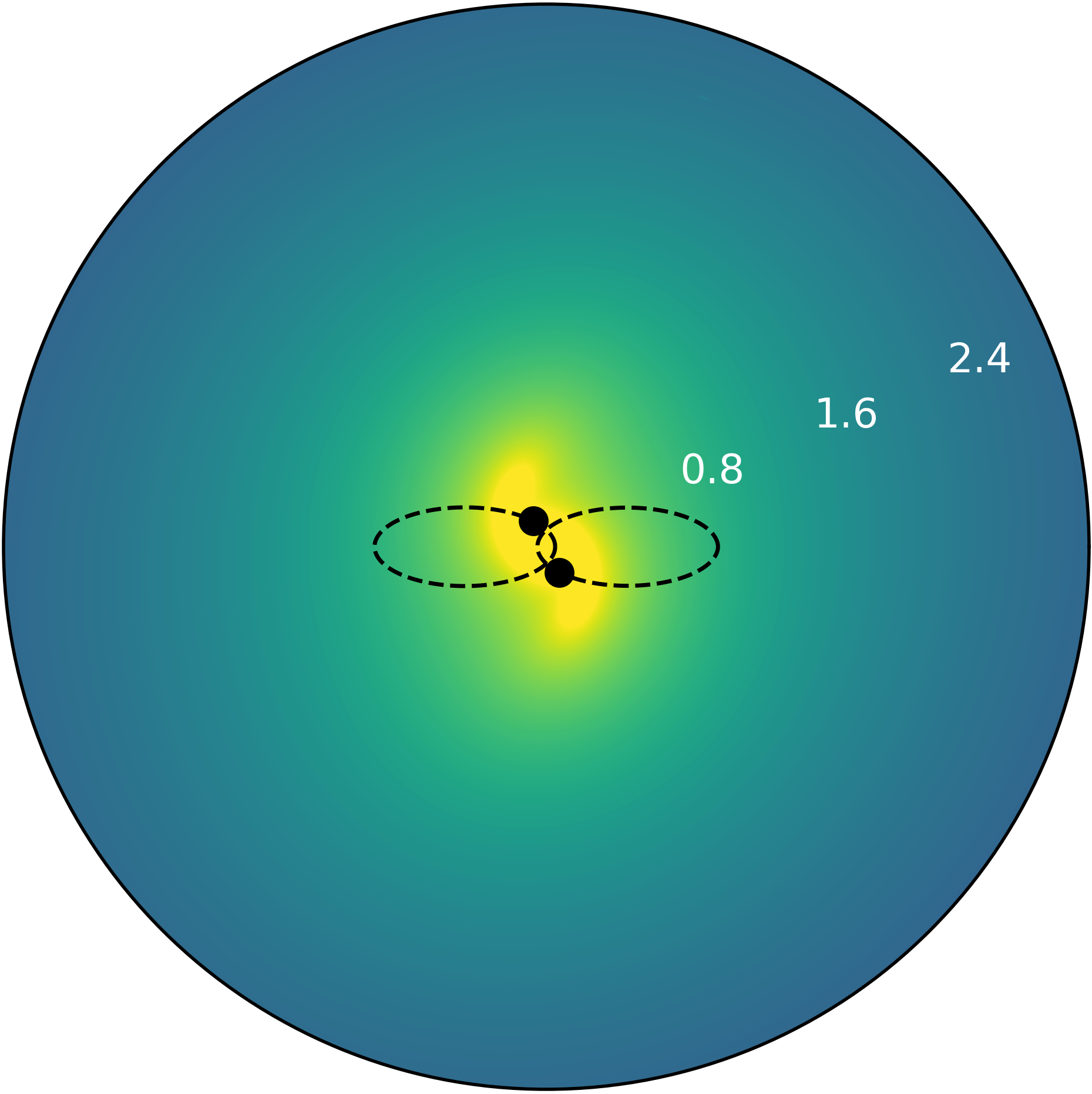}
  \put(-100,160){$\textbf{t = 0.02~P}$}
\endminipage\hfill
\minipage{0.28\textwidth}
  \includegraphics[width=\linewidth]{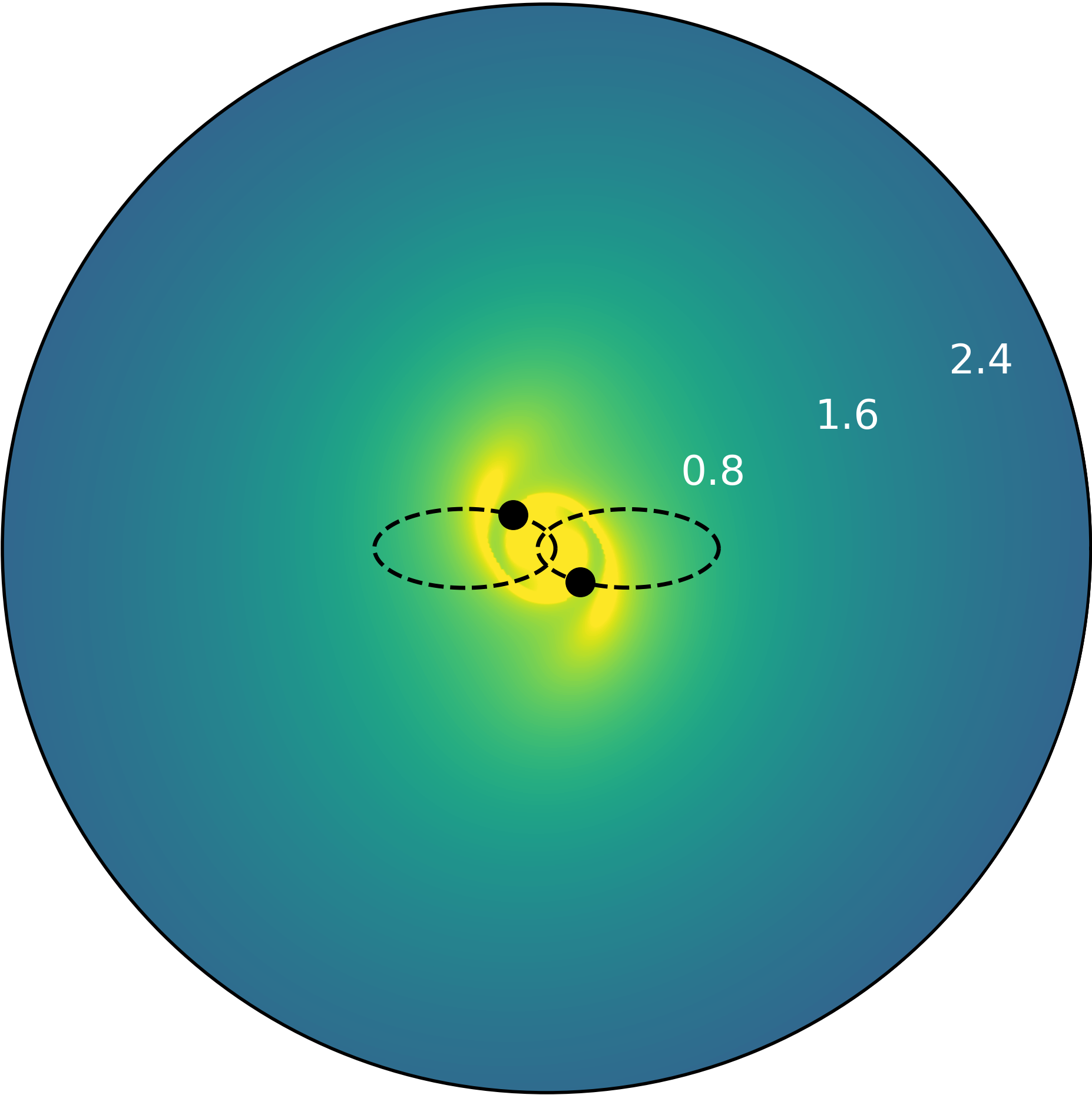}
  \put(-100,160){$\textbf{t = 0.04~P}$}
\endminipage\hfill
\minipage{0.37\textwidth}%
  \includegraphics[width=\linewidth]{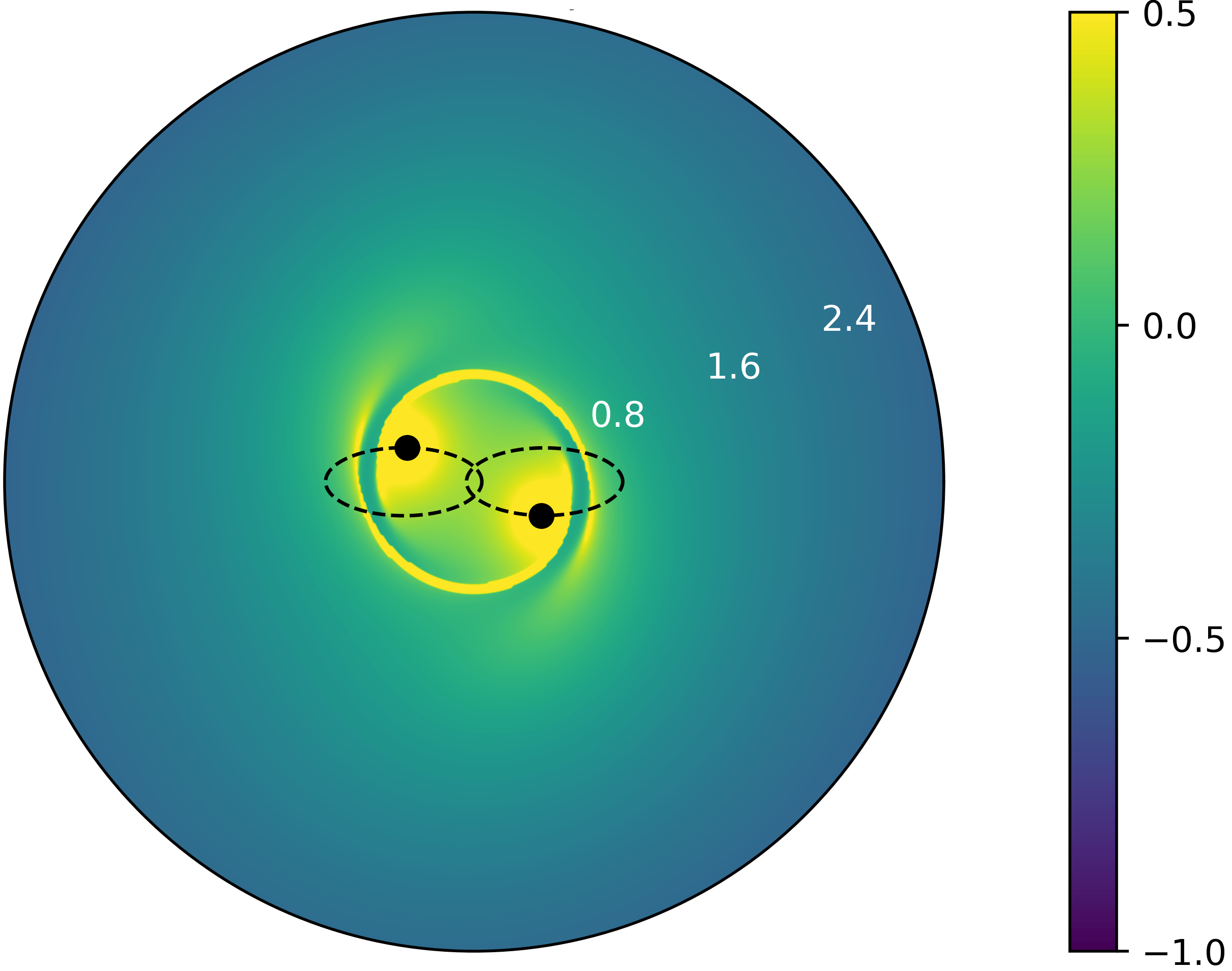}
  \put(-125,160){$\textbf{t = 0.1~P}$}
\endminipage
\vspace{1mm}
\hspace{55mm}\par\medskip
\vspace{1mm}
\minipage{0.28\textwidth}
  \includegraphics[width=\linewidth]{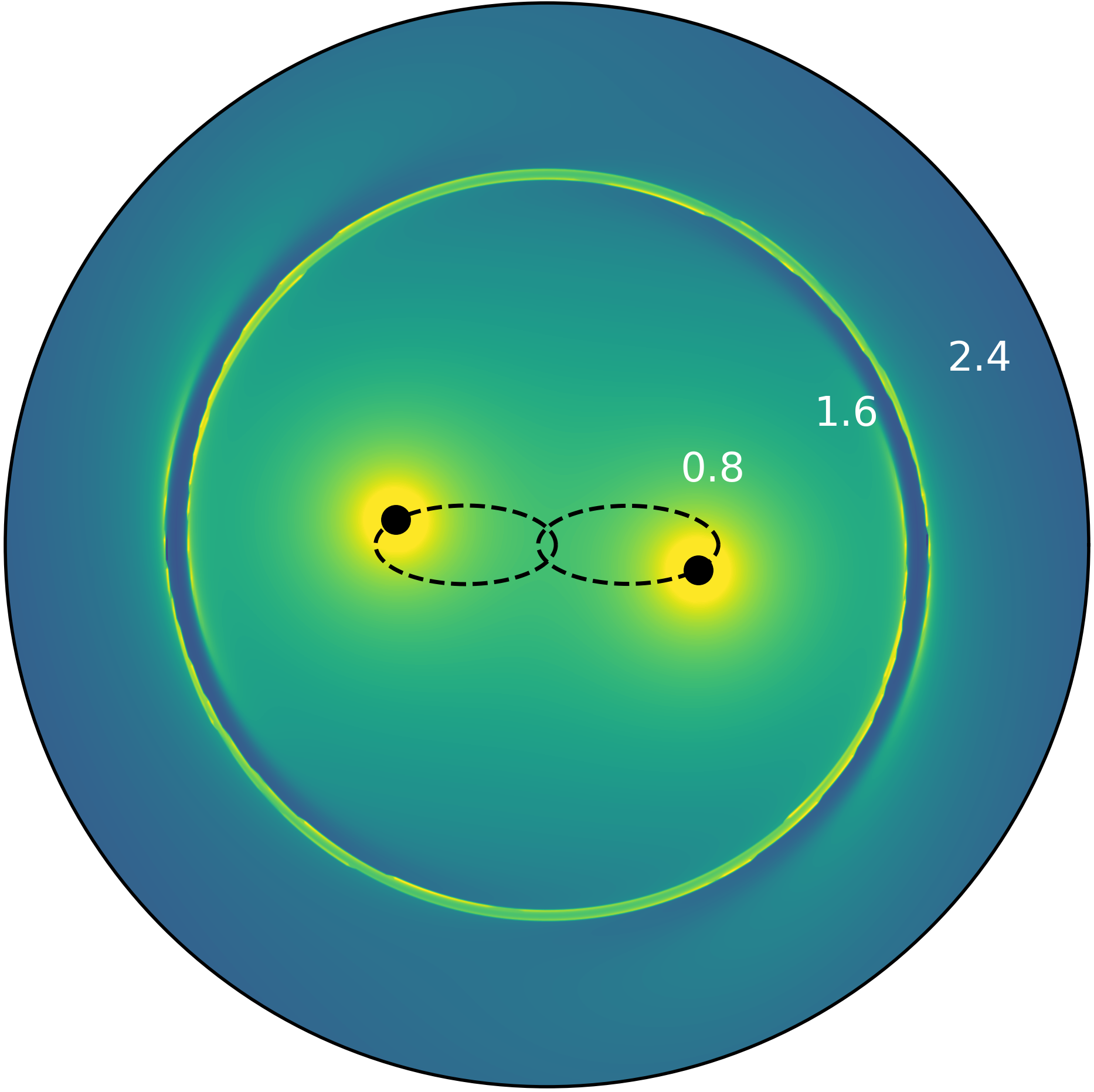}
  \put(-100,160){$\textbf{t = 0.3~P}$}
\endminipage\hfill
\minipage{0.28\textwidth}
  \includegraphics[width=\linewidth]{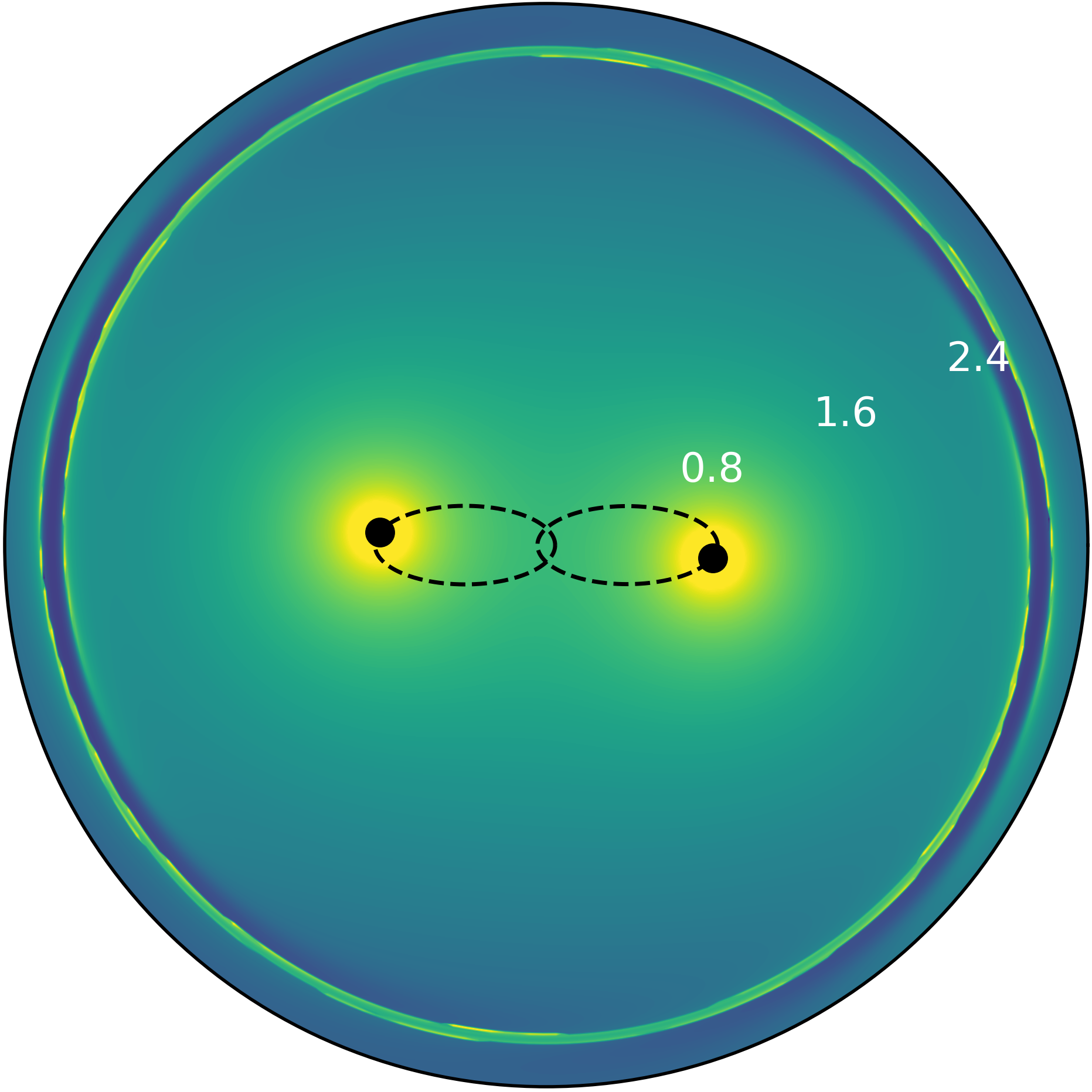}
  \put(-100,160){$\textbf{t = 0.4~P}$}
\endminipage\hfill
\minipage{0.37\textwidth}%
  \includegraphics[width=\linewidth]{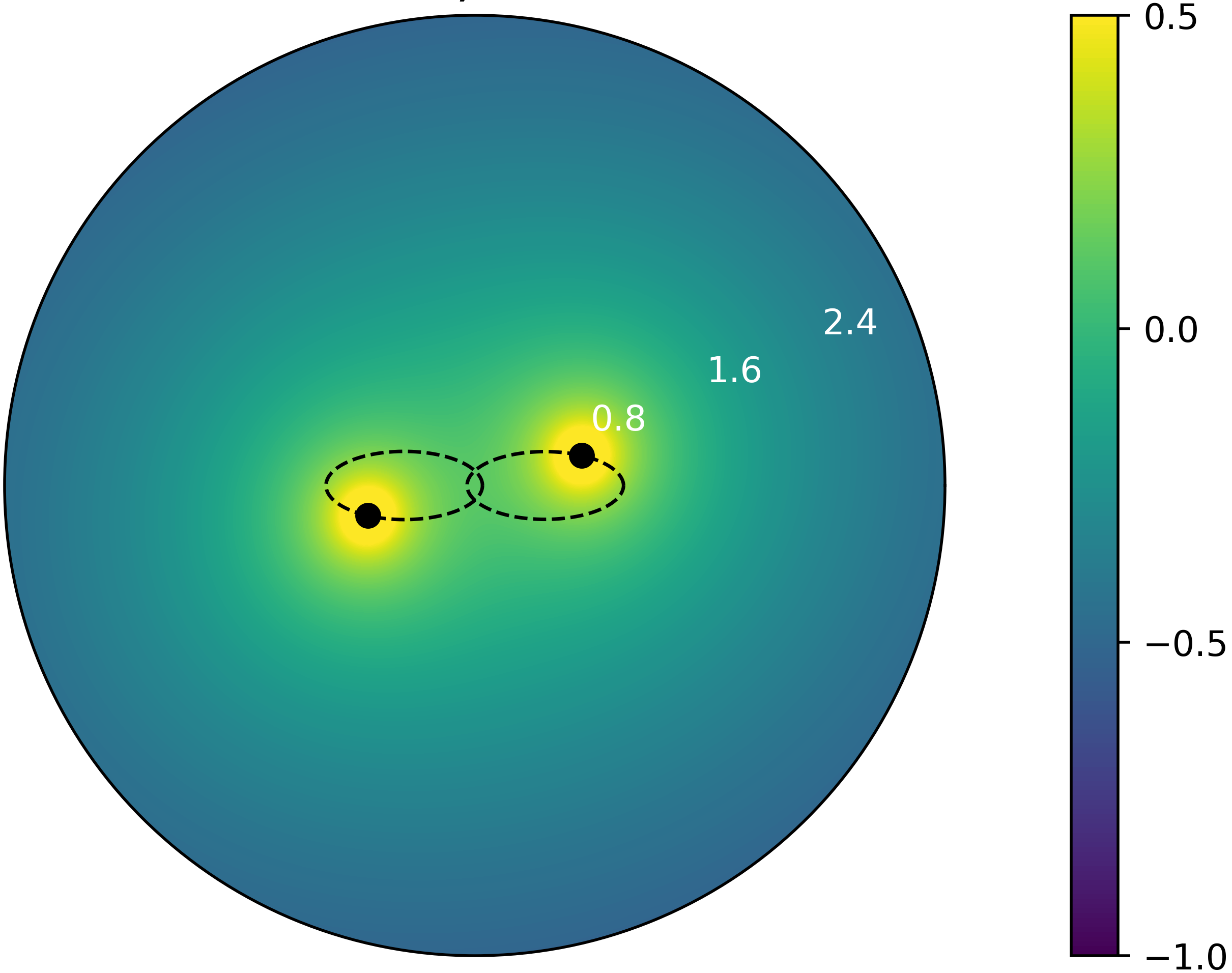}
  \put(-125,160){$\textbf{t = 0.8~P}$}
\endminipage
\caption{The same as Fig.~\ref{BinaryDensities}, except here we plot a single eccentric transonic binary at six different times during the orbit. The top left panel describes the wake just following the binary's pericenter passage $t_1 = 0.02P$. The upper middle and upper right panels depict the deceleration of the binary-- travelling sonically at time $t_2 = 0.04P$ and subsonically at time $t_3 = 0.1P$. The lower-left and middle panels depict the binary expelling a wavefront propagating at the speed of sound, followed by the pericenter return approach (bottom right) at $t = 0.8P$.}
\label{BinaryWakes}
\end{figure}

The transonic crossing will be simultaneous for both components of an equal mass binary -- a simple consequence of their equal orbital speeds. We find it illustrative to plot a time series of the transonic crossing in Fig.~\ref{BinaryWakes} where the thin shock front of high density is almost point symmetric with respect to the barycenter (this is not true for other mass ratios). The emitted shock front in Fig.~\ref{BinaryWakes} is created due to the deceleration of a perturber below the sound speed, wherein the Mach cone can become detached from the perturber. In fact, all supersonic Mach cones (and tail edges) equally represent shock fronts (cf. §3.1.1 \citetalias{KK07}) because they represent discontinuities in gas density, pressure and velocity. In this regard, the thin shock front created during a transonic crossing (Fig.~\ref{BinaryWakes}) is always emitted once per orbit.\\

\begin{figure}[!h]
\hspace{15mm}\textbf{$\mathrm{e} = 0, \Mach_p=0.5$}
\hspace{30mm}\textbf{$\mathrm{e} = 0.2, \Mach_p=0.5$} 
\hspace{30mm}\textbf{$\mathrm{e} = 0.4, \Mach_p=0.5$}
\par\medskip
\minipage{0.28\textwidth}
  \includegraphics[width=\linewidth]{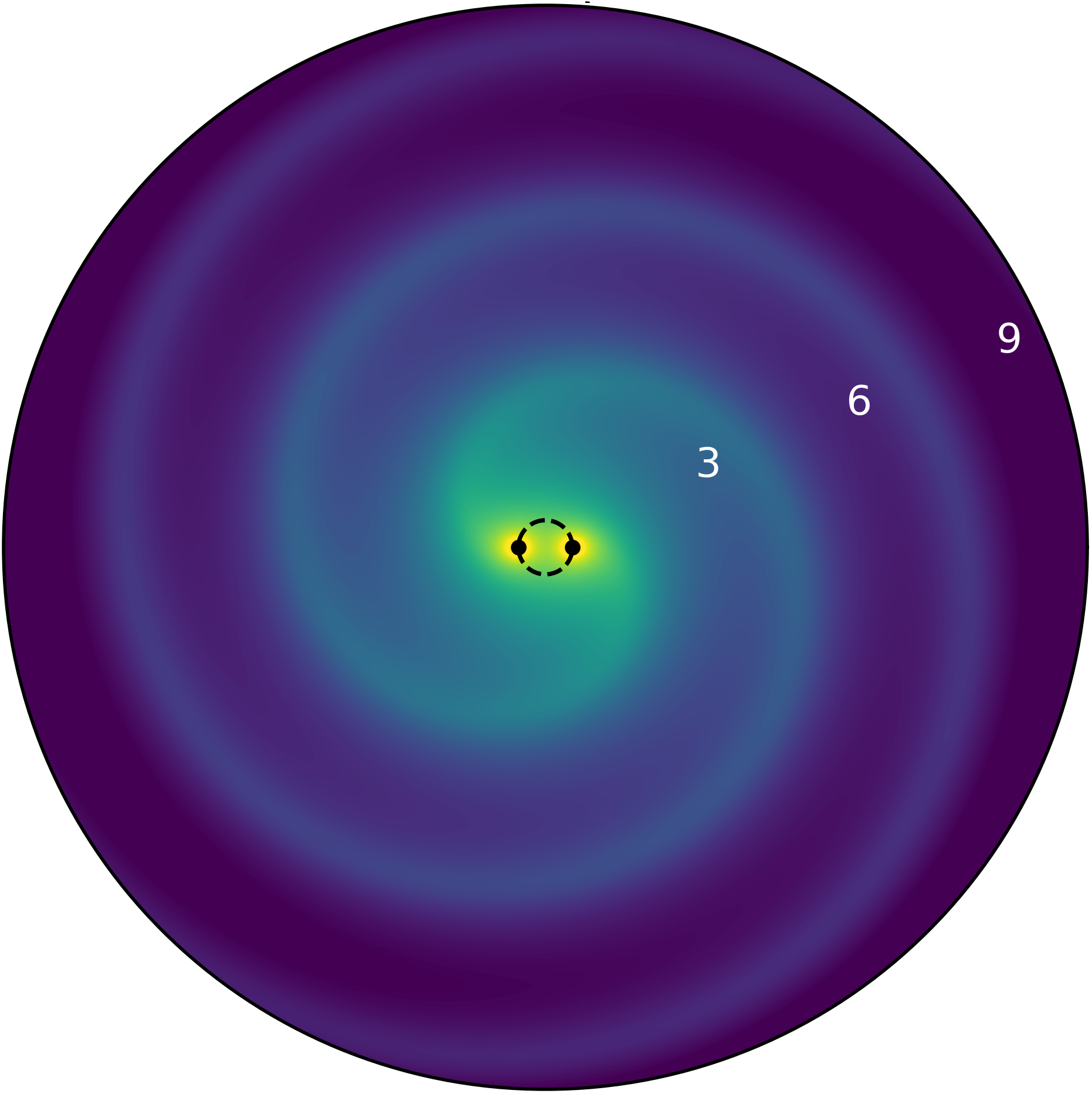}
\endminipage\hfill
\minipage{0.28\textwidth}
  \includegraphics[width=\linewidth]{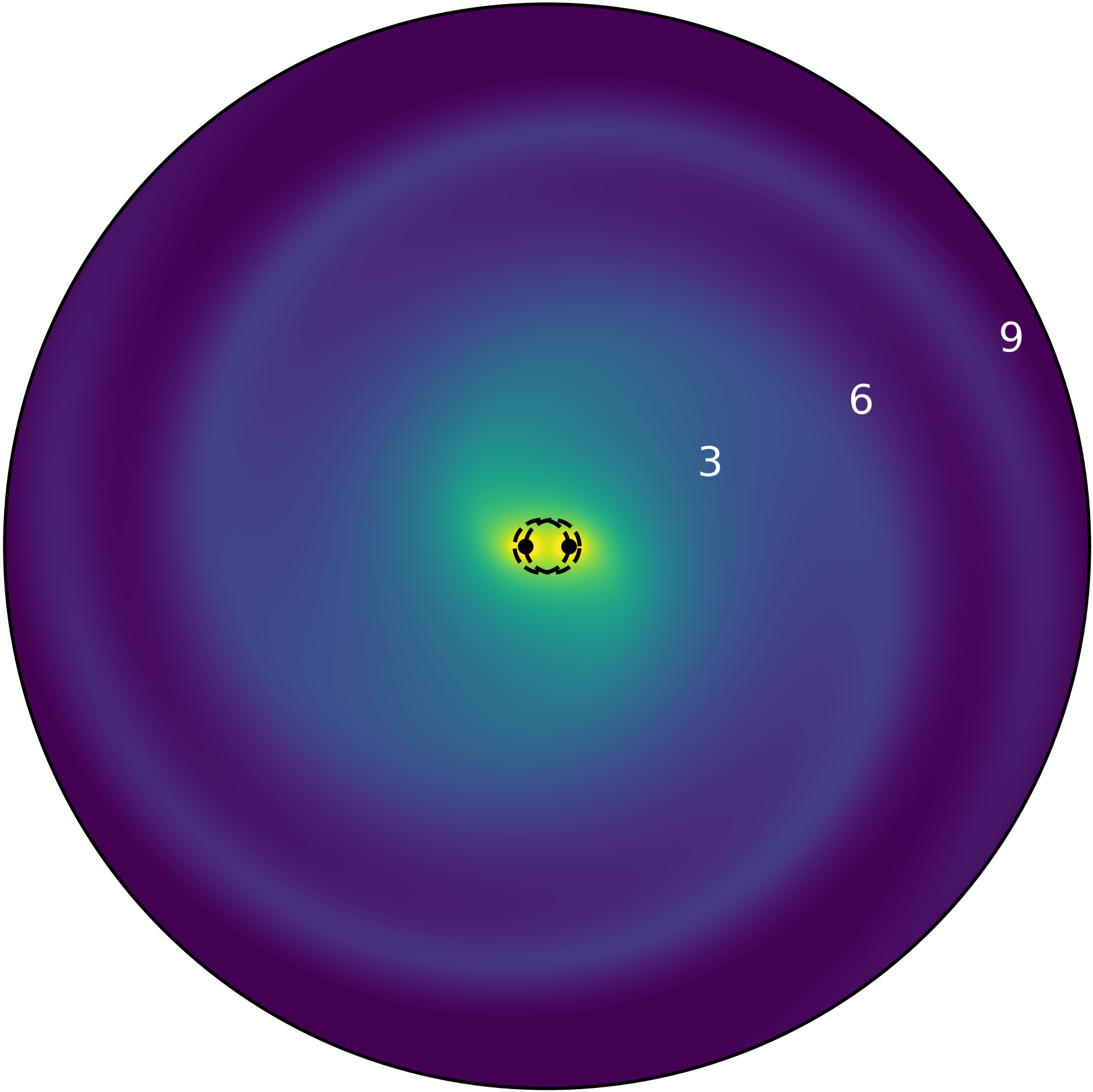}
\endminipage\hfill
\minipage{0.37\textwidth}%
  \includegraphics[width=\linewidth]{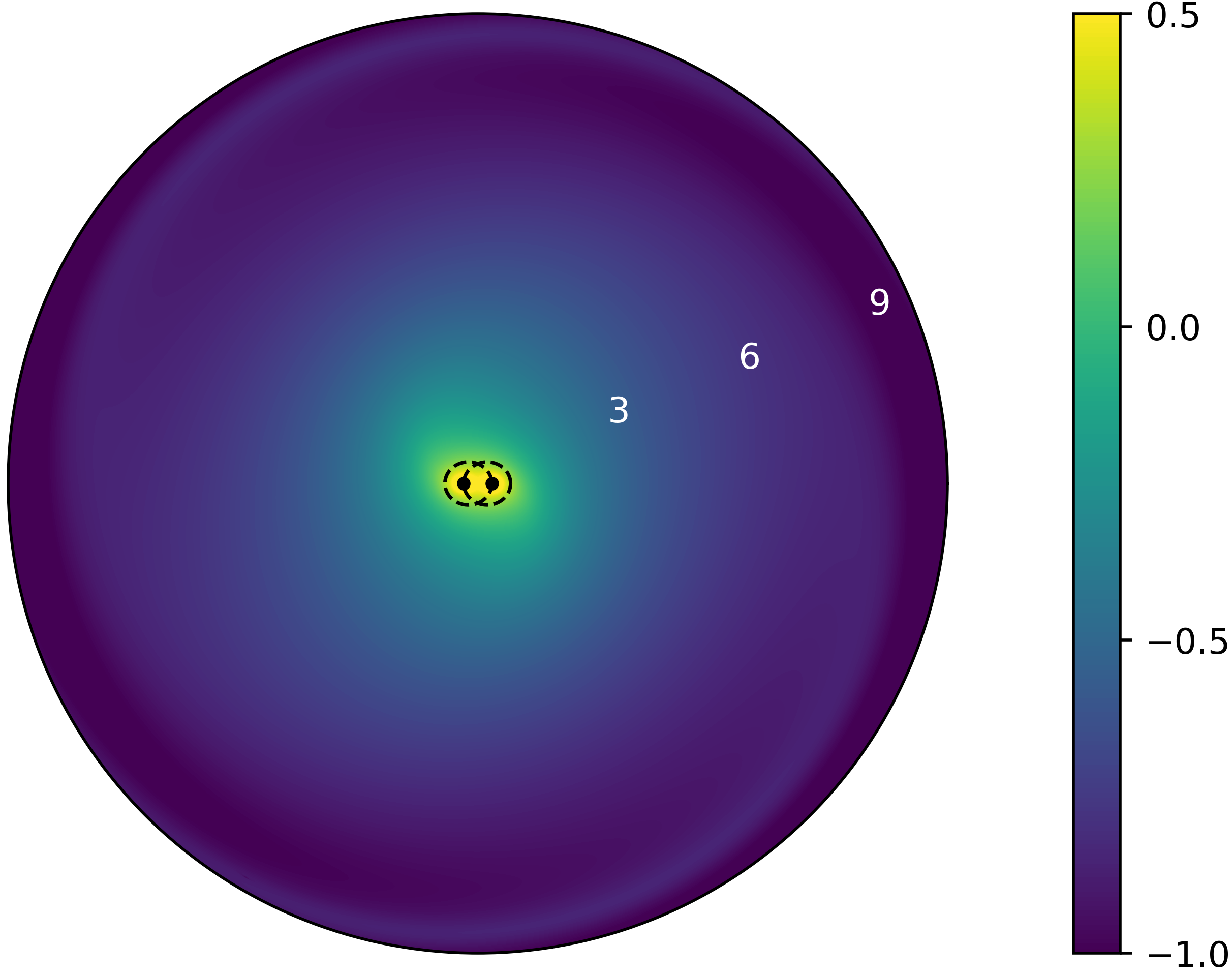}
\endminipage
\caption{The weakening of the spiral density waves with increasing binary eccentricity (see text). Here we plot the logarithm of the dimensionless density perturbation created by an equal mass binary for pericenter Mach number $\Mach_p = 0.5$. The radial resolution is approximately $0.07a_\mathrm{b}$, while the azimuthal resolution is $\pi/120$ as before.}
\label{SpiralWavesBinary}
\end{figure}

As has been noted in Fig.~\ref{WiderSubsonic}, the circular, subsonic perturber created long trailing spiral density arms. Unsurprisingly, the subsonic binary (created by the superposition of two single perturbers) also produces the same spiral density arms, which can be seen in Fig.~\ref{SpiralWavesBinary}. While this bears resemblance to that created by a rigidly rotating external potential, we reiterate that this spiral structure is created due to the precession of ellipsoidal density contours, and not a resonance. As the eccentricity of the orbit increases, the relative precession of density contours decreases (except for a short duration near pericenter). Therefore, the majority of ellipsoidal density contours which have been emitted during the orbit, are aligned with one another and create a wake better approximated by a series of concentric ellipsoids. As a result, the spiral structure begins to weaken at higher eccentricities as seen in Fig.~\ref{SpiralWavesBinary}.\\   

\pagebreak It is of interest to compare the binary results to rectilinear and circular proxy values. In doing so, we must compare to a reasonable \emph{binary proxy}, which we can construct in two ways-- one based on the results of \citetalias{O99} and one based on the results of \citetalias{KK07}: We define the ``rectilinear binary proxy'' to simply be the application of the rectilinear single proxy on both binary components-- namely a binary where each component experiences an individual (but independent) drag force. This means that each binary component feels its own wake but not that of its companion. On the other hand, \cite{DoublePerturbers} give prescriptions for the drag force experienced by \emph{only} the companions wake. Thus, we can consider modelling the total force as coming from the companions wake plus the single \citetalias{KK07} force.\\ 

\begin{figure}[H]
\vspace{0.5mm}
\hspace{65mm}\textbf{\fontsize{13}{17}\selectfont Binary Perturbers}
\vspace{2mm}

\minipage{0.305\textwidth}
  \includegraphics[width=\linewidth]{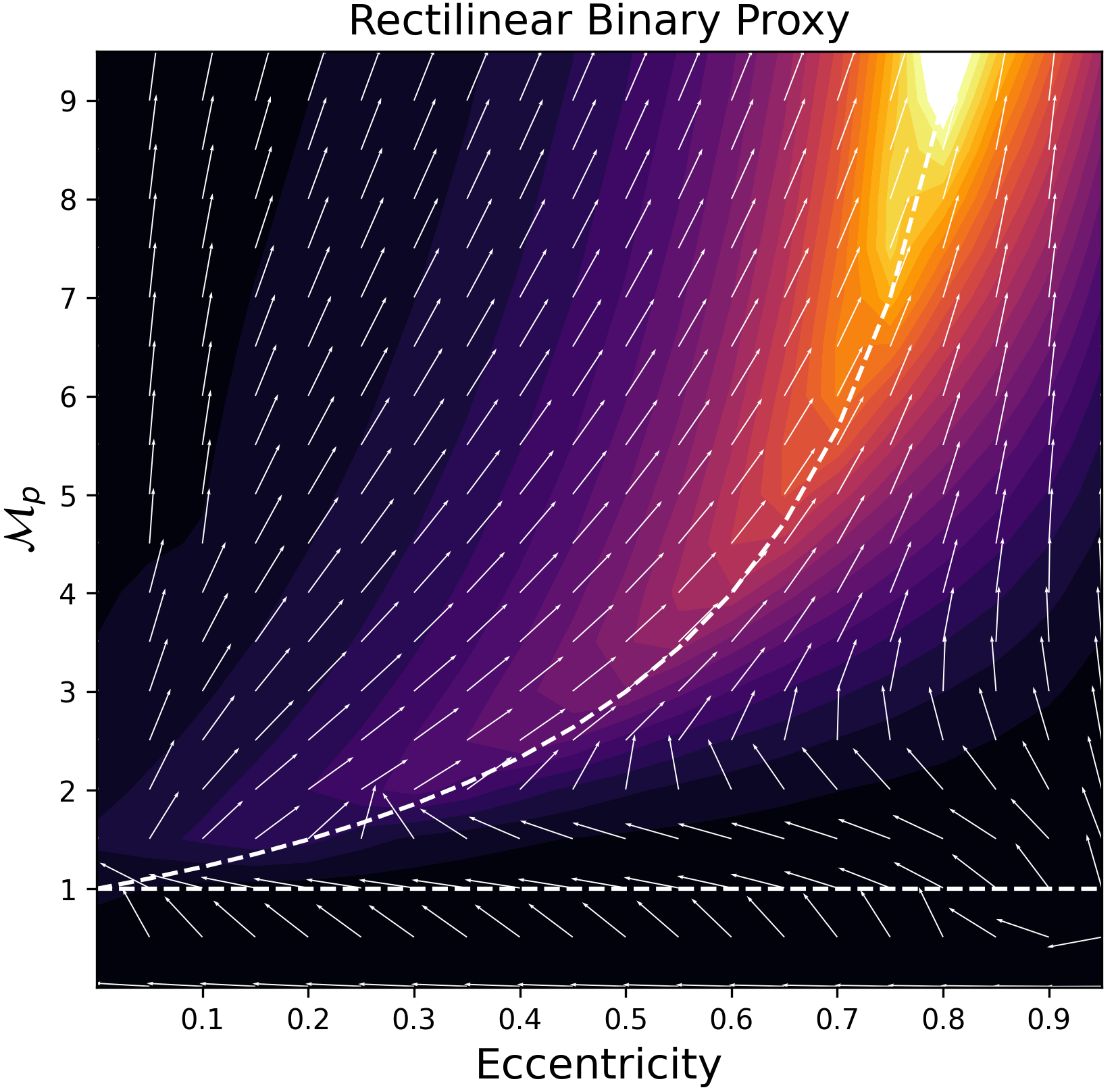}
\endminipage
\minipage{0.29\textwidth}
  \includegraphics[width=\linewidth]{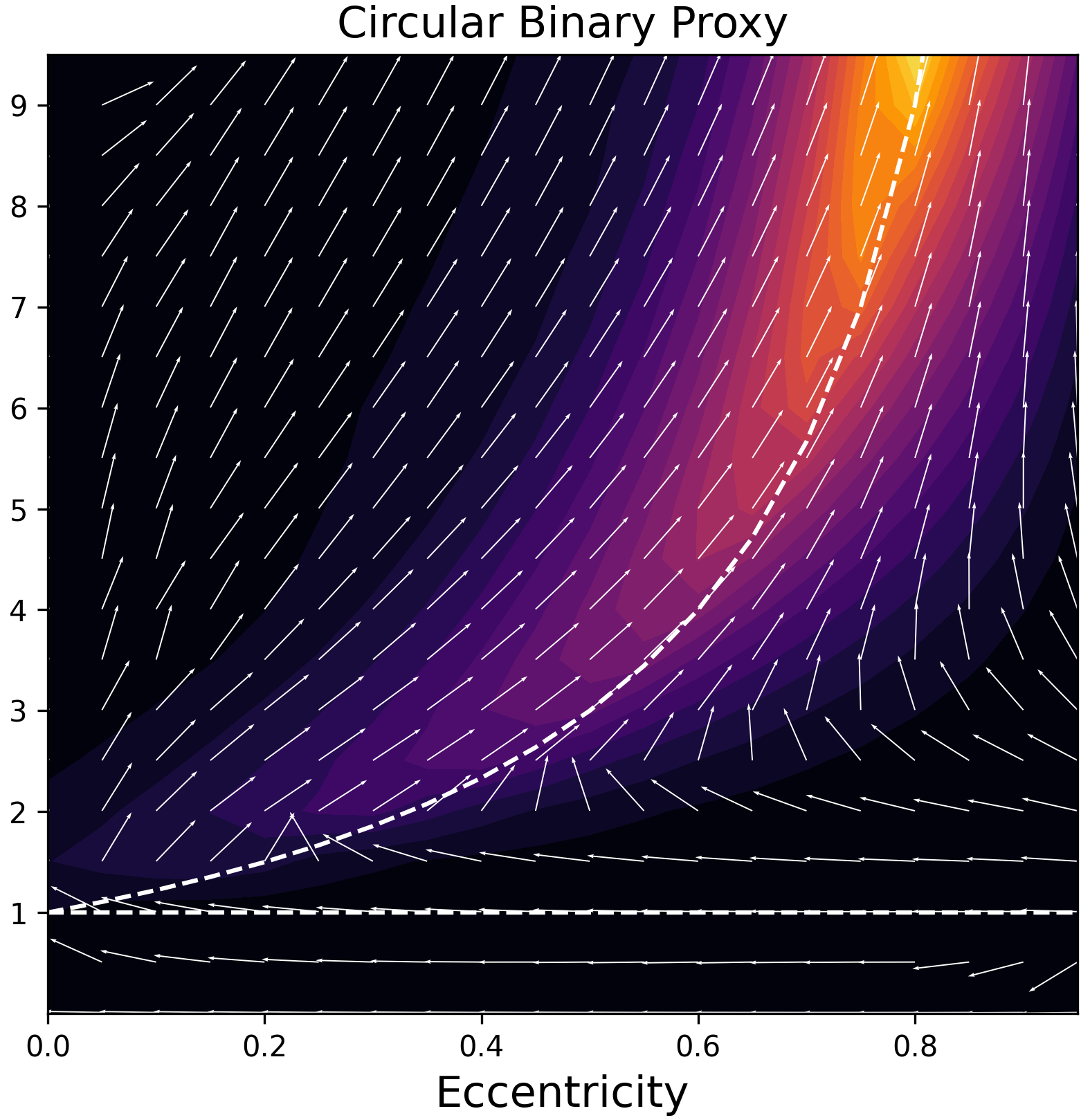}
\endminipage
\minipage{0.335\textwidth}
  \includegraphics[width=\linewidth]{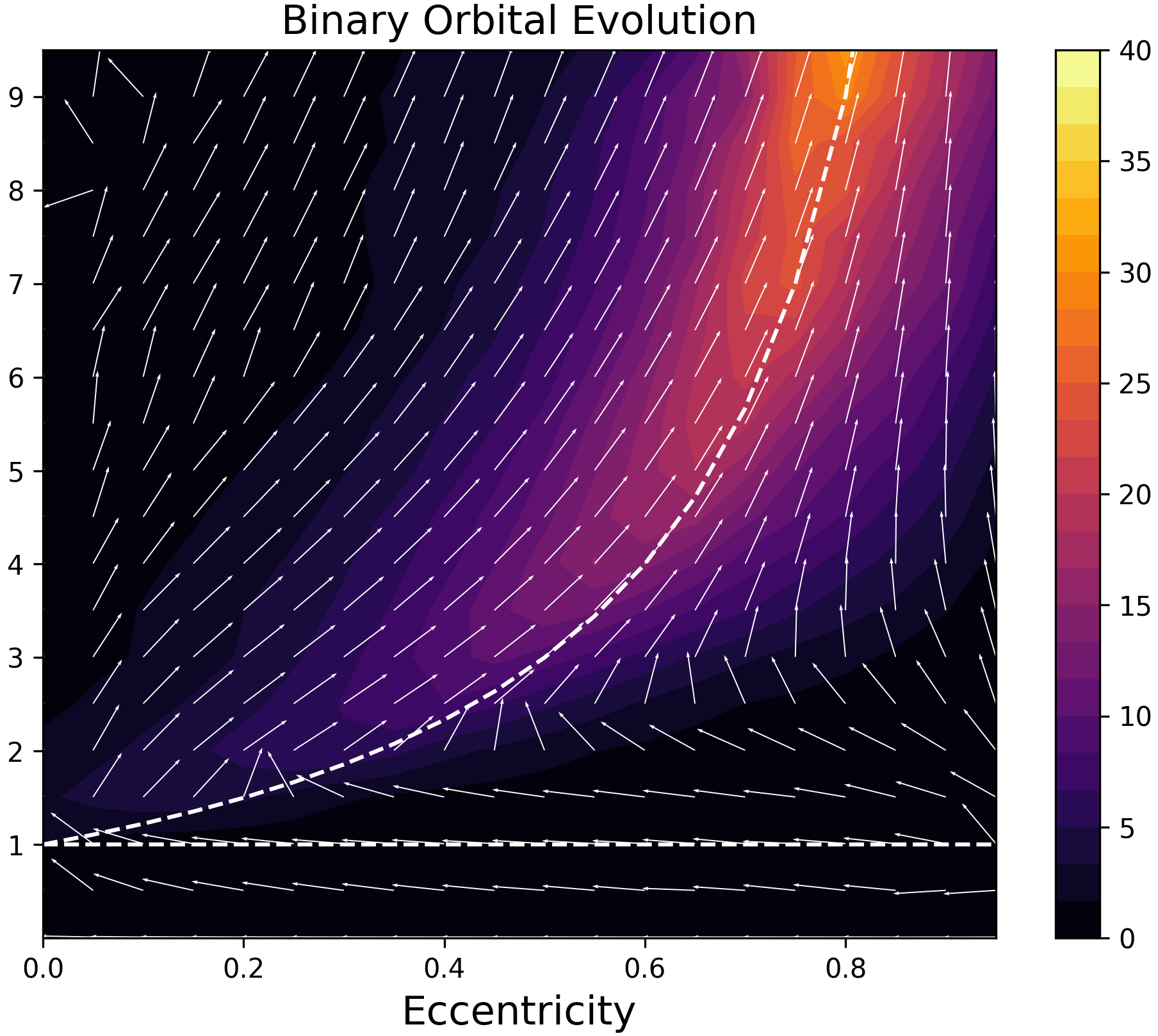}
\endminipage
\minipage{0.34\textwidth}
\hspace{1mm}
\rotatebox{90}{\textbf{\fontsize{8}{8}\selectfont $\sqrt{\dot{e}^2+\dot{\Mach}_p^2}~ \Big[4\pi \tau \mathcal{A} \Omega\Big]$}}
\endminipage
\caption{The same as Fig.~\ref{OrbitTracks} except here we plot the orbital flow of an equal mass binary instead of a single trajectory. Data available at: \url{https://github.com/davidoneill00/Dynamical-Friction-Elliptical-Keplerian-Orbits-Data}}
\label{BinaryOrbitalEvol}
\end{figure}

In analogy to the single perturber, we define the binarys dimensionless quantities $\dot{\mathcal{I}}_\mathrm{a,b}$, $\dot{\mathcal{I}}_\mathrm{e,b}$, and $\dot{\mathcal{I}}_\mathrm{\omega,b}$ exactly the same as in Eqs.~(\ref{EvolutionScalingsA}),~(\ref{EvolutionScalingsE}), (\ref{EvolutionScalingsOmega}) with the only exception being the dimensionless force $\mathcal{F}$ differing from that found on a single trajectory. The binary's change in semi-major axis, eccentricity and precession frequency can be written as
\begin{align}
     \frac{1}{a_\mathrm{b}}\frac{da_\mathrm{b}}{d\left(\Omega t\right)} &= 4\pi\tau\mathcal{A}_{\mathrm{b}}
     \dot{\mathcal{I}}_\mathrm{a,b},\\
     \frac{de}{d(\Omega t)} &= 4\pi\tau\mathcal{A}_{\mathrm{b}} \dot{\mathcal{I}}_\mathrm{e,b},\\
     \frac{d\omega}{d(\Omega t)} &=4\pi\tau\mathcal{A}_{\mathrm{b}} \dot{\mathcal{I}}_\mathrm{\omega,b},
\label{BinaryEvolutionScalings}
\end{align}
where $\tau$ is defined in Eq.~(\ref{TauDef}) and $\mathcal{A}_{\mathrm{b}} \equiv G M_\mathrm{b}/a_\mathrm{b}c_\mathrm{s}^2$, in analogy to Eq.~(\ref{ADef}). We evaluate $\dot{\mathcal{I}}_\mathrm{a,b}, \dot{\mathcal{I}}_\mathrm{e,b}$ and $ \dot{\mathcal{I}}_\mathrm{\omega,b}$ thereby presenting the orbital evolution for a binary system in Fig.~\ref{BinaryOrbitalEvol}.\vspace{-1mm}

The primary contribution of the companion wake is a force in the radial direction. We argue that the companion's wake is strongly localised around the companion's current position, which would naturally lead to a predominantly radial force. This has been observed previously by \cite{DoublePerturbers} for the circular case and is expected to hold at any eccentricity, because there will always be a high density region in the direction of the companion. In terms of orbital decay, the radial force has importance in determining the eccentric evolution of the binary. Hence in Fig.~\ref{BinaryOrbitalEvol} a much stronger circularisation occurs (especially at low Mach numbers). This creates an orbital evolution which is more agnostic of a binary's low Mach initial conditions, and therefore, likely to lose information about it's initial conditions over secular timescales. A binary will have a similar evolutionary path as that of a single trajectory, with no fixed points in the evolutionary flow meaning that the binary will also evolve into being highly eccentric and supersonic.\\

\begin{figure}[!h]
    \centering
    \begin{picture}(0,0)
        \put(90,200){\textbf{\fontsize{13}{21.6}\selectfont $\dot{\omega}~ \Big[4\pi \tau \mathcal{A} \Omega\Big]$}}
    \end{picture}
    \includegraphics[width=0.45\textwidth]{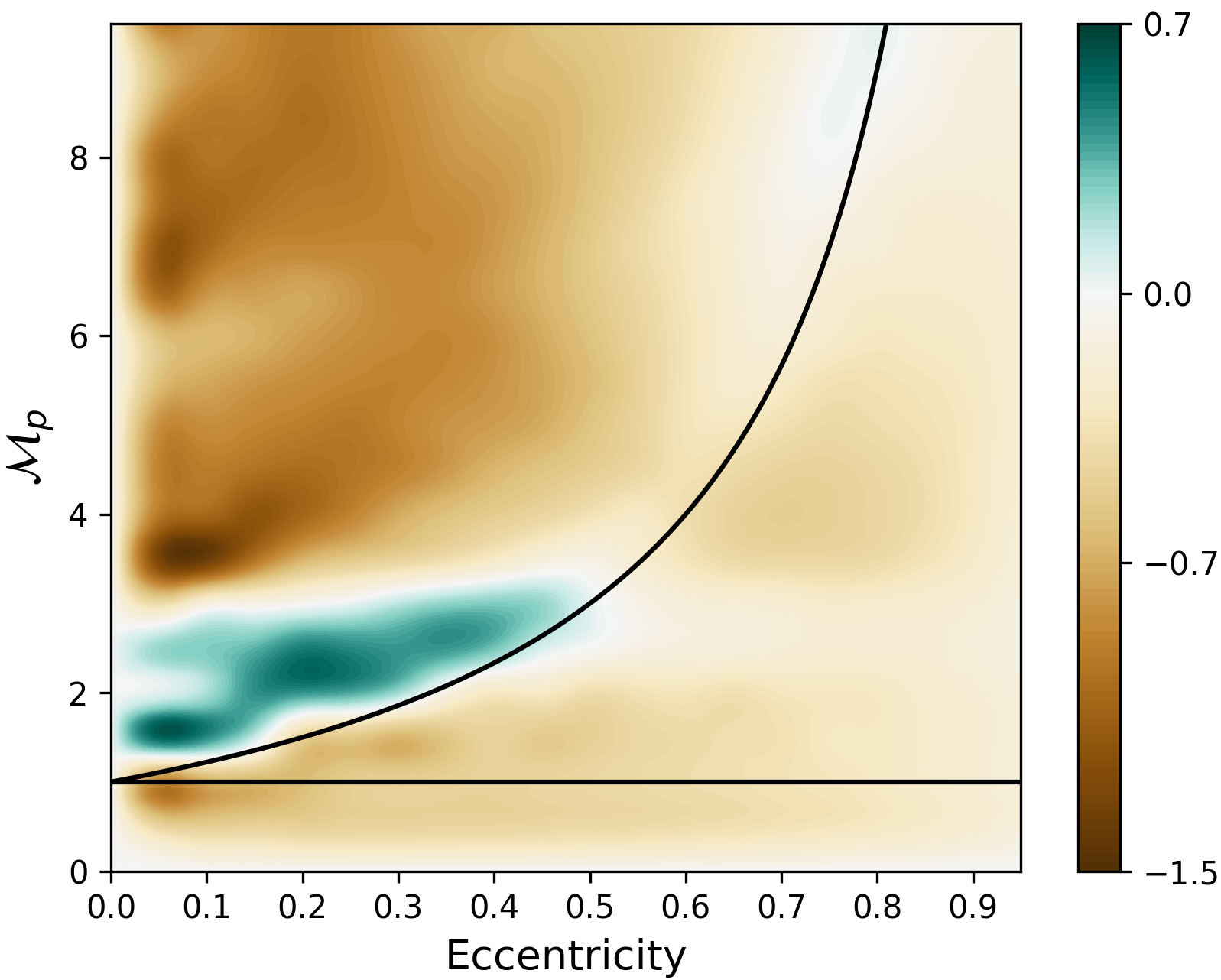}\quad
    \caption{The orbit averaged rate of change of the binary's precession frequency. Once again, the black lines segregate the parameter space into supersonic (upper left), transonic (right) and subsonic (bottom). Data available at: \url{https://github.com/davidoneill00/Dynamical-Friction-Elliptical-Keplerian-Orbits-Data}
    }
    \label{fig:BinaryPrecession}
\end{figure}
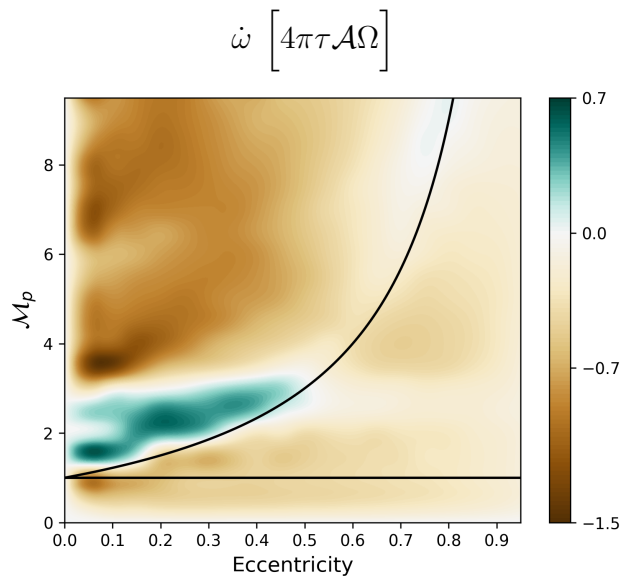

Along with a migration through the parameter space of Fig.~\ref{BinaryOrbitalEvol} determined by changes in semi-major axis and eccentricity, the binary will experience gas induced precession. In Fig.~\ref{fig:BinaryPrecession} we illustrate our results for the binary precession frequency $\dot{\mathcal{I}}_{\omega,\mathrm{b}}$ over our parameter space. We typically observe a negative precession frequency (gold), however for low Mach numbers and low eccentricities, there exist a region of prograde apsidal precession (green).\\

\vspace{9mm}

\subsection{Comparison of Results with the Rectilinear and Circular Proxies}
\begin{figure}[h]
    \centering
    \includegraphics[width = 0.37\linewidth]{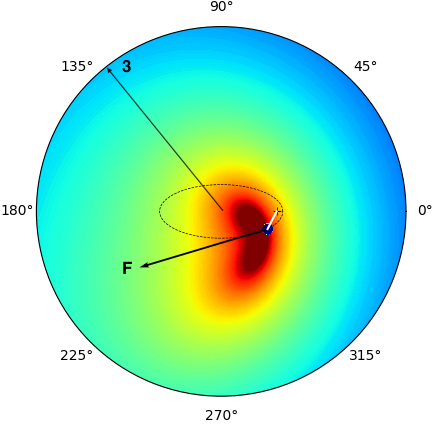}
    \includegraphics[width = 0.45\linewidth]{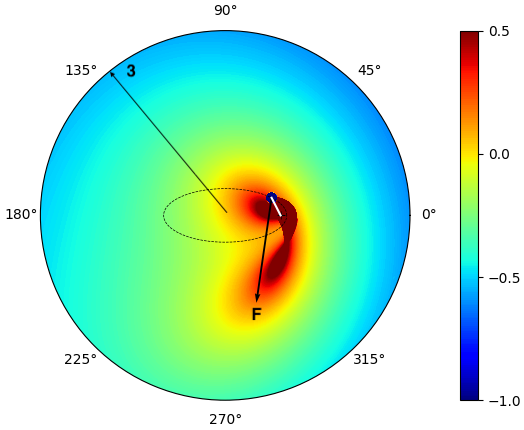}
    \put(-55,198){\rotatebox{0}{$\log_{10}\mathcal{D}(t,\mathbf{x})$}}
    \put(-350,198){\rotatebox{0}{$t = 0.98 P$}}
    \put(-160,198){\rotatebox{0}{$t = 1.02 P$}}

\caption{The force experienced by an orbit with $e = 0.9,~\Mach_p = 3.5$ as it crosses pericenter. The white line connects the current position of the perturber to the focus of the ellipse, while the black arrow is the force experienced by the perturber at that point. We see the sign of the torque changing from negative (left) to positive (right) due to the large curvature of the orbit.}
\label{PositiveTorque}
\end{figure}

The main difference between our findings (for both single trajectories and equal mass binaries) to their respective proxy models is the timescale for evolution (Fig.~\ref{BinaryOrbitalEvol})-- which is found to be longer for self-consistent elliptical trajectories. The first reason for this is the lack of acceleration (along the direction of motion) for the proxies, as only their steady-state values are taken and thus they lack the realization of a transonic crossing. Thereby, the proxies never experience the positive torque from a wavefront leading the perturber which would help to slow the orbital decay. Secondly, the curvature of the orbit around pericenter is so extreme that the direction of the drag force changes over a rapid timescale. Notably, if this change occurs faster than the sound crossing time (which is roughly the timescale required for the wake to reach it's steady state), then the force will not be well approximated by a series of steady state values. To illustrate this, a perturber with eccentricity $e = 0.9$ and pericenter Mach number $\Mach_p = 3.5$ is shown in Fig.~\ref{PositiveTorque}. On the pericenter approach, there is a drag in the positive-radial, negative-azimuthal direction, while post pericenter passage, the force acts in the negative-radial, positive-azimuthal direction. This positive torque from the gas helps to mitigate orbital decay.\\

\begin{figure}[H]
\hspace{35mm}\textbf{
Orbital Evolutionary Tracks: A Comparison to the Proxy Forces
}\par\medskip
\vspace{2mm}
\minipage{0.49\textwidth}
  \includegraphics[width=0.937\linewidth]{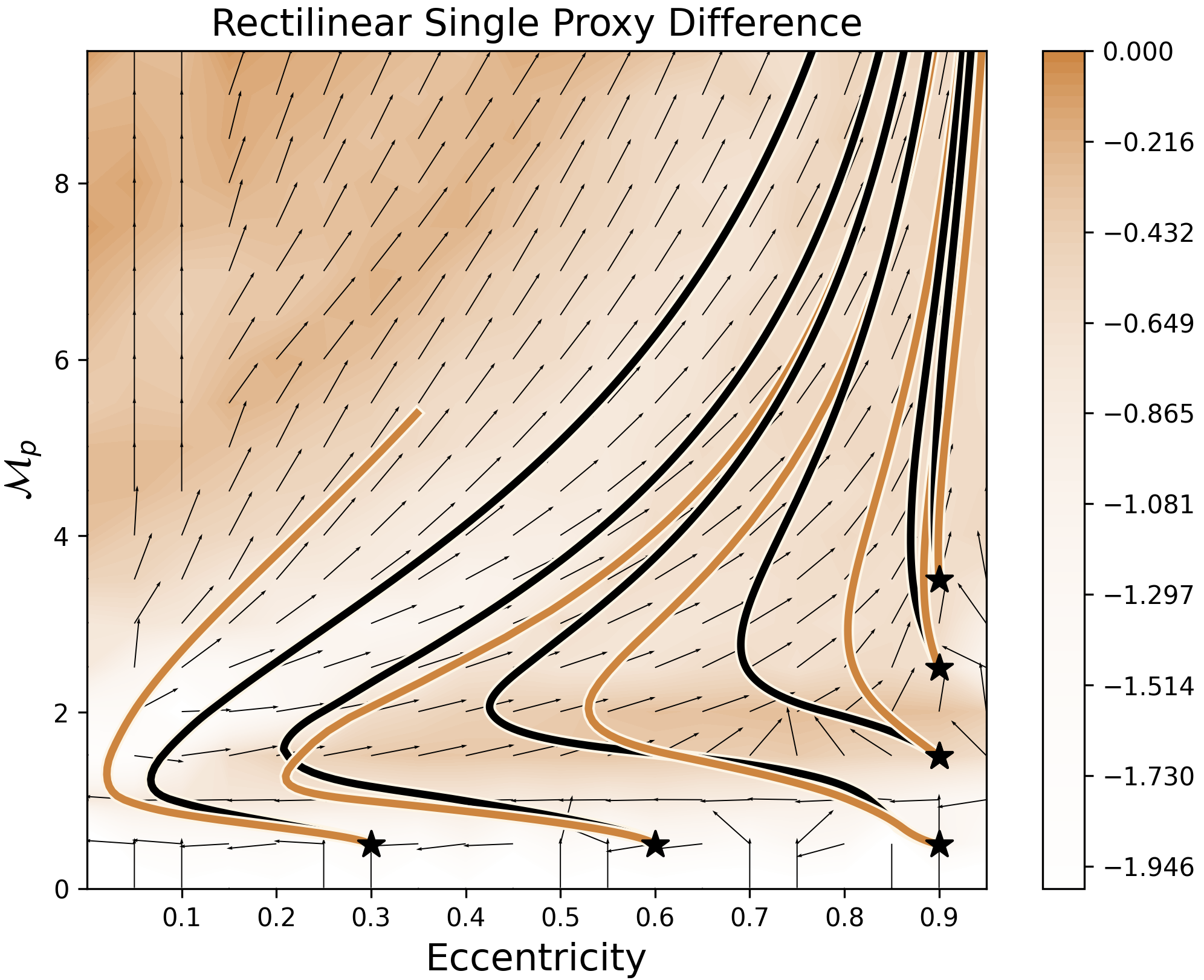}
\endminipage
\put(-52,97){\normalsize $\log_{10}\Delta_R^S$ }
\minipage{0.49\textwidth}
\includegraphics[width=0.956\linewidth]{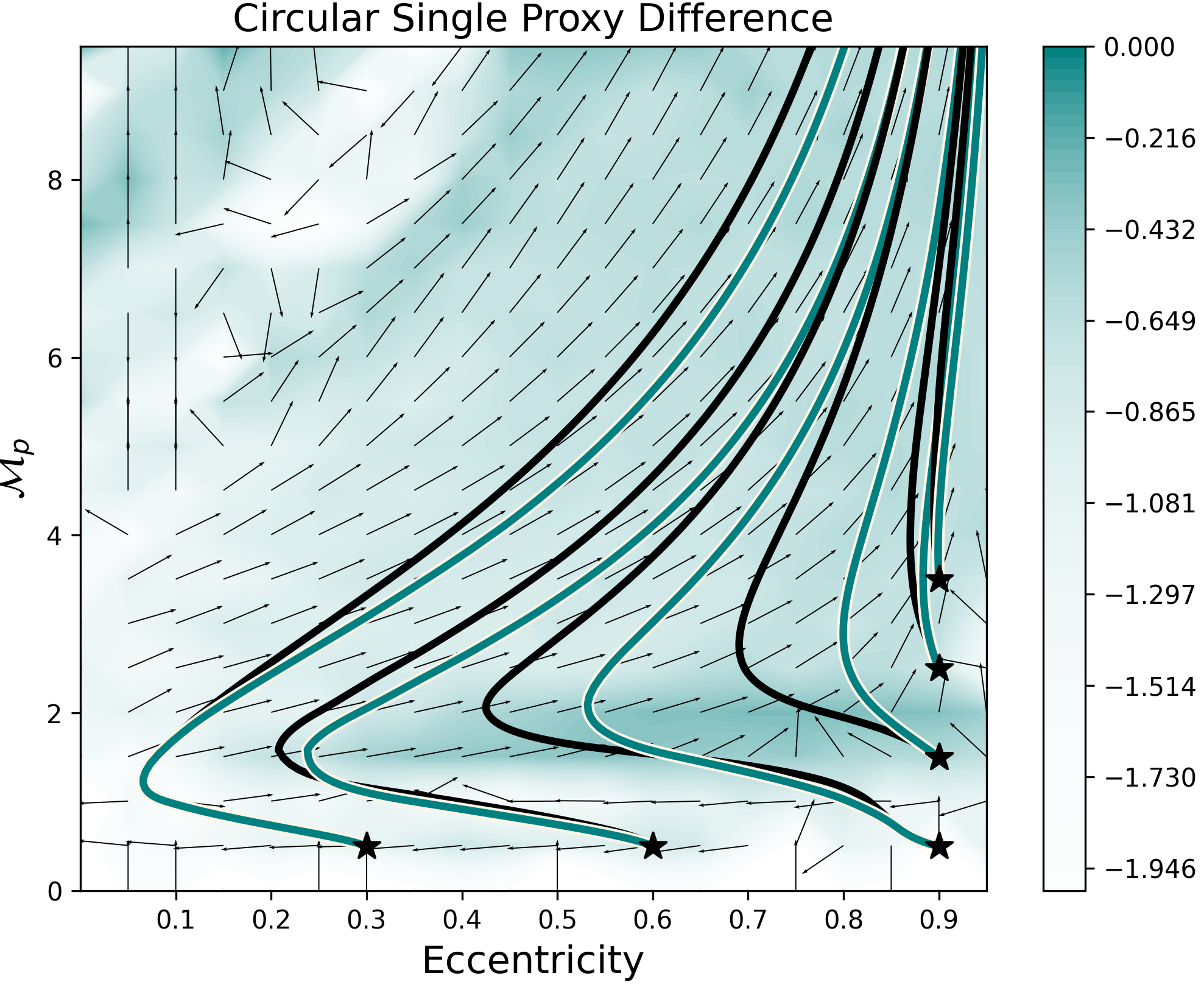}
\put(-40,194){\normalsize $\log_{10}\Delta_C^S$ }
\endminipage

\minipage{0.49\textwidth}
  \includegraphics[width=0.937\linewidth]{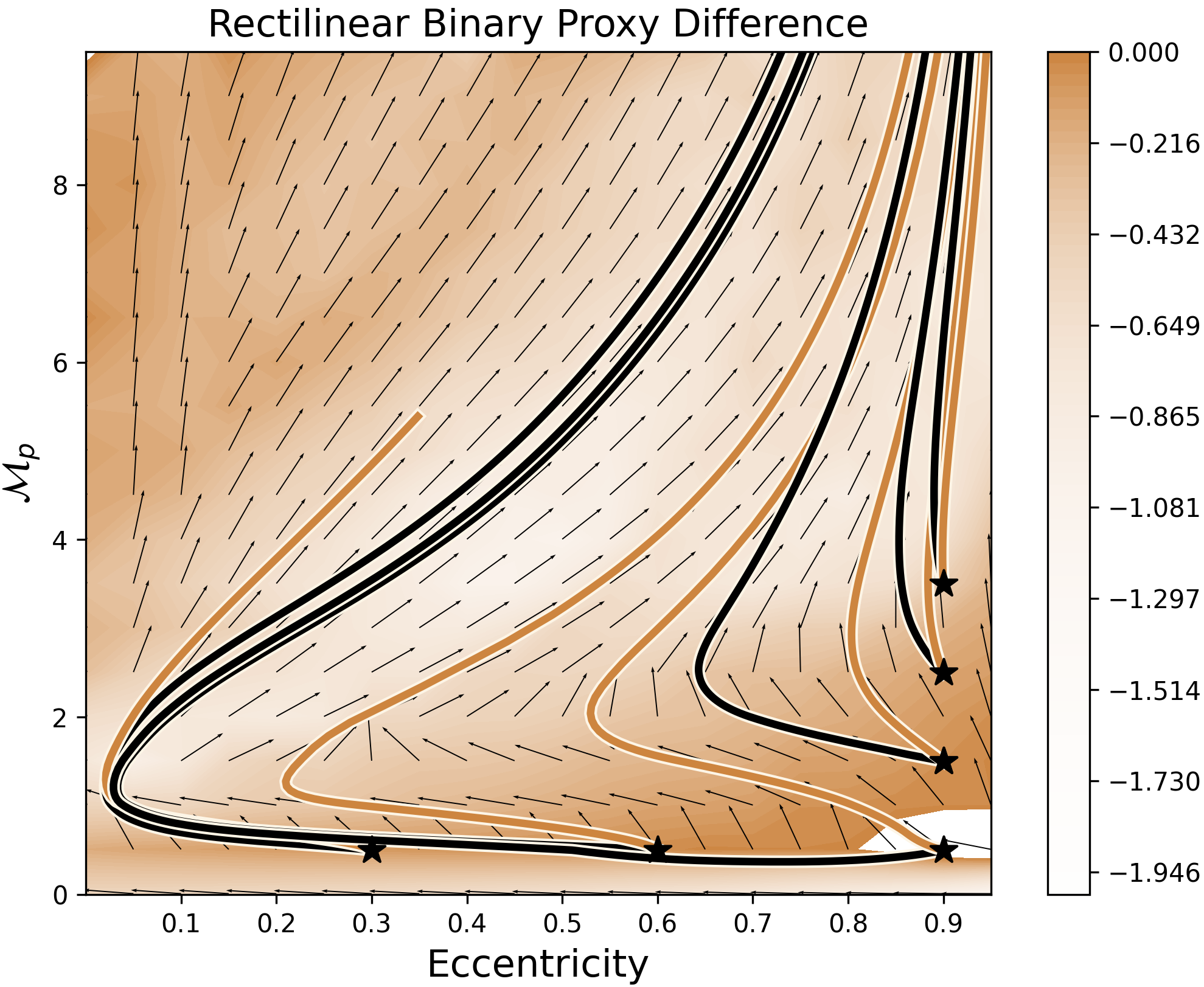}
\endminipage
\put(-52,99){\normalsize $\log_{10}\Delta_R^B$ }
\minipage{0.49\textwidth}
\vspace{-1mm}\includegraphics[width=0.956\linewidth]{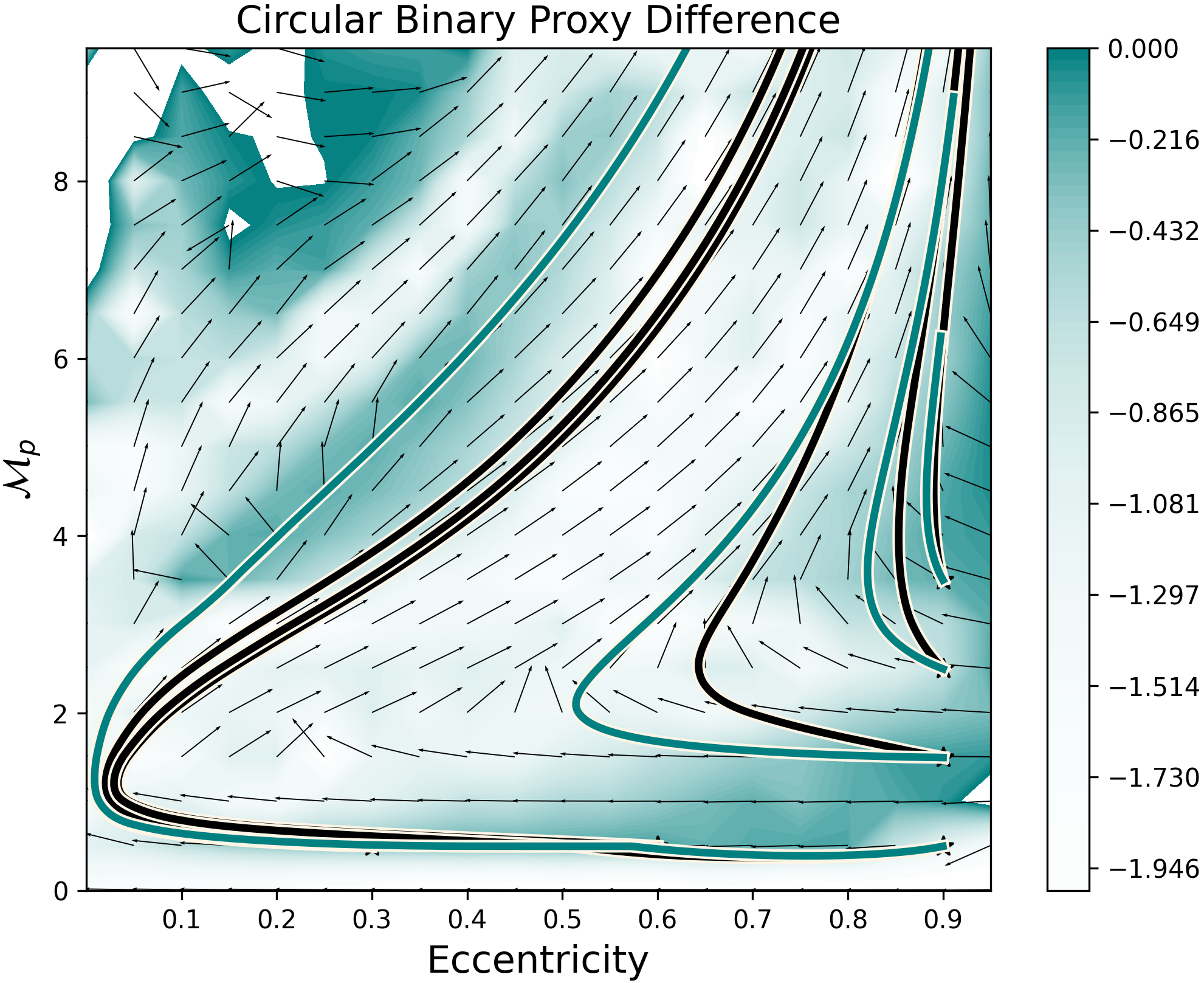}
\put(-40,194){\normalsize $\log_{10}\Delta_C^B$ }
\endminipage
\caption{
Orbital evolution tracks in ($\Mach_p$, $e$) space for specified initial conditions. The black lines are trajectories which have been evolved over secular time according to the results of this work, whereas the orange (left) and teal (right) are those evolved according to the rectilinear and circular proxies. The normalized arrows depict the directional difference in evolution, while the colour scheme illustrates the logarithm of the fractional differences. White regions represent errors at the percent level while darkly colored regions represent errors of order unity. On the lower two panels, we have excluded all regions in which the fractional difference exceeds unity. 
}
\label{fig:IntegrationOfTrajectories}
\end{figure}

The information in Figures~\ref{OrbitTracks} and \ref{BinaryOrbitalEvol} can be used to integrate the orbital evolution over secular timescales. By adopting a variable timestep, we integrate the evolution of a set of initial orbits. In doing so, we can compare $\mathbf{f}$  (defined from Eq.~\ref{VectorDirection}) for the rectilinear and circular proxies to that found using our self-consistent elliptical trajectories. As a measure of accuracy, we plot the logarithm of the fractional difference to each proxy in Fig.~\ref{fig:IntegrationOfTrajectories}, namely
\begin{equation}
    \Delta_{j}^{i} = \frac{\left|\mathbf{f}_{j}^{i} - \mathbf{f}_\mathrm{E}^{i}\right|}{\left|\mathbf{f}_{j}^{i}\right|},
    \label{RelativeErrors}
\end{equation}
at each point in parameter space. Here the index $i$ can represent either single trajectories ($i=\mathrm{S}$) or binary systems ($i=\mathrm{B}$), whereas $j$ represents either rectilinear proxies ($j=\mathrm{R}$) or circular proxies ($j=\mathrm{C}$). The subscript $E$ is used to denote the findings of this work, using elliptical trajectories. We demonstrate two sets of astrophysically motivated initial conditions: the first comprises slowly moving, subsonic trajectories with $\Mach_p = 0.5$. These soft orbits could be created by Jacobi captures inside of AGN disks where one object enters the Hill sphere of a more massive companion and evolves from an initially low Mach number (eg. $\Mach_p\leq{0.5}$ in \cite{GDF_BinaryFormationMechanism}). The second comprise of highly eccentric ($e=0.9$) initial trajectories with Mach numbers $\Mach_p = 1.5, 2.5, 3.5$. These may best describe gas capture events, in which a hyperbolic passing can be aided by GDF in creating a bounded Keplerian orbit, meaning we would expect the orbital dynamics to begin highly eccentric.\\

In Fig.~\ref{fig:IntegrationOfTrajectories} all initial orbits (black stars) evolve into a state of higher eccentricity and higher Mach number. Typically, the proxy models predict a more rapid orbital evolution, reaching this part of parameter space earlier than our findings. We attribute this to the transonic crossing, and the extreme curvature of the orbit near pericenter which can provide a positive torque to the perturber (see Fig.~\ref{PositiveTorque}). Moreover, the circular proxies provide more accurate evolutionary tracks as observed in Fig.~\ref{fig:IntegrationOfTrajectories}. We note that the tangential component of both proxies are in excellent agreement (Fig. 8 \citetalias{KK07}), suggesting that the lack of a perpendicular force component in the rectilinear proxy is the reason for this discrepancy.\\
\vspace{2mm}

\begin{figure}[H]
\hspace{32mm}\textbf{
Orbital Evolutionary Timescales: A Comparison to the Proxy Forces
}\par\medskip
\vspace{2mm}
\minipage{0.49\textwidth}
  \includegraphics[width=0.944\linewidth]{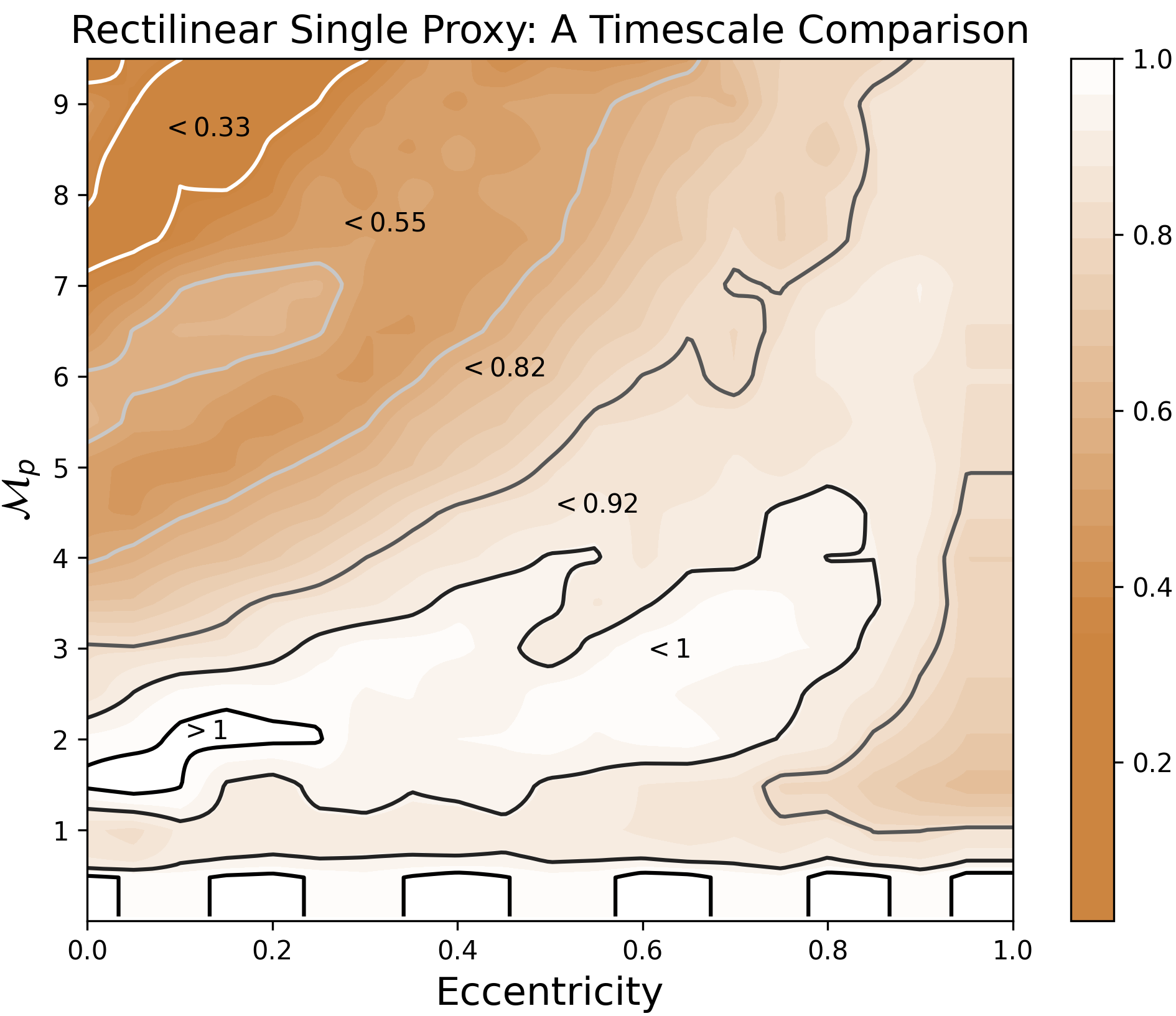}
\endminipage
\put(-32,102){\normalsize $\Lambda_{R}^{S}$ }
\minipage{0.49\textwidth}
\includegraphics[width=0.956\linewidth]{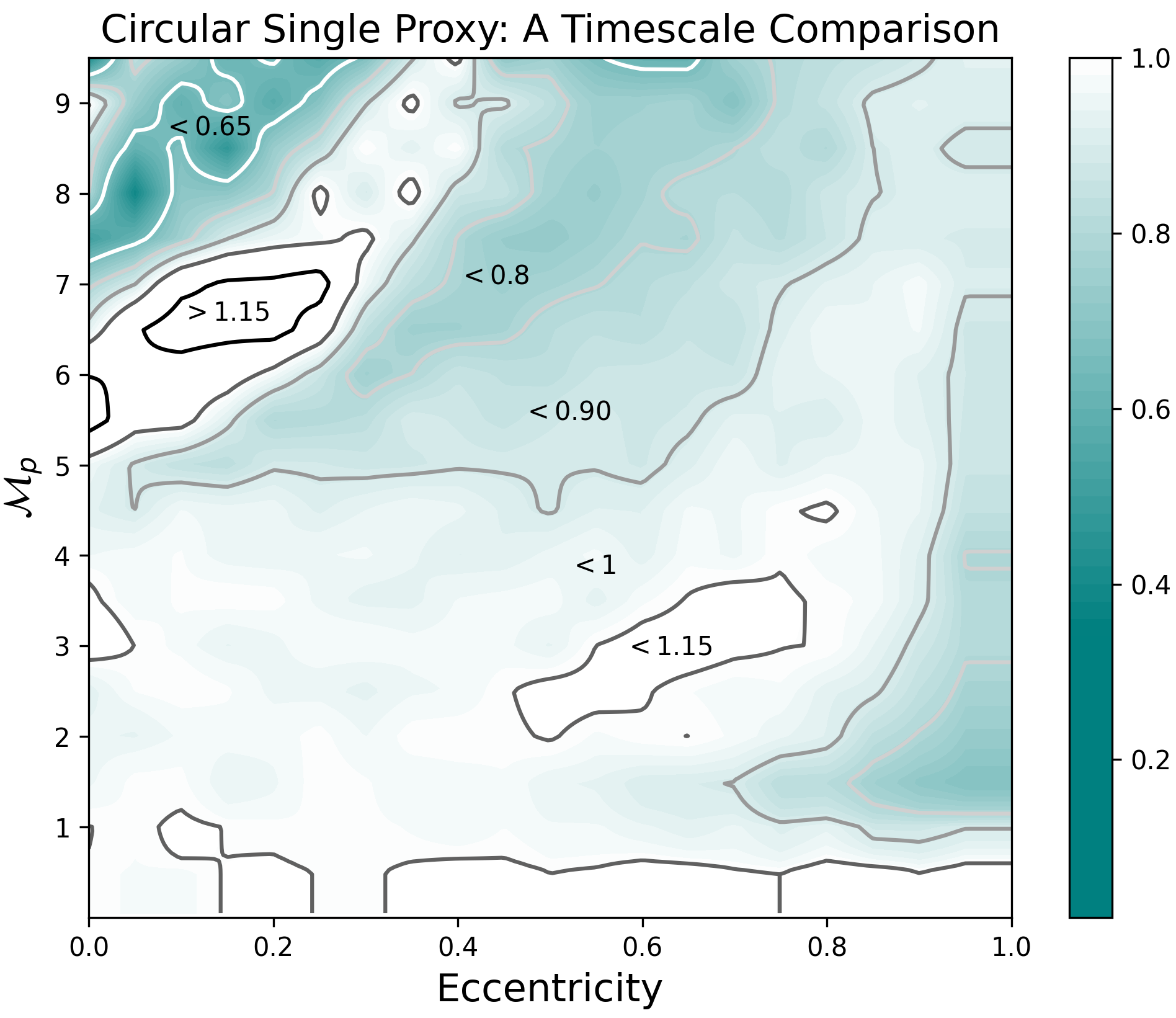}
\put(-20,202){\normalsize $\Lambda_{C}^{S}$ }
\endminipage

\minipage{0.49\textwidth}
  \includegraphics[width=0.944\linewidth]{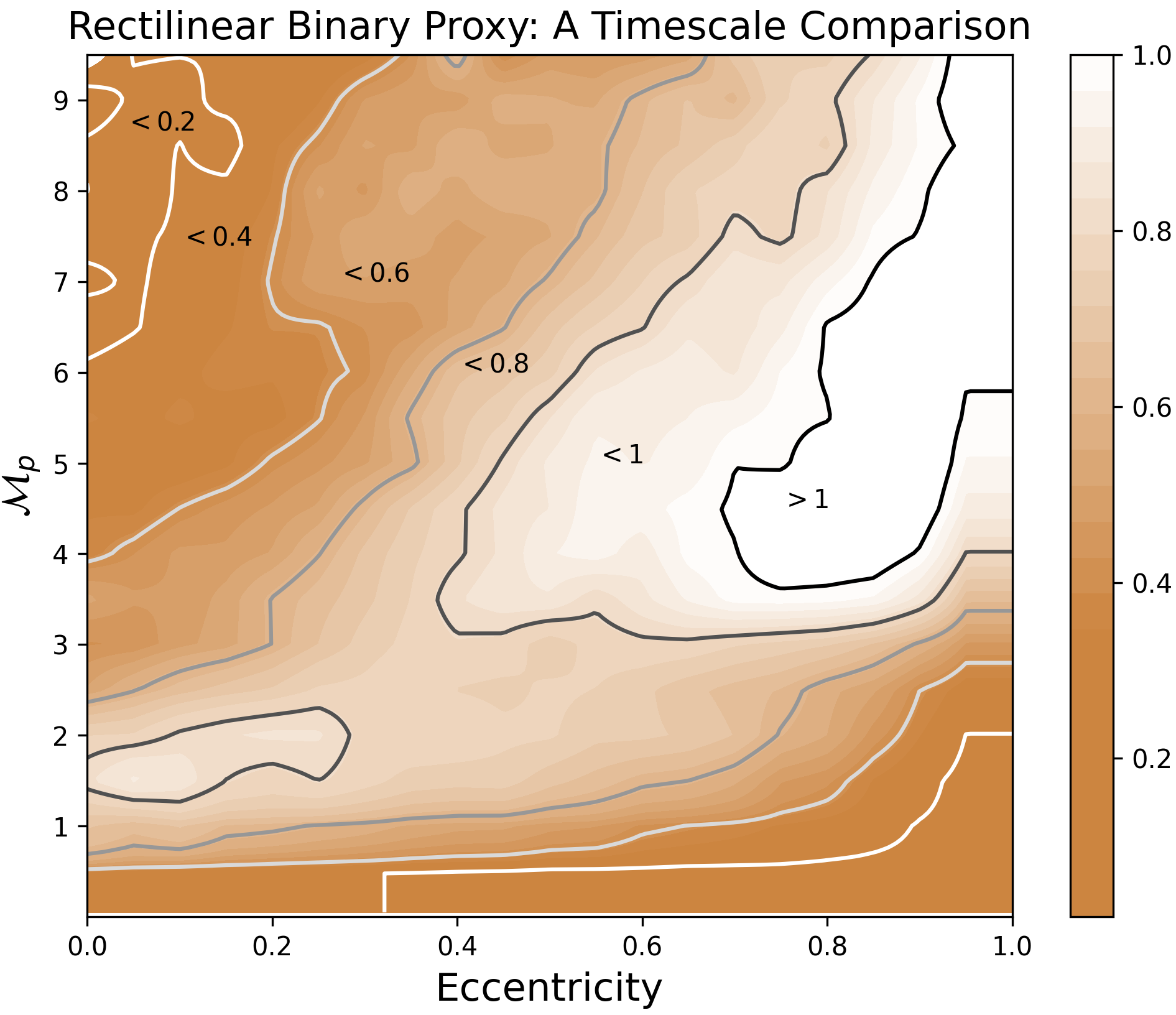}
\endminipage
\put(-32,102){\normalsize $\Lambda_{R}^{B}$ }
\minipage{0.49\textwidth}
\vspace{-1mm}\includegraphics[width=0.956\linewidth]{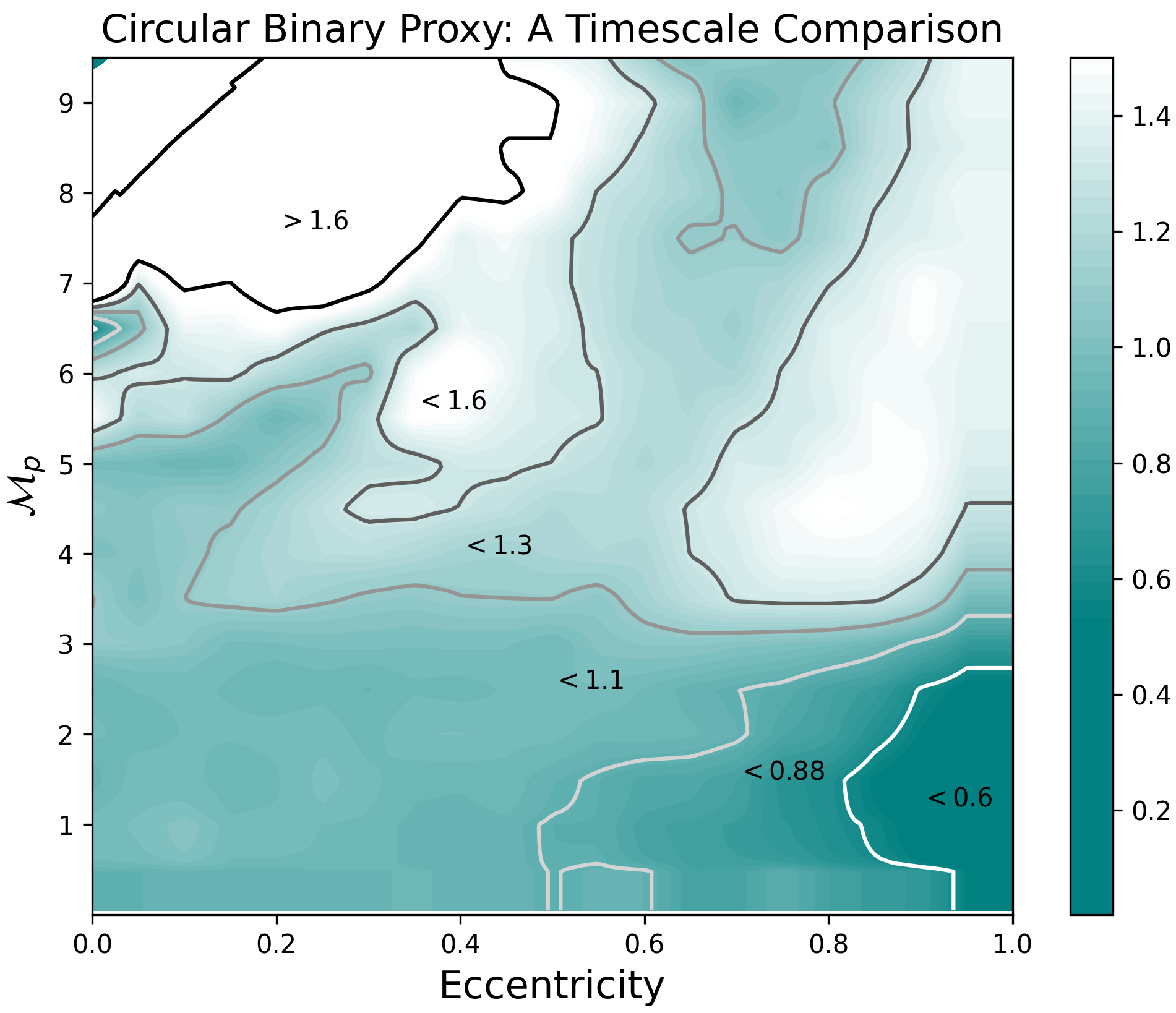}
\put(-20,202){\normalsize $\Lambda_{C}^{B}$ }
\endminipage
\caption{
The ratio of semi-major axis decay timescales between our findings and the proxy models (see text). All proxy models predict the orbital evolution to be too fast ($\Lambda_j^i<1$), with the rectilinear proxies predicting more rapid evolution than the circular proxies. Note the difference in colorbar scales.
}  
\label{fig:TimescalePlot}
\end{figure}

The orbital evolution timescales that we find in this work are lower than that of the proxy values, especially the rectilinear binary proxy. This is likely because the rectilinear binary proxy does not account for the drag force of the companion wake. It has been noted by \cite{DoublePerturbers} that the companion's wake tends to offer a "dynamical boost" and therefore a positive torque which helps to mitigate the orbital decay. By neglecting this effect it may not be surprising that the rectilinear-binary proxy predicts a quicker orbital evolution. Additionally, the sudden sign flip of the torque near pericenter and the supersonic to subsonic crossing will also contribute to a slower orbital decay. To quantify these different evolutionary rates, we plot the ratio of the semi-major axis decay timescales in Fig.~\ref{fig:TimescalePlot}. At each point in parameter space, we evaluate the ratio \begin{equation}
\Lambda_{j}^i=\frac{\dot{a}_{j}^i}{\dot{a}_\mathrm{E}^i},
\label{TimescaleRatioForProxies}
\end{equation}
where $\dot{a}_j^i$ is the rate of change of semi-major axis for a given proxy and $\dot{a}_\mathrm{E}^i$ is the rate measured using self-consistent elliptical trajectories. In this manner, $\Lambda_j^i$ gives an approximate measure of the difference in orbital evolution rates between these models.\\

\vfill

Additionally, in Table~\ref{Tab:Timescales} we present the mean values of $\Lambda_j^i$, averaged over the parameter space. This gives a rudimentary measure of the characteristic timescale for each proxy, normalised to that of the elliptical trajectories. $\bar\Lambda_j^i$ describes how much faster an orbit will decay using the proxies, as compared to the findings of this work.
\begin{table}[H]
\centering
  \begin{tabular}{ c|c|c } 
    \hline
    Proxy Model & $\bar\Lambda_j^i$ & $\sigma_j^i$\\
    \hline
    Rectilinear Single & 0.76 & 0.20\\ 
    Circular Single & 0.91 & 0.11\\
    Rectilinear Binary & 0.62 & 0.28\\ 
    Circular Binary & 1.20 & 0.48\\
    \hline
    \end{tabular}
  \caption{The mean values of $\Lambda_j^i$ along with the standard deviation $\sigma_j^i$ from Fig.~\ref{fig:TimescalePlot}, providing characteristic semi-major axis decay timescales, normalised to the findings of this work. For $\bar\Lambda_j^i<1$ the decay timescales are shorter for the proxies, on average.}
  \label{Tab:Timescales}
\end{table}

\section{Discussion and Conclusion}\label{Conclusion}
\subsection{Summary}
We have implemented a semi-analytical approach to model gaseous dynamical friction for massive perturbers on elliptical Keplerian orbits embedded in a homogeneous, static medium. We have obtained results for both a single perturber and an equal mass binary, embedded in a homogeneous, static gas. Through an analysis of the density wake morphology, we observed distinctive features such as spiral density tails, shock wavefronts and Archimedean spirals within the density wake structure. These features depend on the Mach number and eccentricity of the orbit. In Figs.~\ref{OrbitalChanges},~\ref{BinaryOrbitalEvol} we presented our findings for the gravitational drag force and change in orbital elements culminating in an analysis of how gas drag influences the secular evolution of Keplerian motion. Finally, we compared our findings to proxy force models based on the widely applied prescriptions of \citetalias{O99} and \citetalias{KK07} assessing their validity when applied to eccentric orbits.\\

All Keplerian trajectories are found to evolve into a highly eccentric, supersonic state (given enough time). For supersonic orbits the drag force experienced at apocenter is greater than at pericenter leading to increasing eccentricity, meanwhile the opposite is true for subsonic orbits, the drag force is greater at pericenter than apocenter and eccentricity is damped. We observe the semi-major axis to always decrease due to gas drag, and orbits with mean velocity approximately equal to $\Bar{v}_{\mathrm{orb}} = c_\mathrm{s}$ are most greatly effected (see Fig.~\ref{OrbitalChanges}). A decreasing semi-major axis creates highly supersonic orbits which will drive the orbital eccentricity. In addition to a change in semi-major axis and eccentricity, we observe gas induced apsidal precession, which can be either prograde or retrograde depending on the region of parameter space (see also Fig.~\ref{OrbitalChanges}).\\

We have connected our findings to the preceding works of \citetalias{O99} and \citetalias{KK07} by creating rectilinear and circular proxy force models, respectively. The application of these proxy forces tends to predict a more rapid orbital evolution than found here. In Fig.~\ref{fig:TimescalePlot} we present the ratio of semi-major axis decay timescales between each of the proxies and our findings. The accompanying Table~\ref{Tab:Timescales} presents each proxy's mean timescale, normalised to our results for elliptical trajectories, which can be interpreted as how much shorter (on average) semi-major-axis decay will take for the proxy prediction, compared to the prediction of
our self-consistently computed elliptical trajectory. %In general, the circular proxies are found to be the most accurate proxy models of secular orbital evolution for elliptical orbits. 
The proxies are most accurate along the tracks of orbital evolution in Fig.~\ref{fig:IntegrationOfTrajectories}, suggesting that realistic applications of dynamical friction will return accurate results. Conversely, inaccurate regions of the parameter space are ones in which a trajectory is unlikely to enter.\\

During late stages of this work, we became aware of the study by \cite{AnalyticalElliptical} in which they calculate the gravitational drag force on elliptical single-perturber, transonic orbits. By employing the same analytical method as \cite{Analytical} they obtain results for the orbital evolution of perturbers within the range of Mach numbers $\Mach_p(e) = \Mach_a\sqrt{(1+e)/(1-e)}$ where $\Mach_a\in[0.75,1.05]$. Our results seem consistent in this part of parameter space as we both observe circularisation and semi-major axis decay. However, our results show that the orbits passing through the part of parameter space studied by \cite{AnalyticalElliptical} will ultimately evolve into a region of eccentricity driving leading to highly eccentric orbits as shown in Fig.~\ref{fig:IntegrationOfTrajectories}.\\

\subsection{Additional Physics and Caveats}
\label{sec:Caveats}
The results we have obtained in this study, are based on a number of assumptions and simplified prescriptions. Many of these will need to be relaxed in order to model more realistic astrophysical systems. Below, we include a summary of the caveats associated with the findings of this paper:
\begin{itemize}
    \item \textbf{Background Gas:} 
    Our single trajectory formalism does not account for the density profile of a more massive companion located at the barycenter (which must exist in order for bounded orbits to exist). We note that the case of a small mass ratio binary may alleviate this issue, with the presence of the larger body included. A semi-analytic prescription for the drag force has been derived for perturbers embedded in a linear density gradient by \cite{2022MNRAS.517.4544H}, while \cite{EccentricityEvolution} have used the \citetalias{O99} force as a simple model in power-law density profiles. Likewise, if the perturber was embedded in the flow of an accretion disk, we would expect the shearing velocity flow to form leading and trailing spiral density waves as seen in \cite{2022MNRAS.517.2121F}. Additionally, we have assumed an adiabatic equation of state, however the effects of cooling have been studied by \cite{Velasco_Romero_2018,Velasco_Romero_2020} for an opaque, thermally diffusive gas, accounting for the radiative feedback of a luminous perturber. The results they have obtained are consistent with the findings of \cite{Rad_Feedback} and the linear analysis of \cite{2017MNRAS.465.3175M}. Finally, in the presence of a turbulent medium, GDF becomes less sensitive to the Mach number and less efficient in the transonic ($\Mach \sim 1$) regime \citep{Turbulent}. As a starting point, however, we have chosen a homogeneous, static background to explore the effects of GDF in Keplerian elliptical orbits systematically.
    \item \textbf{Linear Perturbations: } Most semi-analytical models of GDF follow a perturbative approach, discarding higher orders in perturbations. The linear assumption will break down close to the perturber, and the wake solutions will inevitably change as a result. Capturing the non-linearities requires either numerical hydrodynamical studies \citep{ModernDF,KimNonLin,NonLinCirc} or flux conservative laws on a boundary (a distant control sphere, for example \cite{2011MNRAS.418.1238C}) which will provide corrections to the linear GDF forces we have found in this study. For a massive binary, the higher order terms also account for corrections to each component's wake structure due to the background created by it's companion. In this regard, certain non-uniform density backgrounds can be considered by incorporating the non-linear effects of a small mass ratio binary system. These non-linear effects will, therefore, also lead to a different orbital evolution of a massive perturber(s) in a homogeneous gaseous medium  \citep{Antoni_2019, Kaaz_2023}. 
    \item \textbf{Other Forces:} {The dynamical evolution of a massive body is not solely determined by GDF. In addition, an aerodynamic drag would also be expected for any finite extended object, as its surface will interact with the medium \citep{2009ApJ...705L..81V}. Typically compact objects like black holes and neutron stars with sufficiently small surface areas allow this force to be neglected, while stars and planetesimals are often dominated by aerodynamic drag \citep{2015ApJ...811...54G}. Gravitational wave emission can also alter the orbital dynamics along with any other external forces exerted on the perturber. Without considering the impact of gravitational wave emission or aerodynamic drag, our orbital evolution findings may only be directly applicable to compact objects embedded in gaseous media, with separations much larger than their gravitational radii.}
    \item \textbf{Gas Accretion:} The accreted gas will contribute to an increased effective gravitational potential of the perturber, along with a change of the wake due to the presence of a mass and momentum sink, both of which could effect orbital evolution. To understand whether accretion has an important role, one may ask whether the massive perturber would accrete an amount of gas comparable to its initial mass over the course of the trajectory, \textit{i.e.}, comparing mass accretion $m_0/\langle{\dot{m}}\rangle$ and orbital inspiral $a_0/\langle{\dot{a}}\rangle$ timescales\footnote{\cite{Antoni_2019} present numerical results for these timescales in Fig.~11 for a massive, supersonic binary embedded gaseous medium, undergoing Bondi-Hoyle-Littleton accretion.}. Therefore, very soft binaries with long inspiral timescales may acquire significant mass over the course of their trajectory in which case a mass accretion rate $\dot{m}$ should be taken into consideration. The Mach number dependence of dynamical friction for rectilinear motion and considering gas accretion has been investigated by \cite{suzuguchi2024gas}, finding an increase in the drag force for subsonic perturbers, and a decrease for slightly supersonic perturbers. 
    \item \textbf{Closed Keplerian Orbits:}  
    For $\mathcal{A}\tau\ll 1$, the orbital timescale is much shorter than the evolutionary timescale and the assumption of closed, Keplerian orbits remains valid (see Section \ref{sec:Scalings}). If instead the background gas density is exceedingly high, then the orbit will change significantly over the course of a single period and invalidate our assumptions.
    \end{itemize}

\subsection{Potential Applications}
The linear semi-analytical models allow a tractable, analytical approach useful for gaining insight into the larger problem of gas-induced orbital drag. Additionally it represents a benchmark for numerical studies and applications to simplified regimes. We explore some of the potential applications of the results derived in this work, for single perturbers and equal mass binaries.\\

The single trajectory results we have obtained in this paper may be most applicable for extreme mass ratio systems where one of the bodies is nearly stationary with respect to a homogeneous, stationary medium. As a first approach, it is common to incorporate the \citetalias{O99} and \citetalias{KK07} drag forces to model these systems whereas the results obtained in this work can be used to provide a more consistent approach particularly when the evolution of the orbital eccentricity is important. Common examples include the formation and decay of extreme mass ratio inspirals \citep[]{Barausse_2008, 2021MNRAS.506.1007Z}, the evolution of planetesimals in a protoplanetary disk \citep{Planetesimals1} and binary formation mechanisms in accretion disks \citep{GDF_BinaryFormationMechanism}. In such applications, the most important properties to quantify are the timescale to coalescence and the final eccentricity, both of which may require consistent treatment of Keplerian orbits for accurate estimates. Therefore, the results of this work have direct implications to a range of astrophysical systems.\\

Our results for an equal mass binary may hold applications for black holes and planetesimals embedded in dense gaseous environments. The binary setup does not rely on assumptions of a central body, as both components interact with the gas and orbit a common center of mass. This provides a more consistent treatment of the orbital and gaseous dynamics with applications including stellar mass black holes in gas rich globular clusters \citep[e.g.,][]{GDF_inGlobularClusters}, episodes of common envelope evolution in stellar binaries where a giant star has engulfed a compact companion \citep[e.g.,][]{2023arXiv231106332B}, supermassive black hole binaries \citep[e.g.,][]{DoublePerturbers, Escala_2005}, or binary planetesimals in a protoplanetary disk \citep[e.g.,][]{Planetesimals2}. In addition, we note that the orbital evolution of equal mass binaries are less sensitive to initial conditions (Fig.~\ref{fig:IntegrationOfTrajectories}), making their orbital evolution tracks more uniform for initially subsonic binaries. This could hold important implications for black hole binaries, whose eccentric evolution is important in determining its properties when entering the gravitational wave inspiral regime.\\

The practicality of the analytical proxy forces make them straightforward to implement and a useful tool for modelling the gas drag on Keplerian orbits. We have compared our findings with the proxy models, and quantified the proxy accuracies in different regions across the parameter spaces (see Figs.~\ref{fig:IntegrationOfTrajectories}, \ref{fig:TimescalePlot}). On the one hand, we note that along the tracks of orbital evolution, the percentage error is quite minimal, suggesting that the application of the proxies will be accurate. On the other hand, we have identified the regions of parameter space which are best avoided when employing more simplified proxy models. For specific applications of the force, it is now possible to quote the error associated with using the proxy forces as compared to using self-consistently computed elliptical trajectories.\\

The semi-analytical results of this paper represent a tractable approach which should be recoverable for future numerical simulations. Both Figs.~\ref{OrbitalChanges} and \ref{BinaryOrbitalEvol} should be recoverable in the limit of low mass perturbers, representing a valuable test for future numerical studies. Of particular interest may be studies regarding both, non-linear effects and more general mass ratio binaries. These extensions represent important implications for modelling astrophysically realistic environments, and together may alleviate many of the assumptions which have been made in this work, regarding the background gas dynamics (as described in Section~\ref{sec:Caveats}). We leave these extensions to future work.

\acknowledgements
We thank Hagai Perets, Will Farr, and Scott Tremaine for useful discussions. This work was supported by Villum Fonden Grant No. 29466 and ERC Starting Grant no. 101043143. D.J.D. received funding from the European Union's Horizon 2020 research and innovation programme under Marie Sklodowska-Curie grant agreement No. 101029157, and from the Danish Independent Research Fund through Sapere Aude Starting Grant No. 121587. M.E.P. gratefully acknowledges support from the Independent Research Fund Denmark via grant ID 10.46540/3103-00205B and the Institute for Advanced Study, where part of this work was carried out. The Tycho supercomputer hosted at the SCIENCE HPC center at the University of Copenhagen was used to support this work.

\bibliography{bibliography.bib}

\appendix

\section{Numerical Convergence}
\label{sec:Convergence}
We test convergence for the radial size of the domain $r_\mathrm{d}$ (centered on the perturber's current position), the spatial resolutions $N_r,N_\theta$ and illustrate the number of evaluations per orbit $N_t$.\\

\subsection{Spherical Domain}
\begin{figure}[h!]
\hspace{55mm}\textbf{
Convergence over Spherical Domain Radius
}\par\medskip
\vspace{2mm}
\centering
\includegraphics[width=.32\textwidth]{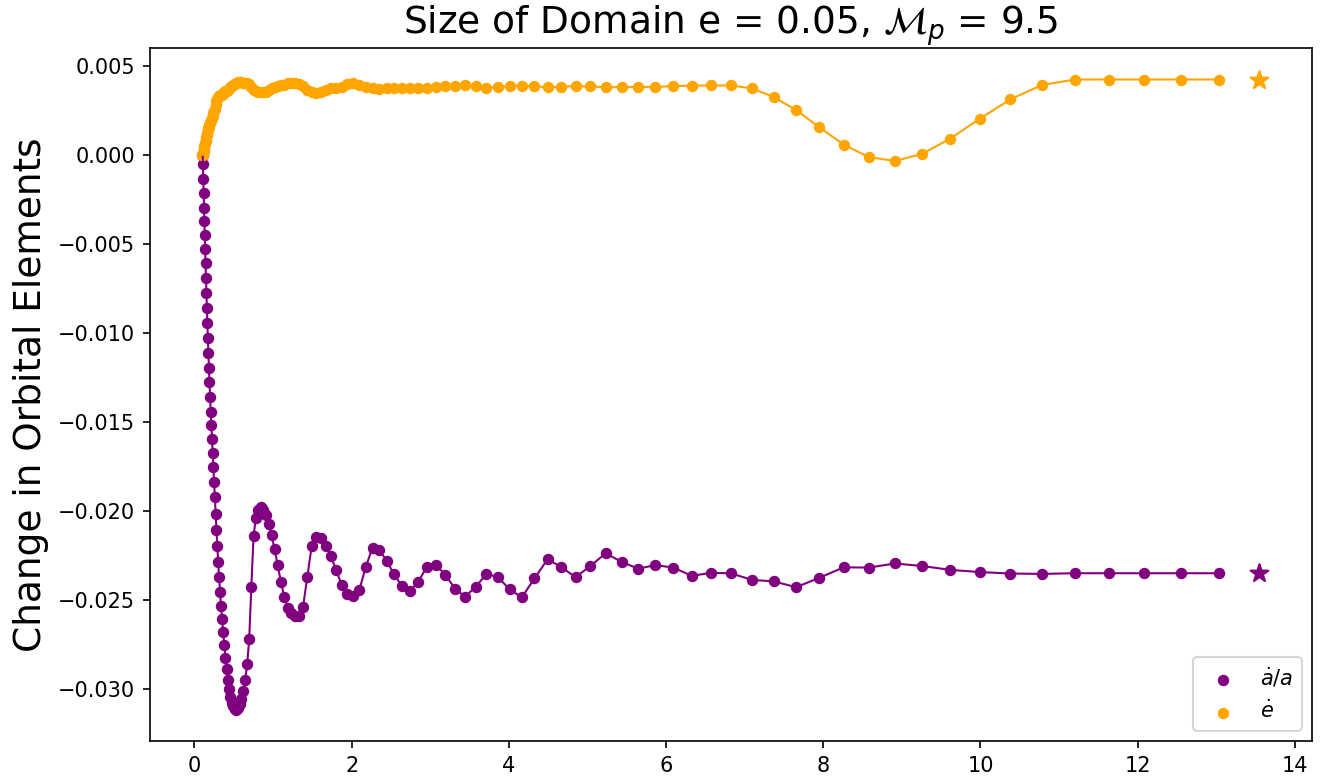}
\includegraphics[width=.31\textwidth]{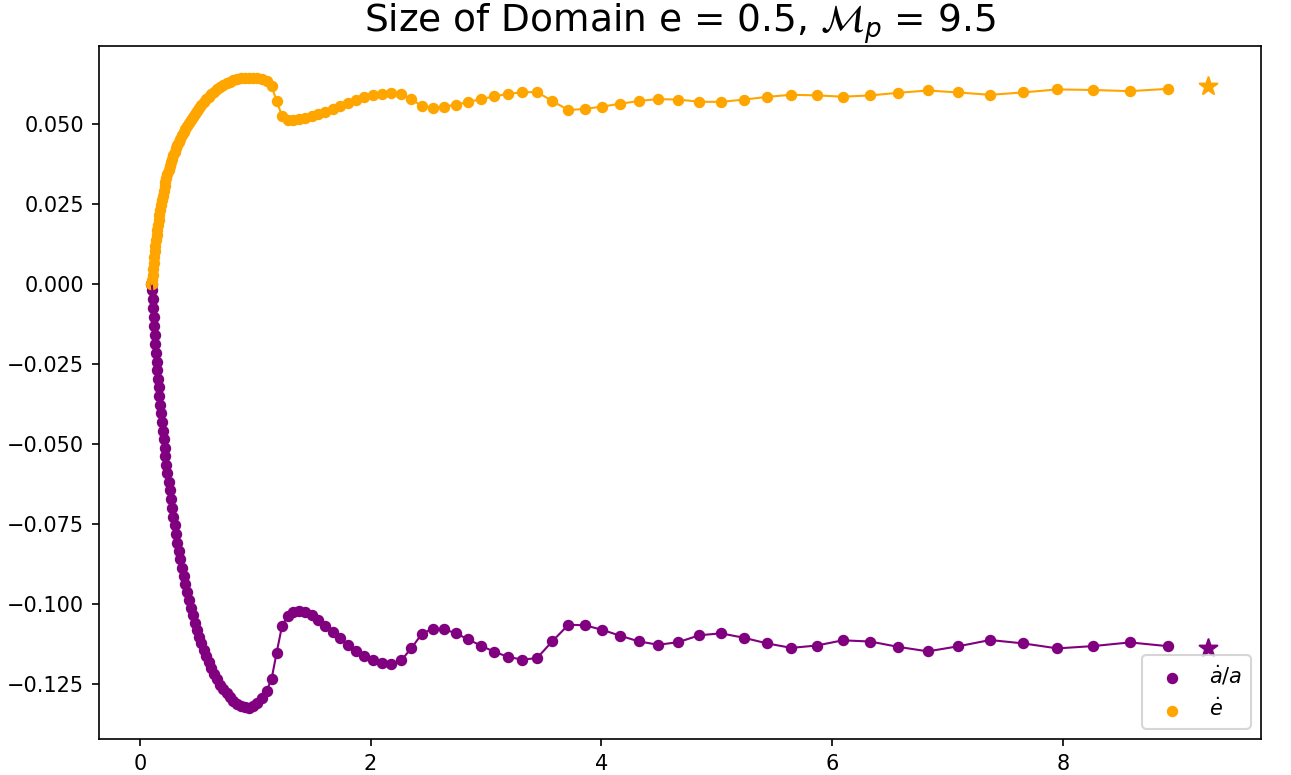}
\includegraphics[width=.305\textwidth]{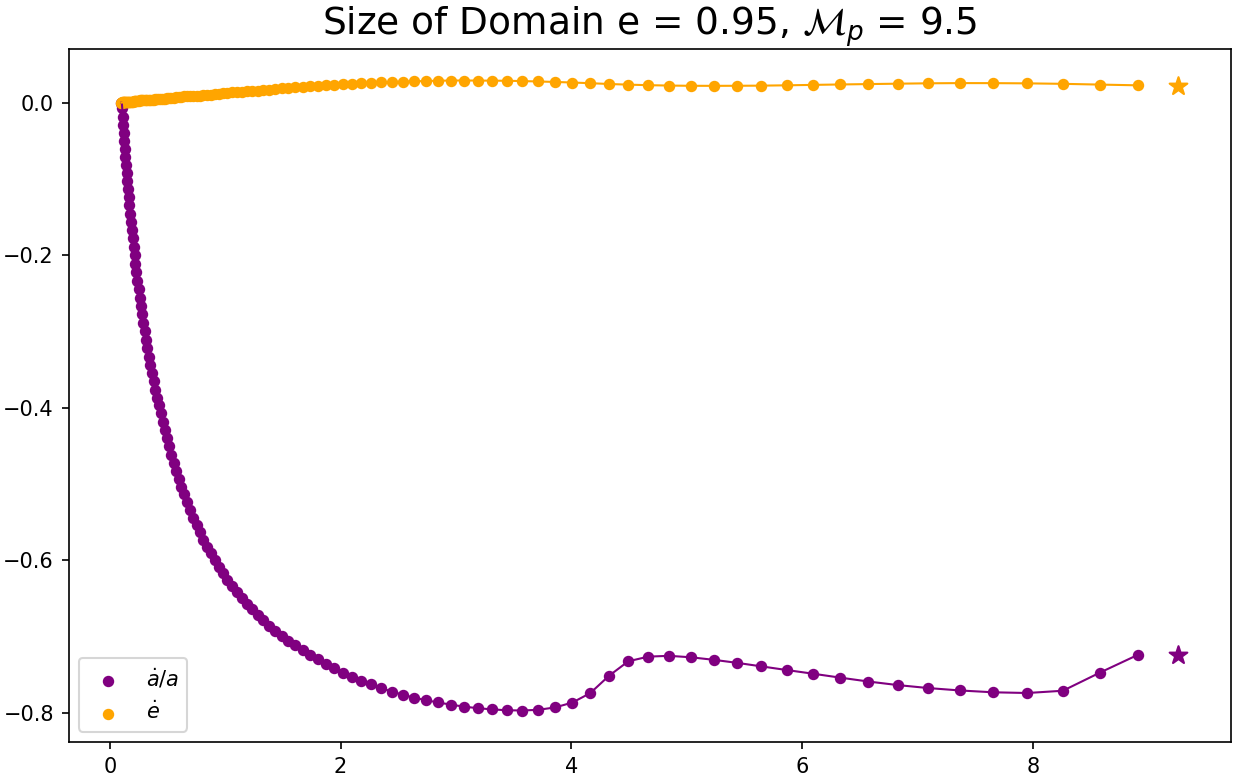}
\medskip

\includegraphics[width=.32\textwidth]{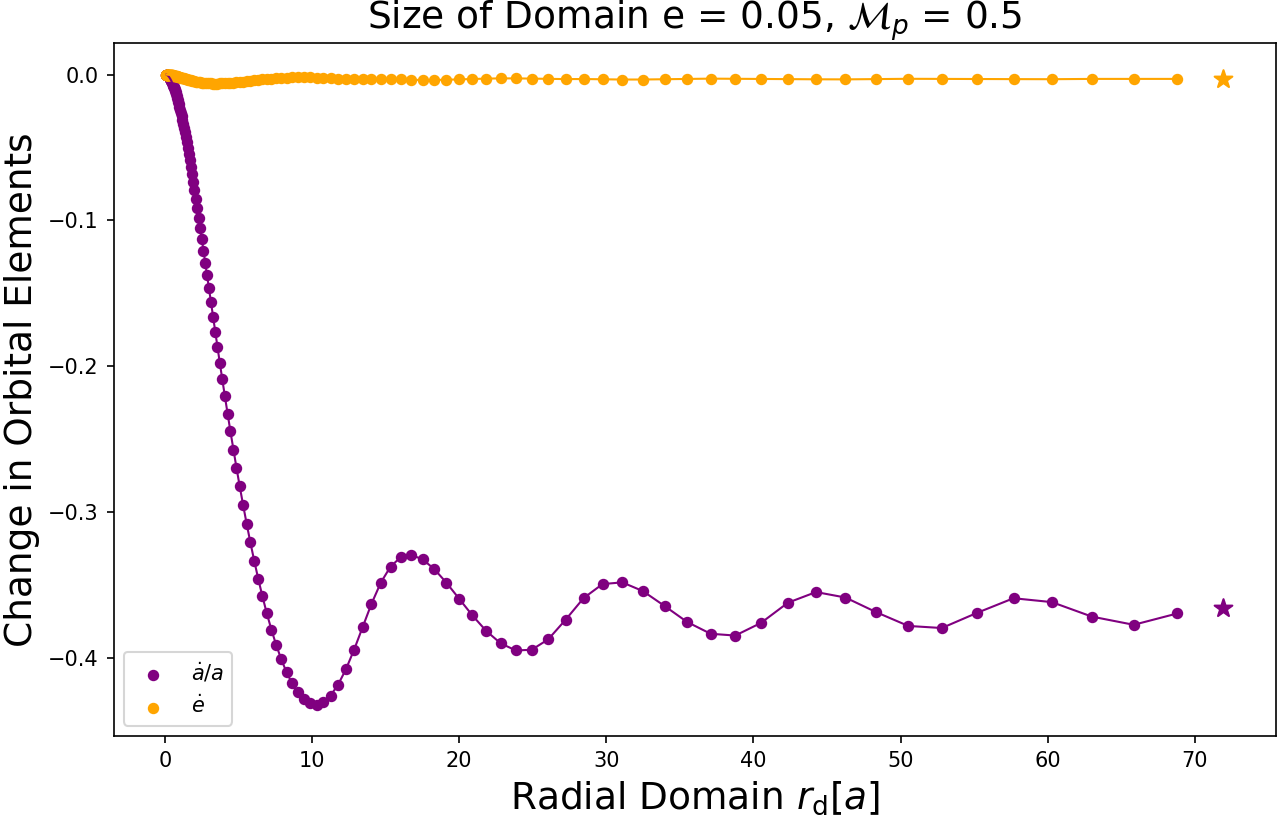}
\includegraphics[width=.31\textwidth]{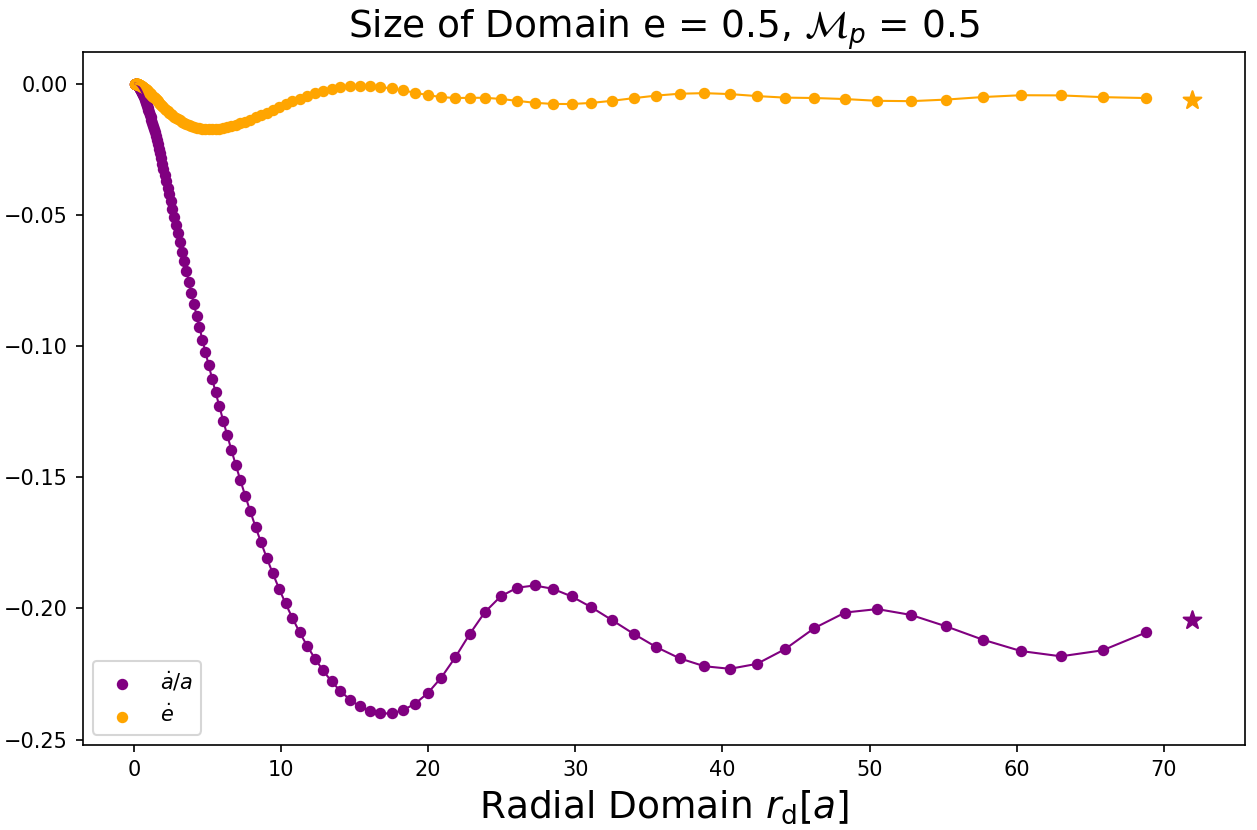}
\includegraphics[width=.305\textwidth]{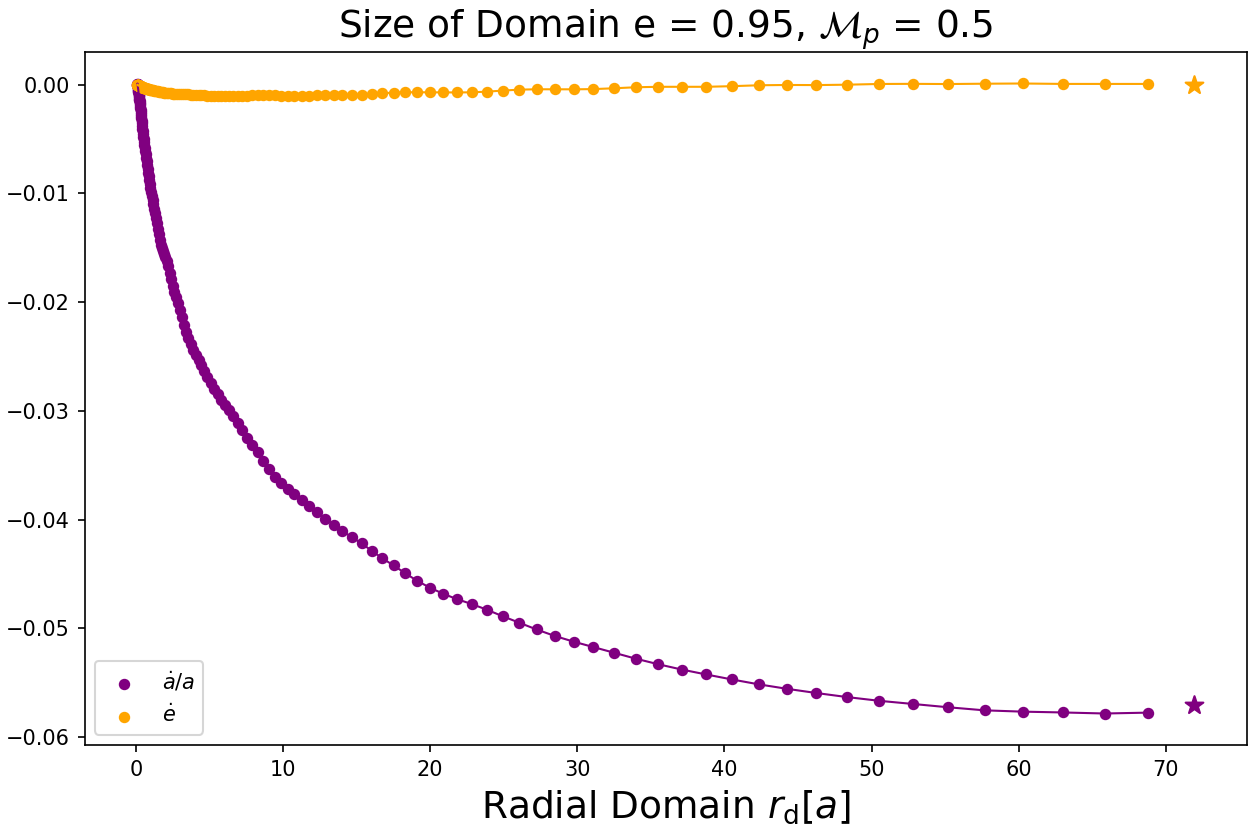}

\caption{The change in $\dot{a}/a, \dot{e}$ as a function of box size for eccentricities $e~=~0.1,~0.5,~0.9$ and pericenter Mach numbers $\Mach_p~=0.5,~9.5$. All values seem to be well converged for our choice of radius $r_\mathrm{d}$. In each panel, the star denotes the value of $r_\mathrm{d}$ which has been used in this study.}
\label{BoxConvergence}
\end{figure}

During each force computation, it is straightforward to construct a series of radially-nested subgrids with radii $r<r_\mathrm{d}$. Over these sub-domains, we re-compute the force, evaluating the sensitivity of $\dot{a}/a$, $\dot{e}$ to the radius of the spatial domain. In Figure \ref{BoxConvergence} we illustrate examples of convergence for selected cases of single trajectories, noting that equivalent equal mass binary cases exhibit identical behaviour.\\

\subsection{Resolution of Spatial Domain}
In a three dimensional logarithmic, spherical grid there are two independent resolution parameters $N_r, N_\theta$\footnote{Where the zenith resolution is defined as $N_\phi \equiv N_\theta/4$ and therefore not an independent parameter.}. We vary both of these quantities-- while keeping the range of the spatial domain and time at which they are evaluated fixed. This allows us to measure the sensitivity of $\dot{a}/a,~\dot{e}$ to the number of points we have sampled in the wake. The wake becomes sharper and more discontinuous at higher Mach numbers suggesting that the highest resolution will be necessary for these trajectories.   
\vspace{10mm}

\begin{figure}[h!]
\centering
\textbf{Single Orbits}
%\hspace{40mm}\textbf{Radial Convergences }

\vspace{4mm}

\includegraphics[width=.315\textwidth]{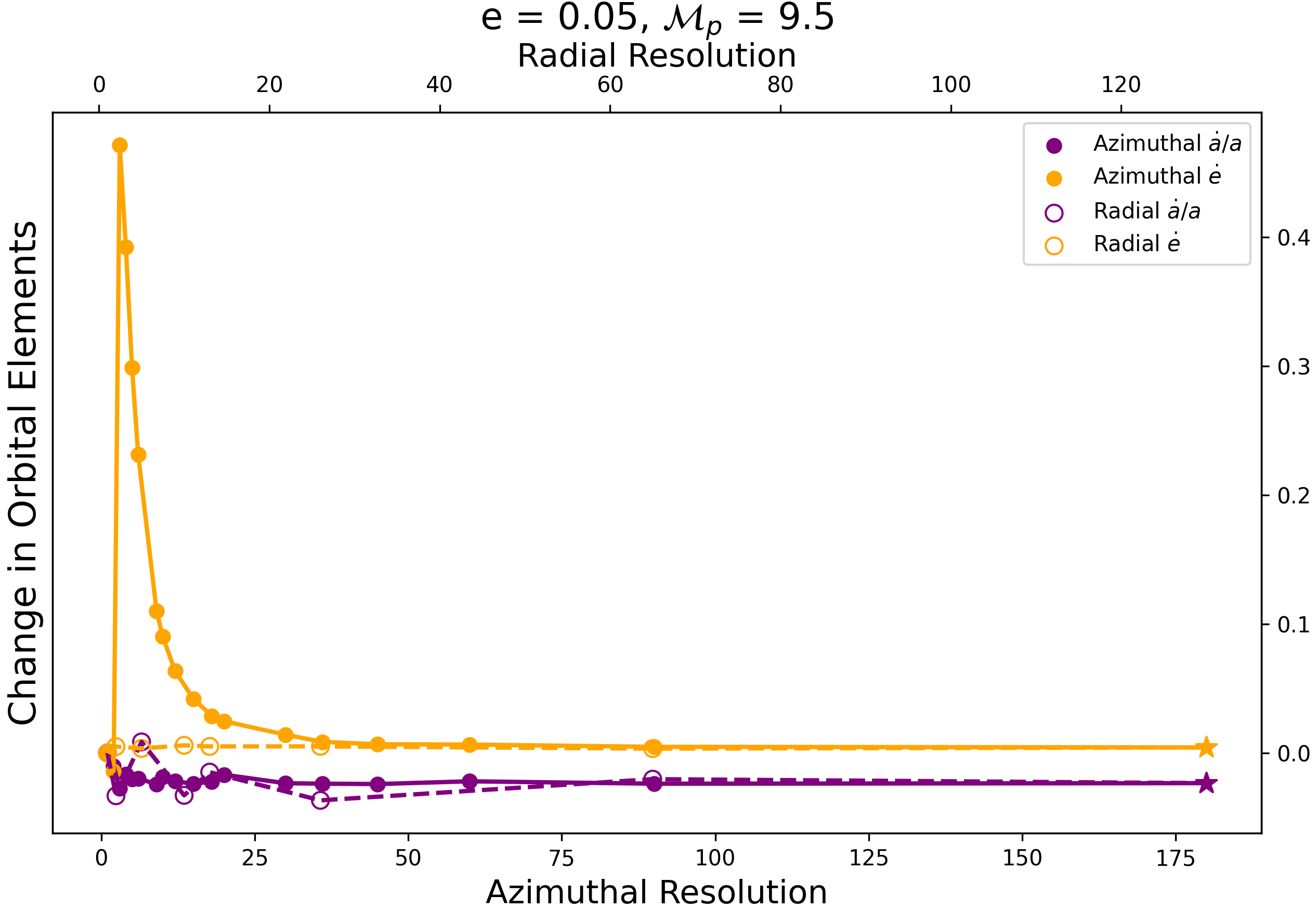}
\includegraphics[width=.31\textwidth]{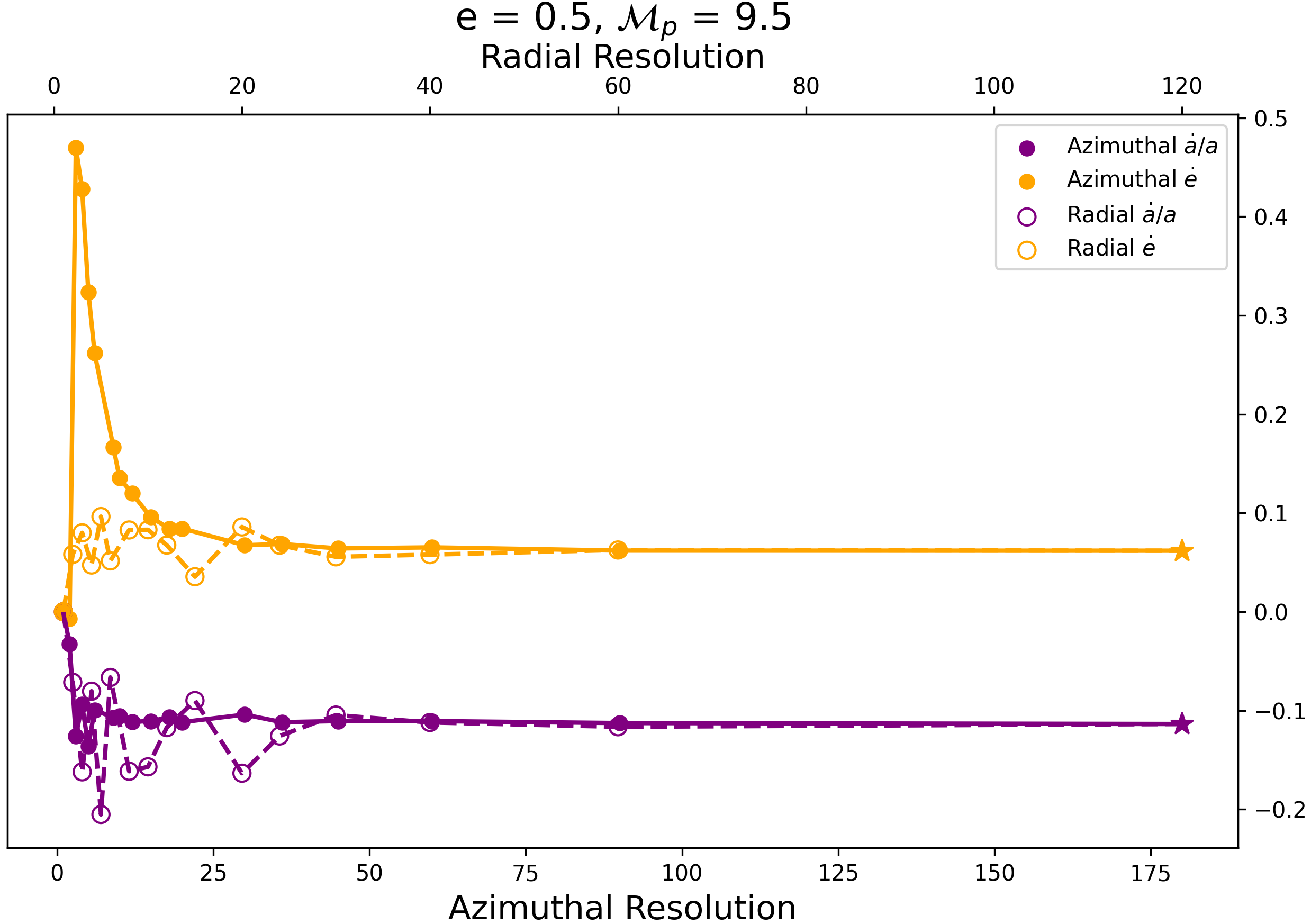}
\includegraphics[width=.31\textwidth]{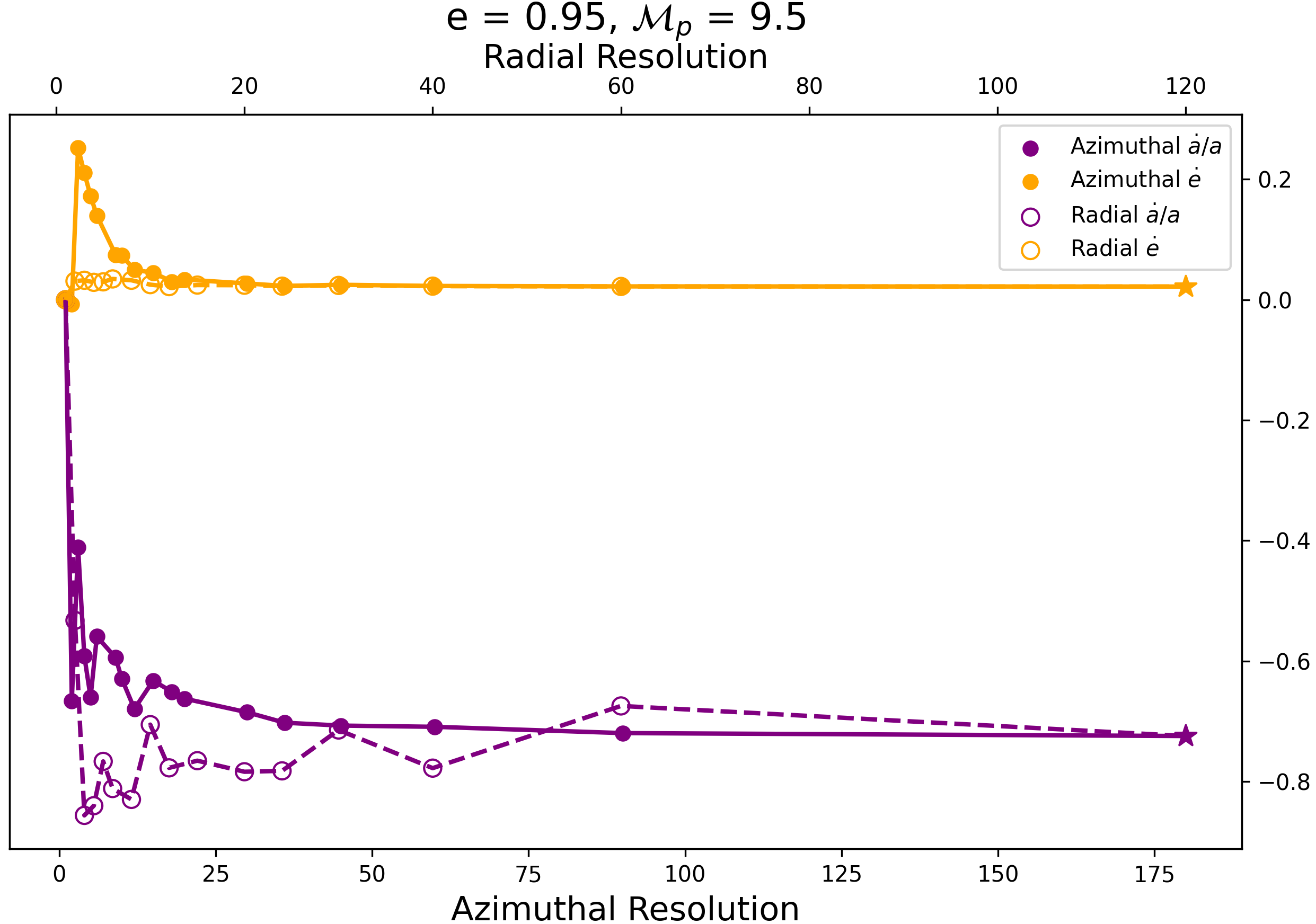}
\medskip

\hspace{-1mm}\includegraphics[width=.324\textwidth]{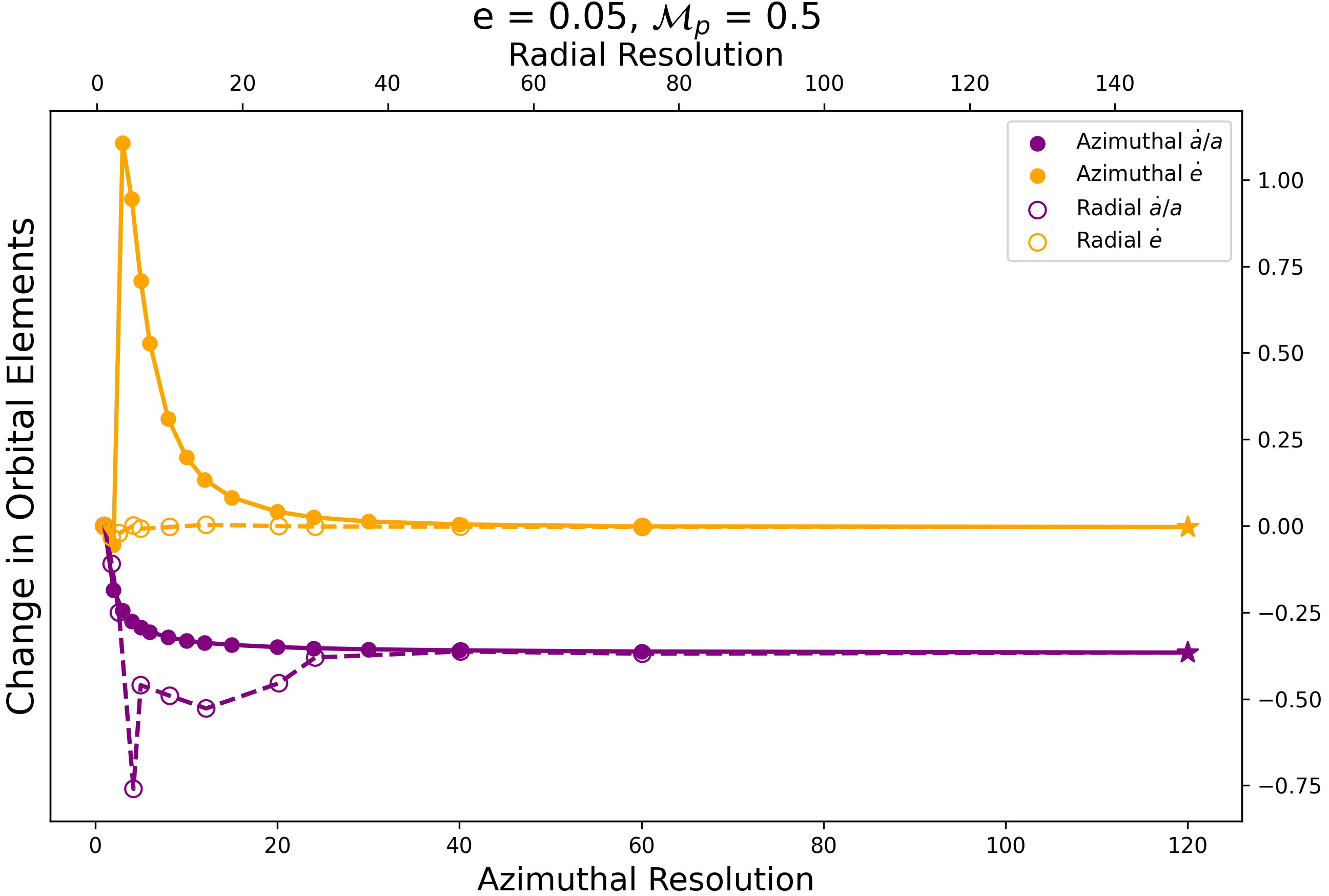}
\includegraphics[width=.31\textwidth]{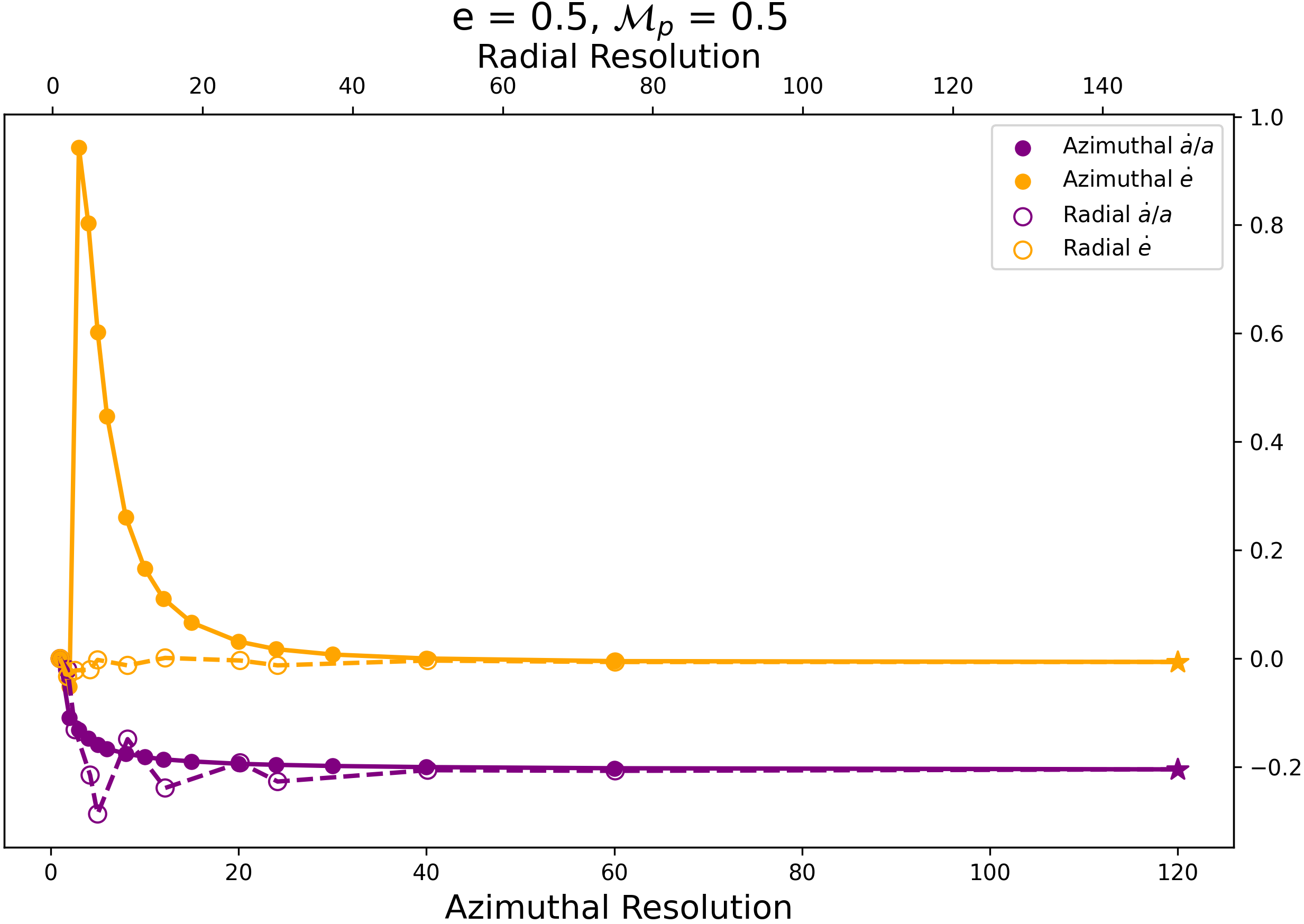}
\includegraphics[width=.313\textwidth]{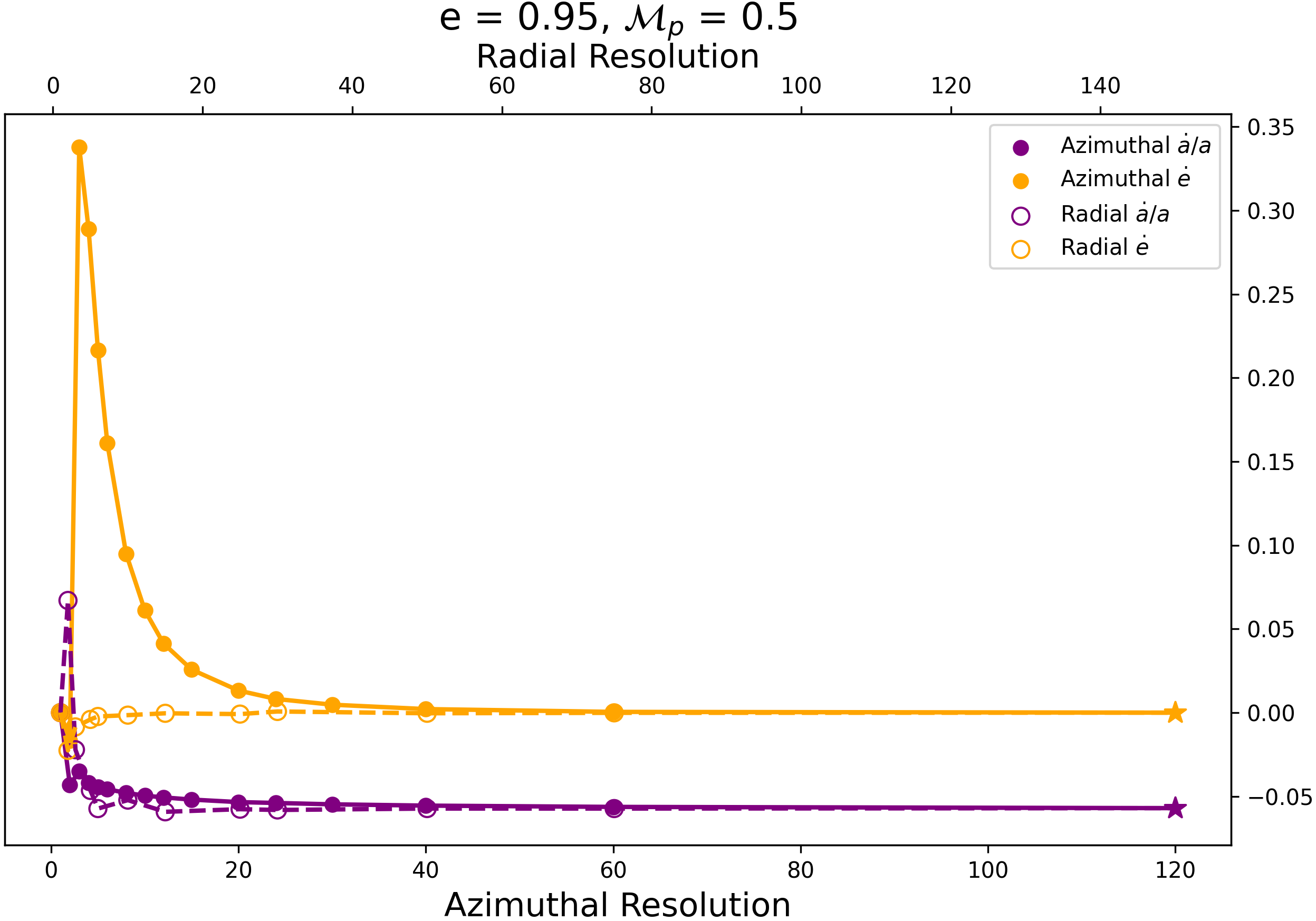}

\vspace{5mm}

\centering
\textbf{Equal Mass Binaries}
\vspace{4mm}

\includegraphics[width=.322\textwidth]{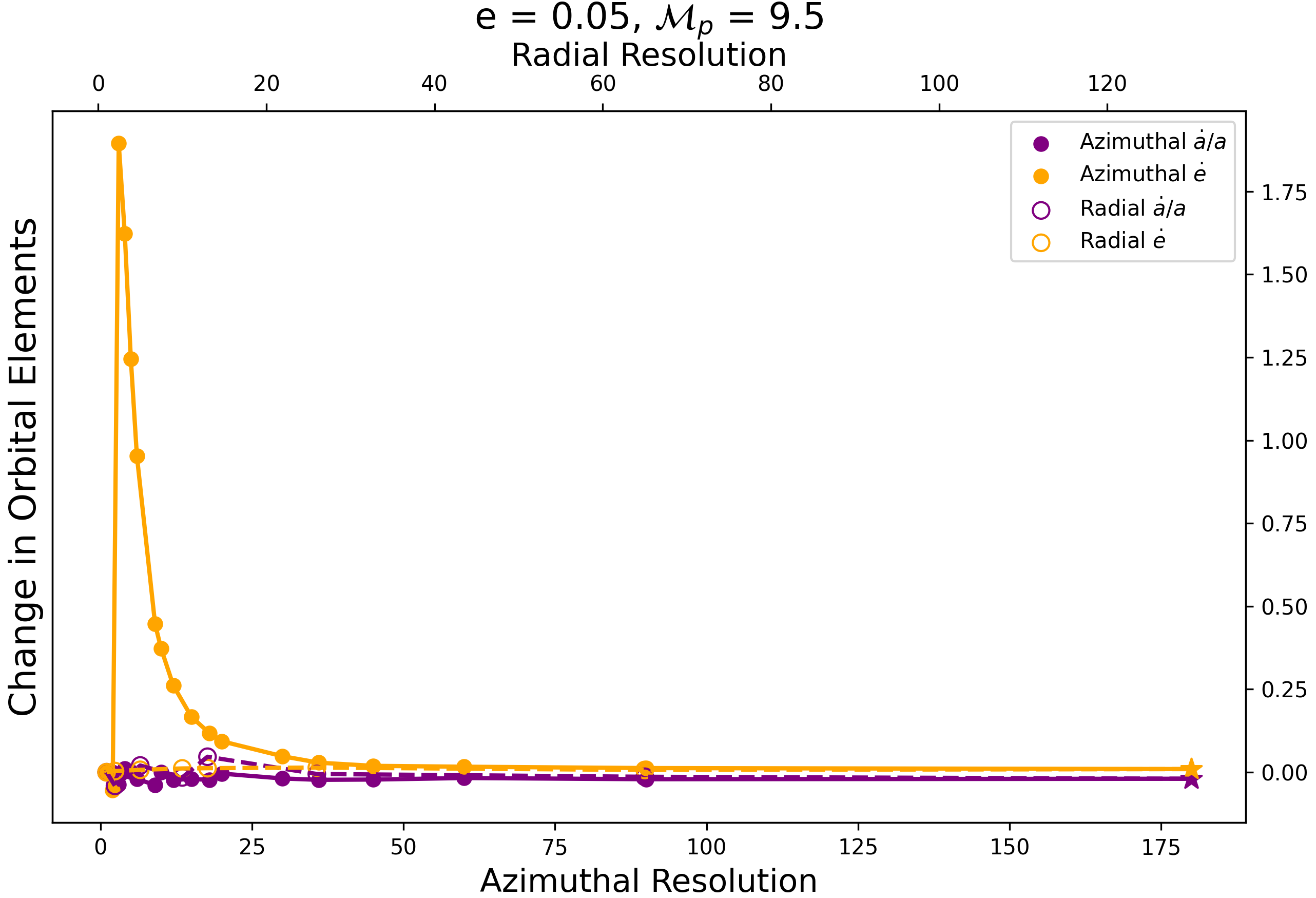}
\includegraphics[width=.316\textwidth]{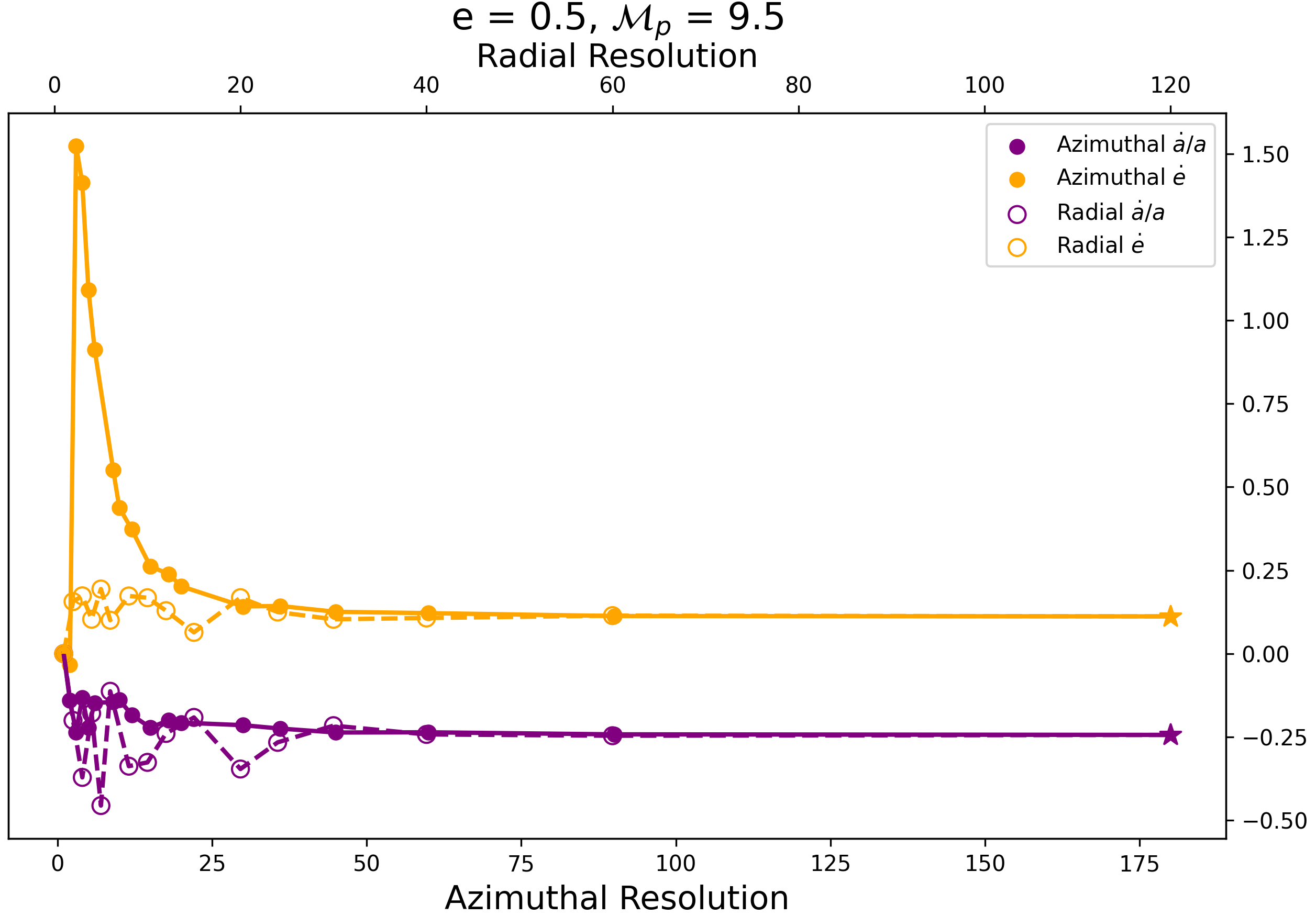}
\includegraphics[width=.314\textwidth]{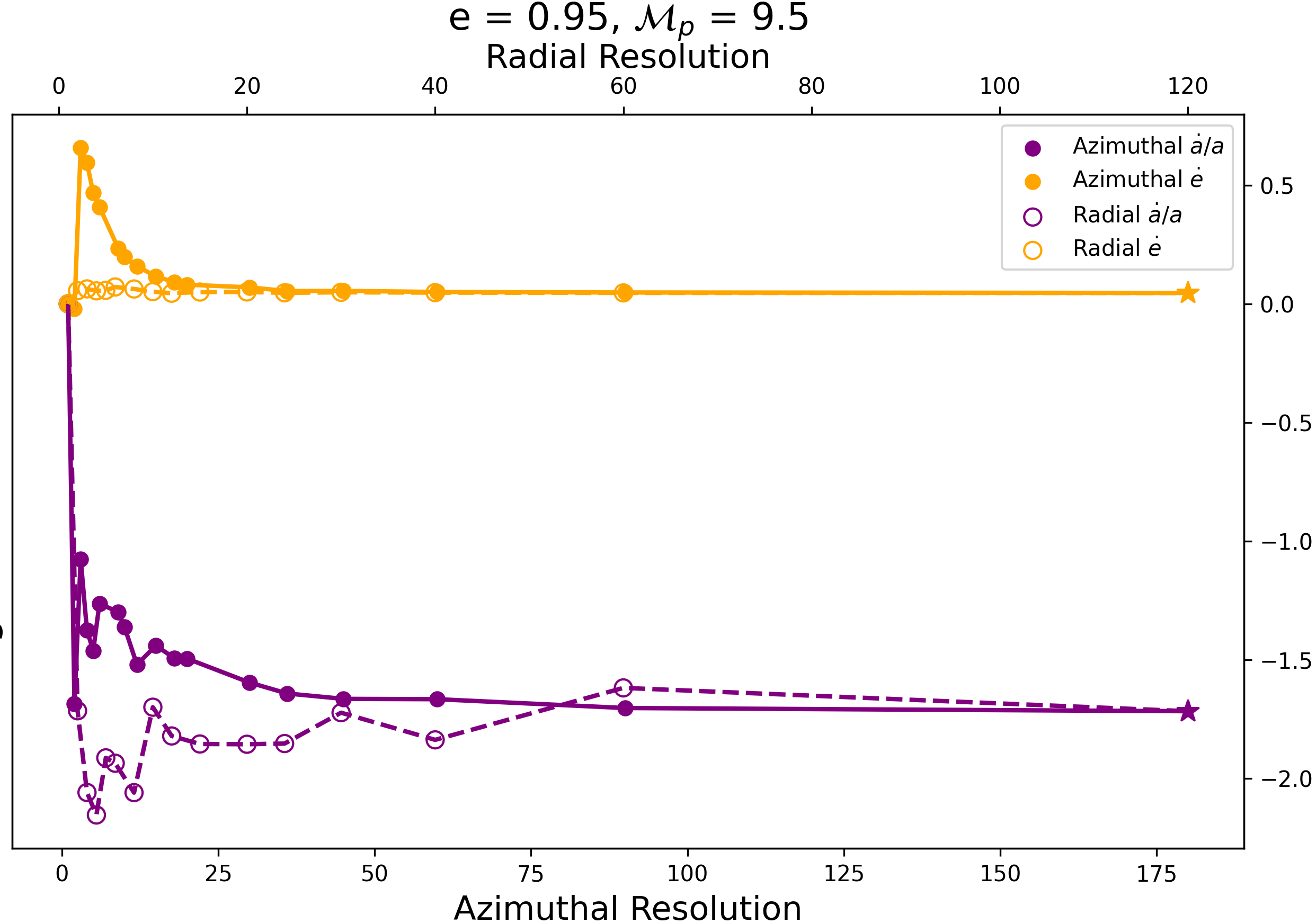}
\medskip

\hspace{-1mm}\includegraphics[width=.325\textwidth]{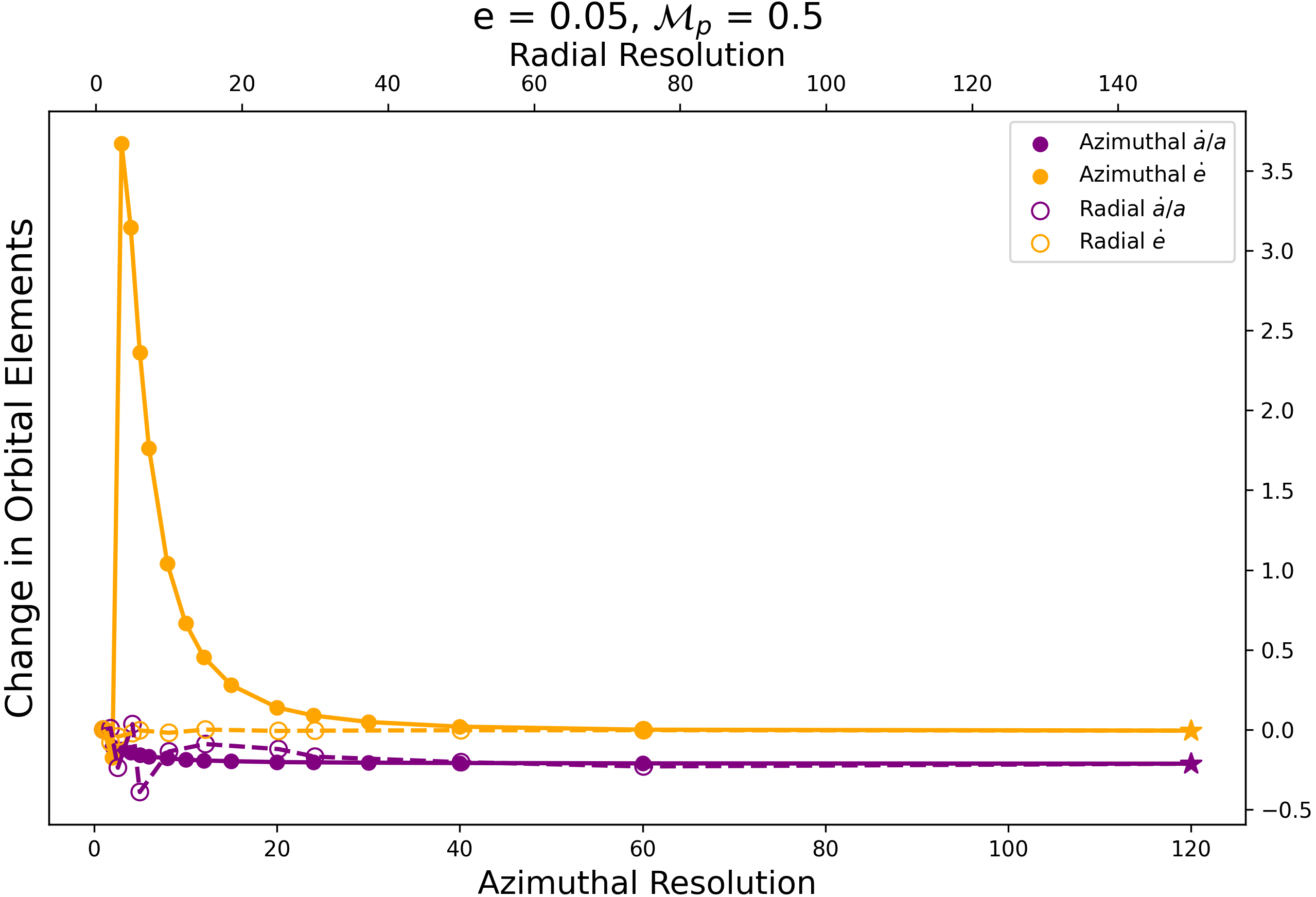}
\includegraphics[width=.317\textwidth]{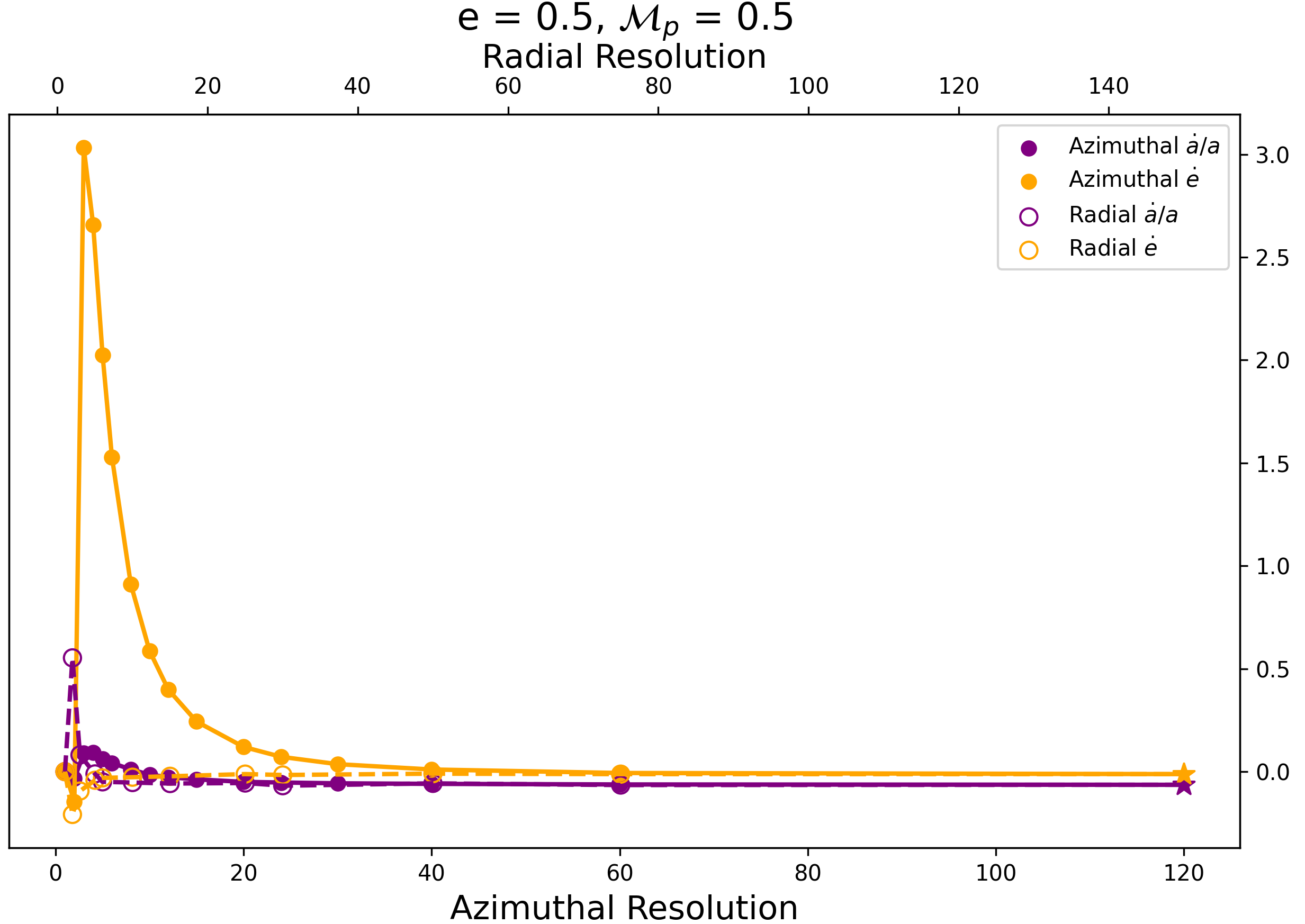}
\includegraphics[width=.314\textwidth]{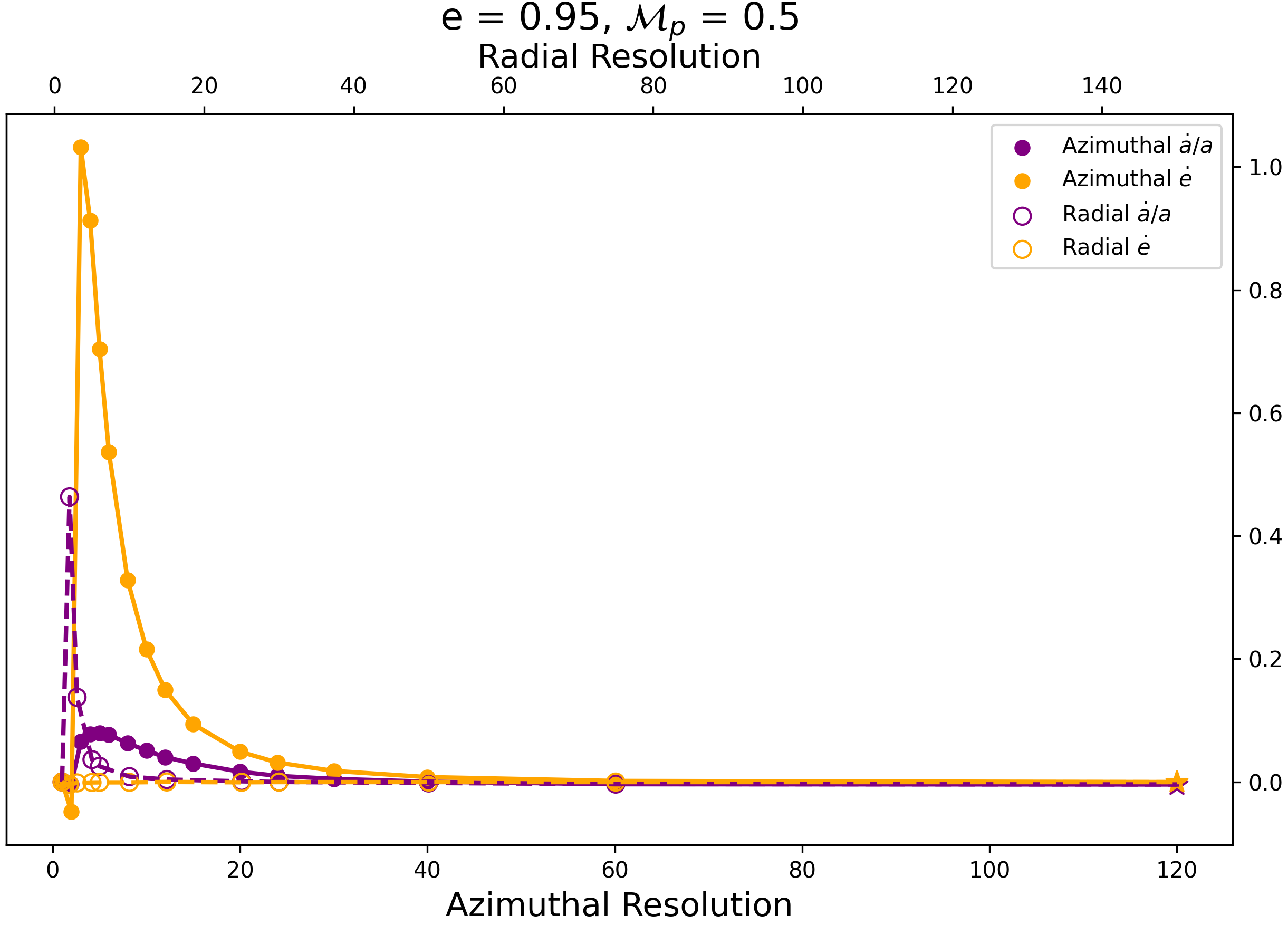}

\caption{Convergences of the orbital elements $\dot a /a, \dot e$ as functions of the number of samples taken in both the azimuthal and radial domain for a fixed box size. We show selected cases for both single orbits and equal mass binaries. Note that the spatial resolutions used in this study (stars) vary for different eccentricities and Mach numbers.}
\label{BoxConvergenceAz}
\end{figure}

\pagebreak

\subsection{Orbital Temporal Sampling}
\begin{figure}[h!]
    \centering
    \includegraphics[width = 0.8\textwidth]{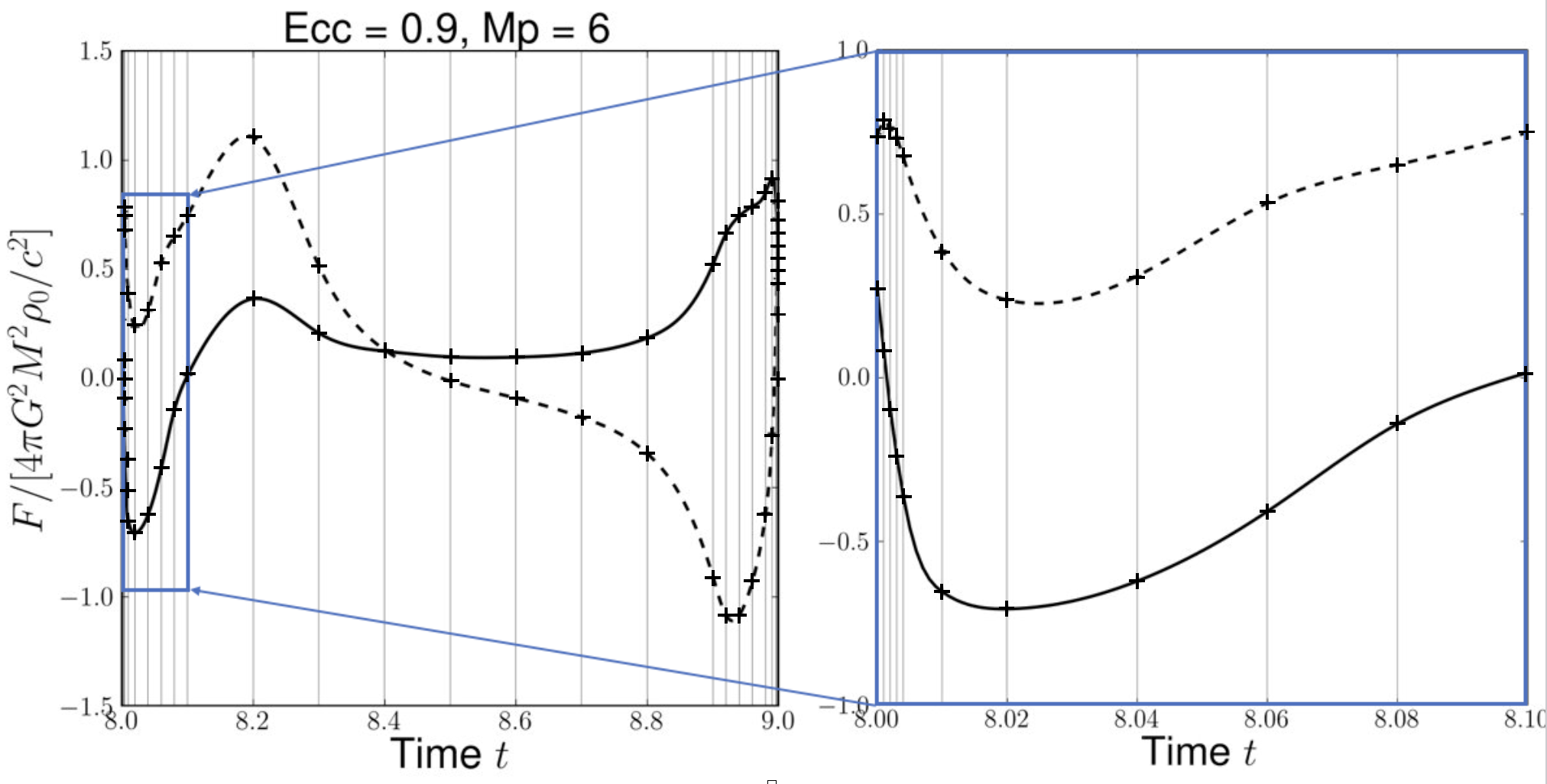}
    \caption{A plot of the interpolated force as a function of time for $e = 0.9, \mathcal{M}_p = 6$. The vertical gray lines are points in time where we evaluate the force and thus represent real data points. The sensitivity is evidently higher near pericenter enabling us to resolve the curve as it undergoes it's most rapid variation. In addition we note that our sampling is symmetric about apocenter.
    }
    \label{TimeConvergence}
\end{figure}

The number of points in time and at which points in the period we sample is also an important factor. This is typically a function of orbital eccentricity: near circular orbits have equal sampling throughout the orbit while eccentric ones are much denser sampled near pericenter. At each point in the parameter space we perform between $20$ and $30$ evaluations. In Fig.~\ref{TimeConvergence} we illustrate the sampling distribution for a highly eccentric orbit.

\section{Circular Trajectory}
\label{CircularAppendix}
\begin{figure}[!h]
    \centering
    \includegraphics[width = 0.4\linewidth]{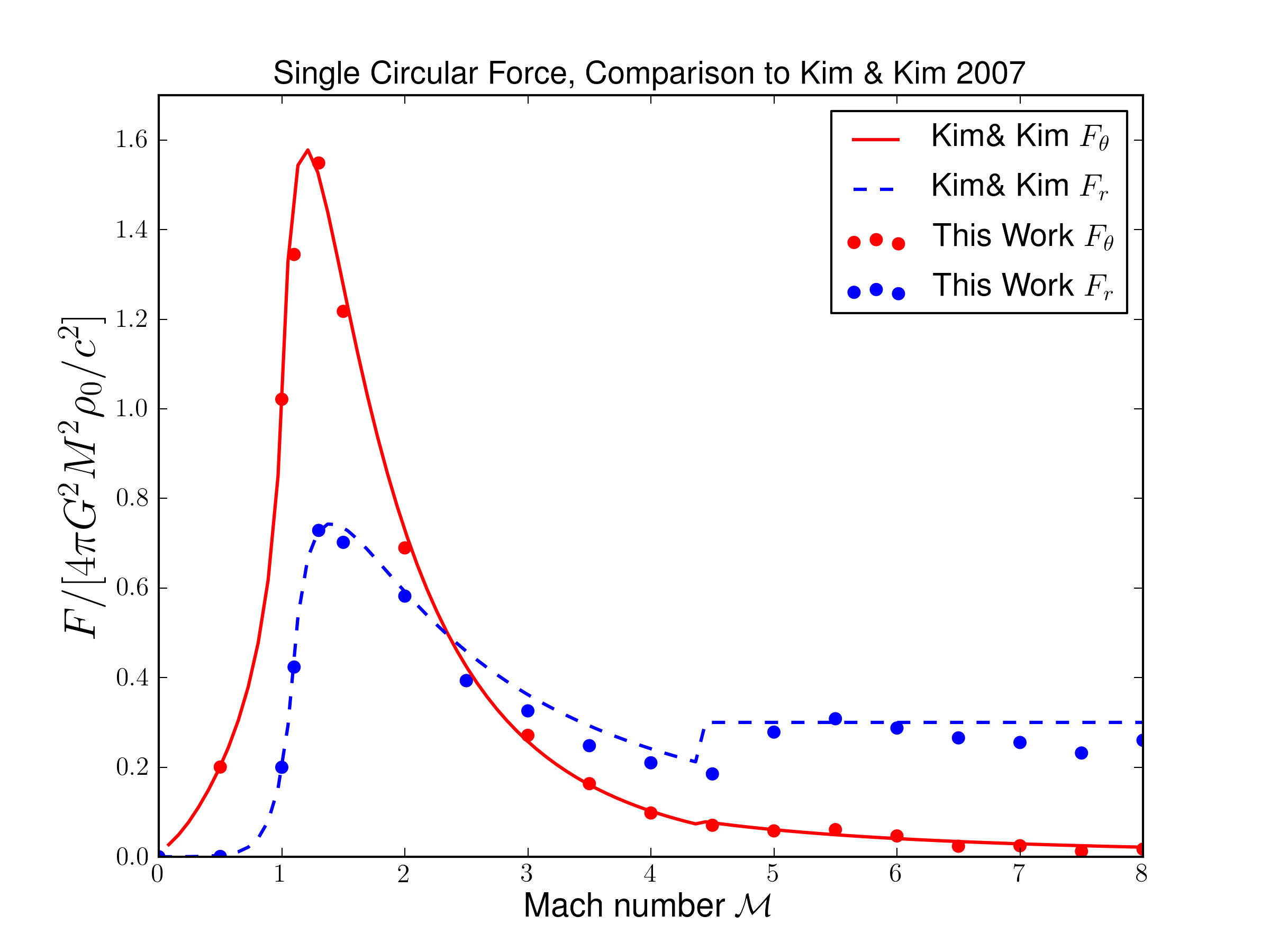}
    \includegraphics[width = 0.41\linewidth]{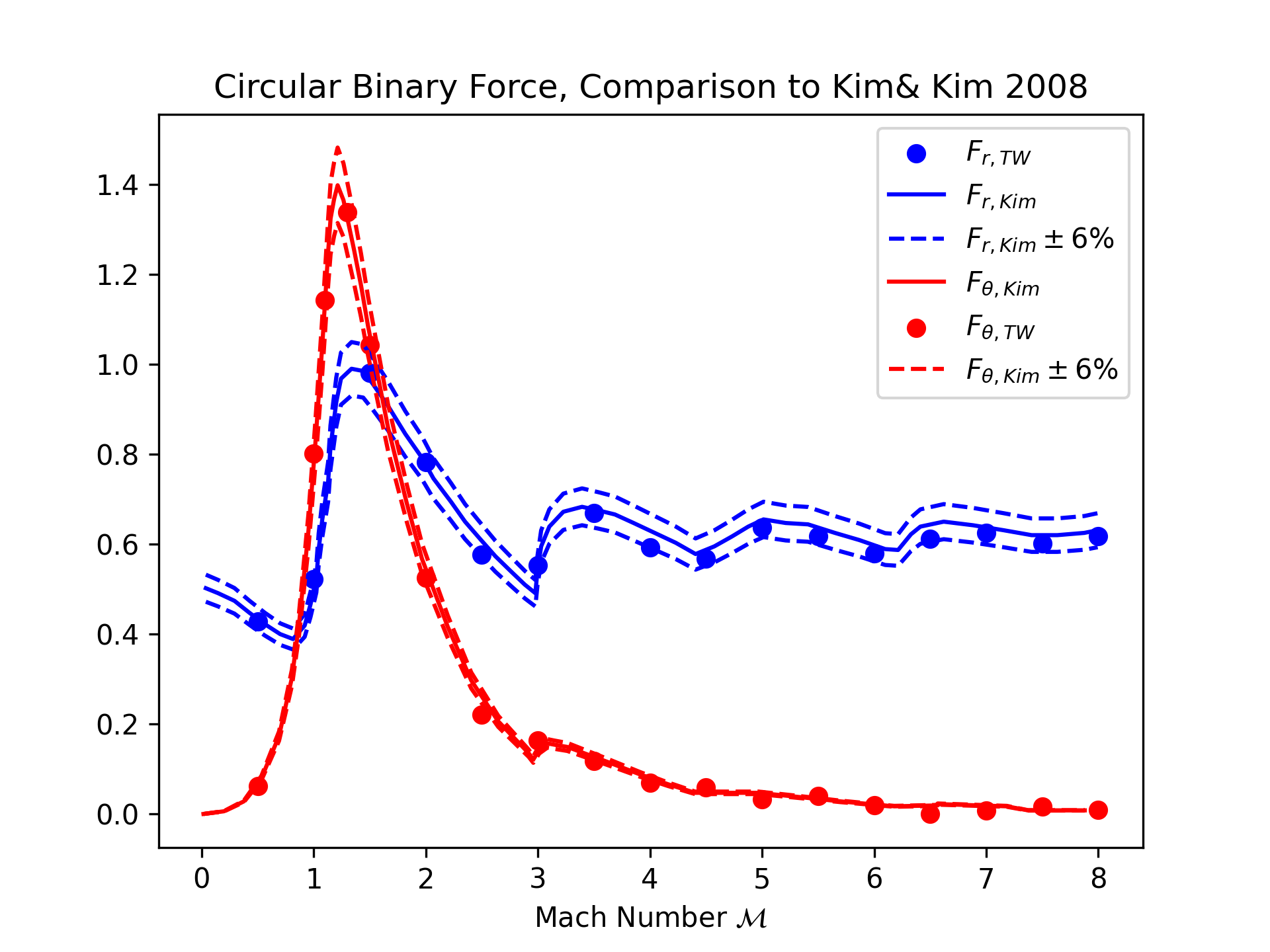}
    \caption{The gravitational drag force experienced by a single massive perturber (left) and a binary (right) on a circular orbit. We compare our data dots to the fitting functions (lines) provided by \citetalias{KK07}, \cite{DoublePerturbers}.}
    \label{CircularAgreement}
\end{figure}

We demonstrate the agreement of our results with preceding studies regarding a single perturber on a circular orbit \cite{KK07} and double perturbers in a circular binary \cite{DoublePerturbers}. On the left panel of Fig.~\ref{CircularAgreement}
 we compare our findings (scatter points) against Eqs.~(13, 14) in \cite{KK07} (solid and dashed lines) and observe excellent agreement up to $\Mach\sim 4.4$. Above this Mach number, however, there are some discrepancies in the radial force profile. Meanwhile, we find very good agreement with the drag force experienced by a binary system in Fig.~\ref{CircularAgreement}. This can be compared to the fitting functions provided in Eqs.~(5,6) in \cite{DoublePerturbers}. They state their fittings are within $6\%$ of their numerical results, so we include a $6\%$ error margin on our plot.
 
\end{document}